# IL CAMPO ELETTROMAGNETICO NEI CRISTALLI FOTONICI UNIDIMENSIONALI

## Studio alle frequenze ottiche tramite la teoria dei "Quasi Normal Modes"

**AUTORE**

*Alessandro Settimi*

UNIVERSITA' DEGLI STUDI DI ROMA
"LA SAPIENZA"
FACOLTA' DI INGEGNERIA ELETTRONICA

TESI DI LAUREA IN ELETTRONICA QUANTISTICA

Relatore

*Prof.ssa Concita Sibilia*

Anno Accademico 2001/2002

# *Indice*

## *Introduzione*

In questo lavoro di tesi, viene studiato il comportamento del campo elettromagnetico, alle frequenze ottiche, nei cristalli fotonici unidimensionali, tramite la teoria dei *"Quasi-Normal-Modes"* (QNM).

Recentemente, le strutture periodiche, ad una o più dimensioni, hanno richiamato l'attenzione nel mondo ottico ed ingegneristico, in quanto possono fornire la chiave per la realizzazione di dispositivi fotonici altamente miniaturizzati. La proprietà sostanziale delle strutture a band-gap fotonico (PBG) è l'esistenza di bande in frequenza permesse e proibite, e quindi con una capacità unica di filtrare e manipolare la luce.

I tipi più semplici di PBG sono unidimensionali (1D), strutture multistrato periodiche o quasi periodiche. Sono state proposte e messe a punto importanti applicazioni per un'ampia classe di dispositivi ottici lineari e non lineari ad una dimensione: un limitatore ottico non lineare [1], un diodo [2], un laser PBG [3], una linea di ritardo "true-time" [4], un amplificatore ottico parametrico a guadagno elevato per una conversione di frequenza non lineare [5], e, più recentemente, strutture metallo-dielettrico [6].

Nei primi lavori teorici, l'1D-PBG veniva approssimato con una struttura infinita; l'inconveniente era che la densità dei modi e.m. (DOM) diveniva formalmente infinita al bordo banda fotonico. Nei lavori successivi, l'1D-PBG veniva trattato come un numero finito di celle base, disposte periodicamente, così rimuovendo la singolarità della DOM al bordo banda; ma la metodologia di analisi era puramente numerica oppure, se analitica, riusciva a sviluppare il campo e.m. in un set di modi normali solo per i picchi di risonanza.

Il seguente lavoro di tesi prende spunto dalla teoria QNM, sviluppata di recente per cavità omogenee aperte solo da un lato. Viene sottolineato che ogni 1D-PBG è una cavità aperta da entrambi lati che consente un confinamento del campo e.m., ma che induce perdite di radiazione; il campo e.m. inizialmente presente al suo interno, allo scorrere del tempo, non può che propagarsi verso l'esterno. In generale, l'1D-PBG non è un sistema conservativo e l'evoluzione naturale del campo e.m. non può essere descritta da un operatore hermitiano: in definitiva, si rinuncia ad una trattazione in termini dei modi normali del campo di radiazione.

Il metodo dei QNM utilizza, come strumenti di analisi, la funzione di Green e gli spazi biortogonali. L'1D-PBG è studiato in modo più realistico: una struttura finita, immersa in uno spazio illimitato. La rinuncia alla conservazione dell'energia per il sistema in esame ha delle conseguenze importanti sia numeriche che concettuali: si passa da autofrequenze reali a quasi autofrequenze complesse. Inoltre, poiché l'operatore che descrive l'evoluzione del sistema nello spazio-tempo non è hermitiano, i modi non sono più normali, ma quasi normali: difatti, solo sotto opportune condizioni, questi non solo costituiscono una base ortogonale, ma la loro evoluzione spazio-temporale è analoga a quella hermitiana dei modi normali.

Nel seguente lavoro di tesi, si intende estendere la trattazione dei QNM, proposta in letteratura, alla descrizione del campo e.m. negli 1D-PBG, sia per incidenza normale che obliqua, distinguendo così il comportamento del campo alle polarizzazioni TE e TM.

Vengono determinate le quasi-autofrequenze con i quasi-modi corrispondenti per i 1D-PBG. Quindi, tramite la teoria dei QNM, si descrive l'evoluzione del campo e.m., in regime monocromatico, all'interno dell'1D-PBG e si caratterizza il campo tramite la densità dei modi; i risultati vengono confrontati con quelli ottenuti tramite metodi numerici.

# *Bibliografia*

# Capitolo 1. Cristalli fotonici unidimensionali.

## Paragrafo 1.1. Introduzione.

I cristalli fotonici (PC, da photonic crystals [1]) sono quelle strutture in cui il campo elettromagnetico ha alcune caratteristiche di propagazione assimilabili a quelle tipiche di un elettrone all'interno di un cristallo. Il problema della propagazione elettromagnetica è stato affrontato con gli strumenti concettuali e matematici tipici della fisica dei solidi [2]. Tuttavia, in un cristallo, le dimensioni della cella elementare sono tali da giustificare l'ipotesi che qualsiasi struttura reale contenga un numero in pratica infinito di periodi, mentre, nel caso delle strutture periodiche, le dimensioni sono paragonabili alla lunghezza d'onda della radiazione che si sta considerando e quindi, per dispositivi con dimensioni submillimetriche, il numero di periodi è così basso da non poter assolutamente considerare la struttura infinita.

I 'dispositivi fotonici' [3] offrono una serie di vantaggi rispetto a quelli elettronici convenzionali, ad esempio tempo di vita e velocità di operazione maggiore, dimensioni ridotte e robustezza alle variazioni dovute a fluttuazioni di temperatura o altri cambiamenti ambientali.

In generale i PC sono strutture in cui la periodicità può aver luogo in tutte le direzioni spaziali. Qui si illustra lo studio dei PC unidimensionali.

## Paragrafo 1.1. Teoria di Bloch.

Una struttura multisrato periodica e infinita rappresenta un modello fisico ideale nel quale una cella di base è ripetuta infinite volte. Si considera una tale struttura unidimensionale in cui la cella base è a sua volta costituita dalla giustapposizione di due o più materiali e gode della proprietà d'invarianza traslazionale; un osservatore che la percorre lungo la direzione di stratificazione, spostandosi per passi pari alla lunghezza del singolo periodo (o suoi multipli interi), non avvertirà alcun cambiamento nelle grandezze fisiche d'interesse: questa proprietà è la caratteristica peculiare dei cristalli, quindi è naturale usare le nozioni di fisica dello stato solido [4].

Le proprietà ottiche di un qualsiasi mezzo isotropo e non magnetico sono descritte dalla sua costante dielettrica $\varepsilon(\vec{r})$.

Se il mezzo è periodico, allora:

$$\varepsilon(\vec{r}) = \varepsilon(\vec{r} + \vec{a}) \qquad (1.2.1)$$

dove si è indicato con $\vec{a}$ un arbitrario vettore di reticolo.

La propagazione di un'onda monocromatica è descritta dalle equazioni di Maxwell:

$$\begin{cases} \nabla \times \vec{E} = -i\omega\mu_0 \vec{H} \\ \nabla \times \vec{H} = i\omega\varepsilon \vec{E} \end{cases} \qquad (1.2.2)$$

Si enuncia il teorema di Bloch. Le soluzioni delle equazioni (1.2.2) devono rispettare la simmetria traslazionale del mezzo. Quindi:

$$\begin{cases} \vec{E} = \vec{E}_{\vec{k}}(\vec{r}) e^{-i\vec{k}\cdot\vec{r}} \\ \vec{H} = \vec{H}_{\vec{k}}(\vec{r}) e^{-i\vec{k}\cdot\vec{r}} \end{cases} \qquad (1.2.3)$$

dove:

$$\begin{cases} \vec{E}_{\vec{k}}(\vec{r}) = \vec{E}_{\vec{k}}(\vec{r} + \vec{a}) \\ \vec{H}_{\vec{k}}(\vec{r}) = \vec{H}_{\vec{k}}(\vec{r} + \vec{a}) \end{cases} \qquad (1.2.4)$$

Il vettore $\vec{k}$, detto vettore d'onda di Bloch, è legato alla pulsazione $\omega$ tramite un'opportuna legge di dispersione:

$$\omega = \omega(\vec{k}) \qquad (1.2.5)$$

Si dimostra il teorema di Bloch. L'equazione delle onde è data da:

$$\nabla \times \nabla \times \vec{E} - \omega^2 \varepsilon \mu_0 \vec{E} = 0 \qquad (1.2.6)$$

Per la proprietà di periodicità spaziale (1.2.1), la costante dielettrica può essere sviluppata in serie di Fourier:

$$\varepsilon(\vec{r}) = \sum_{\vec{G}} \varepsilon_{\vec{G}} e^{-i\vec{G}\cdot\vec{r}} \qquad (1.2.7)$$

dove la sommatoria si estende a tutti i vettori del reticolo reciproco $\vec{G}$.

Il campo elettrico può essere scritto sotto forma d'integrale di Fourier come:

$$\vec{E} = \int A(\vec{k}) e^{-i\vec{k}\cdot\vec{x}} d^3k \qquad (1.2.8)$$

Sostituendo nella (1.2.6) le (1.2.7) e (1.2.8) si ottiene un'equazione che è soddisfatta quando, per ogni $\vec{k}$, vale:

$$\vec{k} \times \left[\vec{k} \times \vec{A}(\vec{k})\right] + \omega^2 \mu_0 \sum_{\vec{G}} \varepsilon_{\vec{G}} \vec{A}(\vec{k} - \vec{G}) = 0 \qquad (1.2.9)$$

La (1.2.9) è un set infinito di equazioni omogenee nei coefficienti $\vec{A}(\vec{k})$, ognuna con un diverso valore di $\vec{k}$. In linea di principio, si potrebbe risolvere l'equazione secolare associata alla (1.2.9) uguagliandone a zero il determinante; si nota però che i coefficienti $\vec{A}(\vec{k})$ sono accoppiati, più precisamente sono accoppiati quelli della forma $\vec{A}(\vec{k} - \vec{G})$. Possiamo dividere l'intero sistema di equazioni (1.2.9) in molti sottosistemi, uno per ogni valore di $\vec{k}$; ciascun sottosistema conterrà, a sua volta, equazioni in $\vec{A}(\vec{k})$ e $\vec{A}(\vec{k} - \vec{G})$ per tutti i possibili valori di $\vec{G}$.

I singoli sottosistemi sono risolvibili separatamente e la soluzione sarà data da:

$$\vec{E}_{\vec{k}} = e^{-i\vec{k}\cdot\vec{r}} \sum_{\vec{G}} \vec{A}(\vec{k} - \vec{G}) e^{i\vec{G}\cdot\vec{r}} = e^{-i\vec{k}\cdot\vec{r}} \vec{E}_{\vec{k}}(\vec{r}) \qquad (1.2.10)$$

L'equazione (1.2.9) è la legge di dispersione cui si è accennato all'inizio di questo paragrafo; si può ricavare $k_x$, purché siano dati la frequenza $\omega$ e $k_y$, $k_z$: si deduce l'esistenza delle bande proibite, intervalli di valori di $\omega$ per i quali $k_x$ diventa un numero complesso e, di conseguenza, l'onda di Bloch diventa evanescente. Una radiazione incidente che abbia frequenza complessa nella banda non si propaga nel mezzo ma è totalmente riflessa: questo fenomeno è anche conosciuto come riflessione di Bragg.

Ora si considera un mezzo periodico unidimensionale.

La costante dielettrica è tale che:

$$\varepsilon(x) = \varepsilon(x + nd) \qquad (1.2.11)$$

dove $d$ è il periodo e $n$ è un intero.

La radiazione incidente sul mezzo, con angolo $\theta$, sarà parzialmente riflessa e rifratta ad ogni interfaccia. Si avrà interferenza costruttiva in riflessione solo quando è soddisfatta la condizione di Bragg:

$$md = 2d\cos\theta \qquad (1.2.12)$$

L'insieme dei vettori del reticolo reciproco diventa:

$$\vec{G} = n\vec{g} = n\frac{2\pi}{d}\hat{x} \qquad (1.2.13)$$

Lo sviluppo in serie di Fourier della costante dielettrica si riesprime come:

$$\varepsilon(x) = \sum_n \varepsilon_n e^{-ingx} = \sum_n \varepsilon_n e^{-in\frac{2\pi}{d}x} \qquad (1.2.14)$$

Nel caso unidimensionale il teorema di Bloch dice:

$$E_k = e^{-ikx}\sum_n A(k - n\frac{2\pi}{d})e^{in\frac{2\pi}{d}x} = e^{-ikx}E_k(x) \qquad (1.2.15)$$

Questa soluzione costituisce un modo normale per la propagazione del campo elettromagnetico all'interno del mezzo: è evidente l'invarianza di $E_k(x)$ per traslazione del passo reticolare *d*.

Il mezzo è omogeneo nelle direzioni *y* e *z*, quindi i modi di Bloch del campo elettrico diventano:

$$E_k = e^{-i(k_z z + k_y y)}e^{-ik_x x}E_k(x) \qquad (1.2.16)$$

dove $E_k(x)$ è una funzione periodica in *x*.

Nel caso unidimensionale si può osservare che vi è una banda proibita associata ad ogni coefficiente dello sviluppo di Fourier $\varepsilon_n$ relativo alla costante dielettrica $\varepsilon(x)$. Per $n = 1$ si ha la prima banda, per $n \neq 1$ si hanno le altre, di ordine superiore che sono relative a frequenze più alte e, in molti casi, sono più piccole rispetto alla prima banda.

## *Paragrafo 1.3. Elementi di propagazione e.m.*

In questo paragrafo, sono introdotti degli strumenti analitici per lo studio del problema elettromagnetico nelle strutture stratificate unidimensionali.

### *1.3.1. Onde piane e monocromatiche TE e TM.*

Si considera un'onda piana monocromatica che si propaga lungo un mezzo stratificato. Qualsiasi onda piana arbitrariamente polarizzata è la sovrapposizione di due onde indipendenti, una TE e l'altra TM. Nel caso di onda polarizzata linearmente, si parla di onda TE (trasversa elettrica) quando il campo elettrico è perpendicolare al piano di incidenza e di onda TM (trasversa magnetica) quando il campo magnetico è perpendicolare al piano di incidenza. Le equazioni di Maxwell rimangono invariate quando si scambiano simultaneamente $E$ con $H$ ed $\varepsilon$ con $\mu$; segue che qualsiasi teorema relativo alle onde TM è dedotto dal risultato corrispondente per le onde TE tramite questo scambio [5]. Inoltre qualsiasi onda è la sovrapposizione di onde monocromatiche; per una data pulsazione $\omega$, la dipendenza del campo dal tempo sta nel fattore $e^{-i\omega\tau}$. Il mezzo sia isotropo non magnetico, quindi le sue proprietà elettrodinamiche sono contenute nella costante dielettrica $\varepsilon(\vec{r},\omega)$ e nella permeabilità magnetica $\mu(\vec{r},\omega)$ [6].

Si suppone che il piano di incidenza sia il piano $yz$ e che la direzione di stratificazione sia la direzione $z$.

a)TE

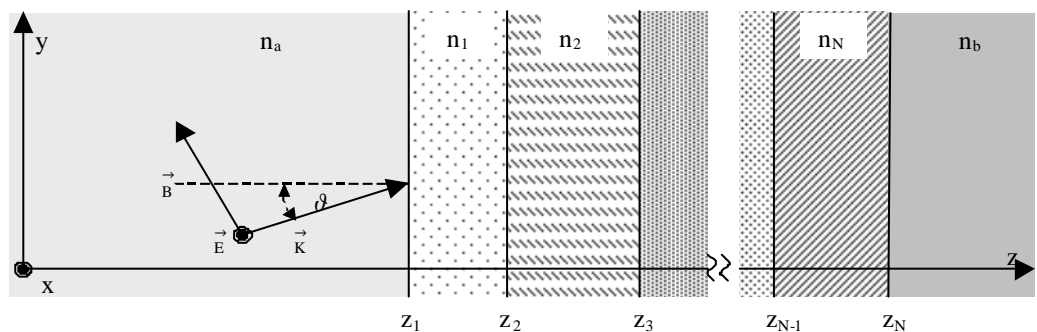

b)TM

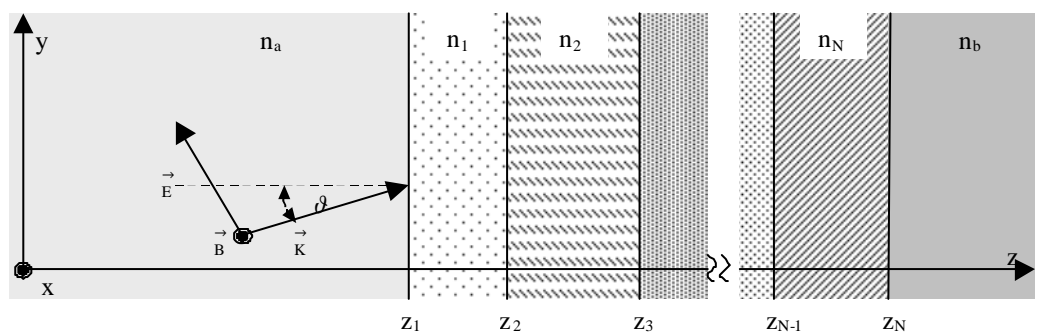

Figura 1.1: a) onda elettromagnetica con polarizzazione TE che incide su una struttura non omogenea con indice di rifrazione costante a tratti;
b) onda elettromagnetica con polarizzazione TM che incide sulla stessa struttura.

Per un'onda TE ($E_y = E_z = 0$) monocromatica, con pulsazione $\omega$, le equazioni di Maxwell si riducono alle seguenti sei equazioni scalari:

$$\frac{\partial H_z}{\partial y} - \frac{\partial H_y}{\partial z} + \frac{i\omega\varepsilon}{c} E_x = 0 \tag{1.3.1.1}$$

$$\frac{\partial H_x}{\partial z} - \frac{\partial H_z}{\partial x} = 0 \tag{1.3.1.2}$$

$$\frac{\partial H_y}{\partial x} - \frac{\partial H_x}{\partial y} = 0 \tag{1.3.1.3}$$

$$\frac{i\omega\mu}{c} H_x = 0 \tag{1.3.1.4}$$

$$\frac{\partial E_x}{\partial z} - \frac{i\omega\mu}{c} H_y = 0 \tag{1.3.1.5}$$

$$\frac{\partial E_x}{\partial y} + \frac{i\omega\mu}{c} H_z = 0 \tag{1.3.1.6}$$

Queste equazioni mostrano che $H_y$, $H_z$, e $E_x$ sono funzioni solo di $y$ e $z$. Se si eliminano $H_y$ e $H_z$ tra le (1.3.1.1), (1.3.1.4), (1.3.1.6) si perviene alla:

$$\frac{\partial^2 E_x}{\partial y^2} + \frac{\partial^2 E_x}{\partial z^2} + n^2 k_0^2 E_x = \frac{d(\log\mu)}{dz}\frac{\partial E_x}{\partial z} \tag{1.3.1.7}$$

dove:

$$n^2 = \mu\varepsilon \tag{1.3.1.8}$$

$$k_0 = \frac{\omega}{c} = \frac{2\pi}{\lambda_0} \tag{1.3.1.9}$$

Per risolvere l'equazione (1.3.1.7), si prende come soluzione il prodotto di due funzioni, una che dipende solo da $y$ e l'altra solo da $z$:

$$E_x(y,z) = Y(y)E(z) \tag{1.3.1.10}$$

L'equazione (1.3.1.7) diventa:

$$\frac{1}{Y}\frac{d^2 Y}{dy^2} = -\frac{1}{E}\frac{d^2 E}{dz^2} + \frac{d(\log\mu)}{dz}\frac{1}{E}\frac{dE}{dz} - n^2 k_0^2 \tag{1.3.1.11}$$

Se si osserva, il termine alla sinistra è una funzione solo di $y$ mentre i termini alla destra dipendono solo di $z$.

La (1.3.1.11) vale se ciascun membro è uguale ad una costante:

$$\frac{1}{Y}\frac{d^2 Y}{dy^2} = -K^2 \tag{1.3.1.12}$$

$$\frac{1}{E}\frac{d^2 E}{dz^2} - \frac{d(\log \mu)}{dz}\frac{1}{E}\frac{dE}{dz} + n^2 k_0^2 = K^2 \tag{1.3.1.13}$$

dove conviene porre:

$$K^2 = k_0^2 \alpha^2 \tag{1.3.1.14}$$

Se si risolve la (1.3.1.12) si ottiene:

$$Y(y) = C e^{i k_0 \alpha y} \tag{1.3.1.15}$$

Il campo elettrico ha la forma:

$$E_x = E(z) e^{i k_0 \alpha y} \tag{1.3.1.16}$$

dove $E(z)$ è una funzione che soddisfa la:

$$\frac{d^2 E}{dz^2} - \frac{d(\log \mu)}{dz}\frac{dE}{dz} + k_0^2(n^2 - \alpha^2)E = 0 \tag{1.3.1.17}$$

Secondo la regola di sostituzione, che è una conseguenza della simmetria delle equazioni di Maxwell, segue immediatamente che, per un'onda TM ($H_y = H_z = 0$), il campo magnetico ha la forma:

$$H_x = H(z) e^{i k_0 \alpha y} \tag{1.3.1.18}$$

dove $H(z)$ è una funzione che soddisfa la:

$$\frac{d^2 H}{dz^2} - \frac{d(\log \varepsilon)}{dz}\frac{dH}{dz} + k_0^2(n^2 - \alpha^2)H = 0 \tag{1.3.1.19}$$

Si può porre per la permeabilità magnetica $\mu \cong \mu_0$ e quindi per l'induzione magnetica $B \cong \mu_0 H$.

Allora [6], l'equazione (1.3.1.17) si semplifica nella:

$$\frac{d^2 E}{dz^2} + q^2(z) E(z) = 0 \tag{1.3.1.20}$$

dove $q(z)$ è la componente del vettore d'onda lungo la direzione $z$ ed è tale che:

$$q^2(z) = k_0^2(n^2(z) - \alpha^2) \tag{1.3.1.21}$$

Inoltre, l'equazione (1.3.1.19), dopo opportune manipolazioni, diviene:

$$\frac{d}{dz}\left(\frac{1}{n^2}\frac{dB}{dz}\right) + \left(\frac{q(z)}{n(z)}\right)^2 B(z) = 0 \tag{1.3.1.22}$$

Si indicano genericamente con $U(z)$ le funzioni $E(z)$ e $B(z)$; la $U(z)$ è una funzione complessa:

$$U(z) = |U(z)| e^{i\phi(z)} \qquad (1.3.1.23)$$

Le superfici di ampiezza costante per il campo sono date da:

$$|U(z)| = cost \qquad (1.3.1.24)$$

mentre le superficie di fase costante hanno equazione:

$$\varphi(z) + k_0 \alpha y = cost \qquad (1.3.1.25)$$

In generale i due insiemi di superfici non coincidono cosicché il campo è un'onda disomogenea. Se si differenzia la (1.3.1.25), si ottiene:

$$\phi'(z)dz + k_0 \alpha dy = 0 \qquad (1.3.1.26)$$

Se si indica con $\theta(z)$ l'angolo che la normale ad una superficie di fase forma con l'asse *z*, allora:

$$\tan \theta(z) = -\frac{dz}{dy} = \frac{k_0 \alpha}{\phi'(z)} \qquad (1.3.1.27)$$

Si suppone che l'onda oltre che piana sia anche omogenea; vale la legge di Snell generalizzata per mezzi stratificati:

$$\alpha = n_0 \cdot \sin \theta_0 = cost \qquad (1.3.1.28)$$

Dall'equazione (1.3.1.27) si ottiene:

$$\phi(z) = k_0 z n_0 \cos \theta_0 \qquad (1.3.1.29)$$

### *1.3.2. Matrice di trasferimento.*

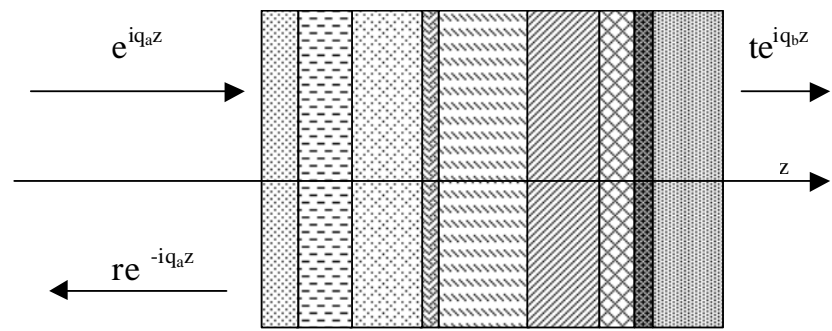

Figura 1.2. Schema delle onde (separate in componente propagante e contropropagante) in ingresso ed in uscita da una generica struttura multistrato.

In questo paragrafo si illustra un metodo per determinare la matrice di trasferimento che permette di calcolare il campo ad un'interfaccia se si conosce il valore del campo all'interfaccia precedente; il metodo sarà utile per il calcolo del coefficiente di trasmissione *t* e di riflessione *r* [6].

E' necessario distinguere tra le due polarizzazioni del campo elettromagnetico. Dapprima si prende in considerazione la polarizzazione TE.

L'equazione differenziale del secondo ordine (1.3.1.20) può essere spezzata in una coppia di equazioni differenziali accoppiate del primo ordine:

$$\begin{cases} \dfrac{dE}{dz} = D(z) \\ \dfrac{dD}{dz} = -q(z)^2 E(z) \end{cases} \qquad (1.3.2.1)$$

Se $q(z)$ assume il valore di $q_n$ in un intervallo dell'ascissa di stratificazione pari a $(z_n, z_{n+1})$, $E_n$ e $D_n$ sono i valori noti del campo e della sua derivata in $z_n$, la soluzione delle (1.3.2.1) in $z_n \leq z \leq z_{n+1}$ è:

$$\begin{cases} E(z) = E_n \cos[q_n(z - z_n)] + \dfrac{D_n}{q_n} \sin[q_n(z - z_n)] \\ D(z) = D_n \cos[q_n(z - z_n)] + E_n q_n \sin[q_n(z - z_n)] \end{cases} \qquad (1.3.2.2)$$

Poiché sia $D(z)$ che $E(z)$ sono continui in $z_{n+1}$, (la continuità segue dall'equazione differenziale scritta per il campo $E(z)$, poiché $q(z)$ non presenta singolarità), è possibile ricavare il campo e la sua derivata all'interfaccia immediatamente successiva:

$$\begin{pmatrix} E_{n+1} \\ D_{n+1} \end{pmatrix} = \begin{pmatrix} \cos \delta_n & \dfrac{\sin \delta_n}{q_n} \\ -q_n \sin \delta_n & \cos \delta_n \end{pmatrix} \begin{pmatrix} E_n \\ D_n \end{pmatrix} = M_n \begin{pmatrix} E_n \\ D_n \end{pmatrix} \qquad (1.3.2.3)$$

dove:

$$\delta_n = q_n(z_{n+1} - z_n) \qquad (1.3.2.4)$$

è l'incremento di fase dovuto alla propagazione del campo da $z_n$ a $z_{n+1}$.

La matrice $M_n$ ha determinante pari a uno.

In modo analogo si procede per la polarizzazione TM, considerando, per ragioni di continuità, il campo induzione magnetica *B* e la sua derivata *C* lungo la direzione *z*.

In questo caso:

$$\begin{pmatrix} B_{n+1} \\ C_{n+1} \end{pmatrix} = \begin{pmatrix} \cos\delta_n & \dfrac{\sin\delta_n}{Q_n} \\ -Q_n \sin\delta_n & \cos\delta_n \end{pmatrix} \begin{pmatrix} B_n \\ C_n \end{pmatrix} = M_n \begin{pmatrix} B_n \\ C_n \end{pmatrix} \qquad (1.3.2.5)$$

dove $Q_n = q_n / \varepsilon_n$.

Anche qui la matrice $M_n$ ha modulo unitario.

Un sistema a $N$ strati è caratterizzato da una matrice quadrata:

$$M = \begin{pmatrix} m_{11} & m_{12} \\ m_{21} & m_{22} \end{pmatrix} = M_N M_{N-1} \ldots M_n \ldots M_2 M_1 \qquad (1.3.2.6)$$

che cambia secondo la polarizzazione che si considera TE o TM.

Si fa riferimento alla polarizzazione TE. Si definiscono le ampiezze dei campi in ingresso (*a*) e in uscita (*b*) dalla struttura:

$$E_1 = e^{i\alpha} + re^{-i\alpha} \quad , \quad D_1 = iq_a\left(e^{i\alpha} - re^{-i\alpha}\right) \qquad (1.3.2.7)$$

$$E_{N+1} = te^{i\beta} \quad , \quad D_{N+1} = iq_b te^{i\beta} \qquad (1.3.2.8)$$

con $\alpha = q_a z_1$ e $\beta = q_b z_{N+1}$.

Si trova che:

$$\begin{pmatrix} te^{i\beta} \\ iq_b te^{i\beta} \end{pmatrix} = \begin{pmatrix} m_{11} & m_{12} \\ m_{21} & m_{22} \end{pmatrix} \begin{pmatrix} e^{i\alpha} + re^{-i\alpha} \\ iq_a\left(e^{i\alpha} - re^{-i\alpha}\right) \end{pmatrix} \qquad (1.3.2.9)$$

Si ricavano i coefficienti:

$$r = e^{2i\alpha} \frac{q_a q_b m_{12} + m_{21} - iq_b m_{11} + iq_a m_{22}}{q_a q_b m_{12} - m_{21} + iq_b m_{11} + iq_a m_{22}} \qquad (1.3.2.10)$$

$$t = e^{i(\alpha-\beta)} \frac{2iq_a}{q_a q_b m_{12} - m_{21} + iq_b m_{11} + iq_a m_{22}} \qquad (1.3.2.11)$$

e anche la riflettanza e la trasmittanza; quando gli elementi di matrice e $q_a$, $q_b$ sono reali:

$$R = |r|^2 = \frac{(q_a q_b m_{12} + m_{21})^2 + (q_b m_{11} - q_a m_{22})^2}{(q_a q_b m_{12} - m_{21})^2 + (q_b m_{11} + q_a m_{22})^2} \qquad (1.3.2.12)$$

$$T = \frac{q_b}{q_a}|t|^2 = \frac{4q_a q_b}{(q_a q_b m_{12} - m_{21})^2 + (q_b m_{11} + q_a m_{22})^2} \qquad (1.3.2.13)$$

Per il caso TM si procede allo stesso modo e si ottengono i coefficienti di riflessione e trasmissione:

$$-r = e^{2i\alpha} \frac{Q_a Q_b m_{12} + m_{21} - iQ_b m_{11} + iQ_a m_{22}}{Q_a Q_b m_{12} - m_{21} + iQ_b m_{11} + iQ_a m_{22}} \qquad (1.3.2.14)$$

$$\left(\frac{\varepsilon_b}{\varepsilon_a}\right)^{1/2} t = e^{i(\alpha-\beta)} \frac{2iQ_a}{Q_a Q_b m_{12} - m_{21} + iQ_b m_{11} + iQ_a m_{22}} \qquad (1.3.2.15)$$

dove $Q_a = q_a/\varepsilon_a$ ed $Q_b = q_b/\varepsilon_b$.

Per quanto riguarda la riflettanza e la trasmittanza, in assenza di riflessione totale e se gli elementi di matrice sono reali:

$$R = |r|^2 = \frac{(Q_a Q_b m_{12} + m_{21})^2 + (Q_b m_{11} - Q_a m_{22})^2}{(Q_a Q_b m_{12} - m_{21})^2 + (Q_b m_{11} + Q_a m_{22})^2} \qquad (1.3.2.16)$$

$$T = \frac{q_b}{q_a}|t|^2 = \frac{4 Q_a Q_b}{(Q_a Q_b m_{12} - m_{21})^2 + (Q_b m_{11} + Q_a m_{22})^2} \qquad (1.3.2.17)$$

## *Paragrafo 1.4. PBG unidimensionali.*

In generale, un PBG unidimensionale (1D PBG) è una ripetizione finita nello spazio di una cella elementare in cui l'indice di rifrazione dipende dalla sola variabile lungo la quale ha luogo la periodicità.

### *1.4.1. Metodo della matrice di scattering.*

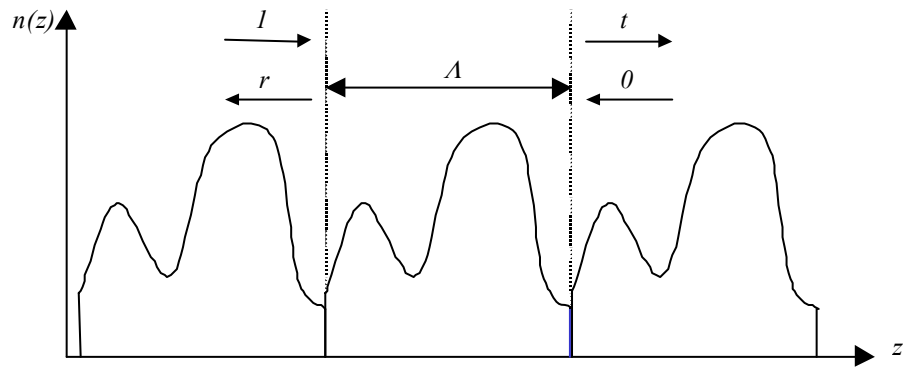

Figura. 1.3. Profilo d'indice generico: cella elementare e sue ripetizioni lungo z.
In figura è anche illustrato il processo di scattering che subisce
il campo ad opera della cella elementare, con *t* ed *r* coefficienti
di trasmissione e riflessione per le ampiezze.

La matrice di scattering di una struttura è definita come quella matrice che lega le ampiezze dell'onda incidente e riflessa in ingresso; in generale, i suoi elementi sono numeri complessi che descrivono sia le variazioni di ampiezza che di fase subite dal campo nella sua propagazione [7].

Se si vuole introdurre la matrice di scattering, si deve risolvere il problema elettromagnetico nella cella elementare.

Nel caso unidimensionale, tale problema è dato dall'equazione di Helmholtz:

$$\frac{d^2 a_k(z)}{dz^2} + \frac{\omega_k^2}{c^2} n^2(z) a_k(z) = 0 \tag{1.4.1.1}$$

con le condizioni al contorno in figura 1.3.

L'equazione esprime un problema di autovalori; il pedice $k$ sulla pulsazione $\omega$ evidenzia che esiste una relazione di dispersione $k = k(\omega)$ o meglio che, per un certo vettore d'onda, non tutte le pulsazioni sono permesse.

Le soluzioni possono essere espresse come sovrapposizione di onde che viaggiano nel verso positivo e negativo dell'asse $z$, opportunamente pesate per un fattore di ampiezza, anch'esso dipendente dall'ascissa $z$.

Ciò posto, si può introdurre la matrice di scattering:

$$\begin{pmatrix} a^+ \\ a^- \end{pmatrix}_{z=z_0} = M \cdot \begin{pmatrix} a^+ \\ a^- \end{pmatrix}_{z=z_0+\Lambda} \tag{1.4.1.2}$$

Si è supposto che l'indice di rifrazione è reale quindi la cella elementare è priva di perdite, la conservazione dell'energia può essere espressa nella forma $|r|^2 + |t|^2 = R + T = 1$ e, per quanto riguarda la matrice di scattering, $\det M = 1$.

Per un profilo di indice reale il processo non cambia con l'inversione dell'asse dei tempi.

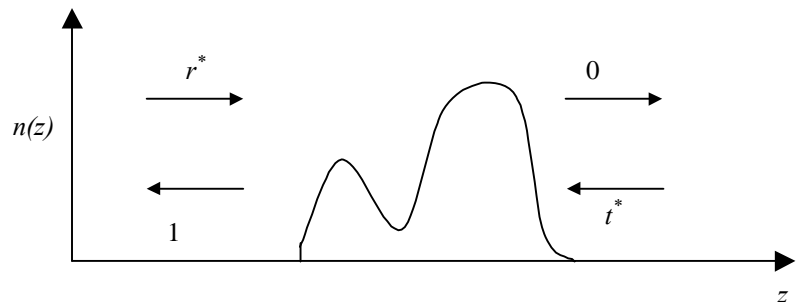

Figura 1.4. Processo di inversione dell'asse dei tempi.

La matrice di scattering consistente con questa proprietà assume la forma:

$$M = \begin{pmatrix} \dfrac{1}{t} & \dfrac{r^*}{t^*} \\ \dfrac{r}{t} & \dfrac{1}{t^*} \end{pmatrix} \qquad (1.4.1.3)$$

Per una struttura a $N$ periodi,

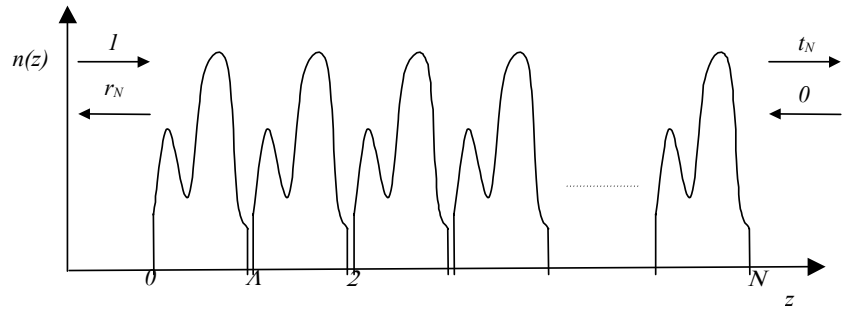

Figura 1.5. Problema di scattering per la struttura con N periodi.

la matrice di scattering è data da:

$$M_N = M^N = \begin{pmatrix} \dfrac{1}{t_N} & \dfrac{r_N^*}{t_N^*} \\ \dfrac{r_N}{t_N} & \dfrac{1}{t_N^*} \end{pmatrix} \qquad (1.4.1.4)$$

Si può determinare l'autovalore $\mu$ della matrice $M$ risolvendo l'equazione caratteristica associata:

$$\mu^2 - 2\mu\, Re(1/t) + 1 = 0 \qquad (1.4.1.5)$$

Ora è opportuno chiarire il significato degli autovalori della matrice di scattering della singola cella: il modo più semplice è quello di operare nelle condizioni in cui sono verificate le ipotesi del Teorema di Bloch; è necessario rifarsi ad una struttura periodica infinita.

Sotto queste ipotesi, l'effetto della propagazione del campo elettromagnetico lungo la cella elementare, fenomeno descritto dall'applicazione della matrice di scattering al vettore che descrive il campo in uscita, è quello di cambiare il campo stesso solo per un fattore di fase che prende il nome di fase di Bloch ed è indicata di solito con $\beta$ [2]. Quindi si ha:

$$M u_B = e^{\pm i\beta} u_B = \mu_B^{\pm} u_B \qquad (1.4.1.6)$$

dove *u* è il vettore colonna che descrive il campo in ingresso e in uscita, il pedice *B* afferma che si è nell'ipotesi di validità del teorema di Bloch e l'ultimo membro sta ad indicare che la propagazione nella cella elementare coincide con il problema della ricerca degli autovalori della matrice *M*.

E' evidente il significato cercato per il problema agli autovalori dell'equazione (1.4.1.5): gli autovalori della matrice di scattering della cella elementare sono assimilabili agli esponenziali complessi della fase di Bloch per la struttura infinita.

Sostituendo la (1.4.1.6) nell'equazione (1.4.1.5), si ottiene una relazione molto importante che lega la trasmissione della cella unitaria e la fase di Bloch:

$$Re\left(\frac{1}{t}\right) = \cos\beta \qquad (1.4.1.7)$$

Sulla base del Teorema di Cayley-Hamilton [8], secondo cui ogni matrice annulla il proprio polinomio caratteristico, si ha:

$$M^2 - 2M\,Re(1/t) + I = 0 \qquad (1.4.1.8)$$

Introducendo la (1.4.1.7) nella (1.4.1.8), è possibile ricavare, con il metodo induttivo e con l'aiuto di semplici identità trigonometriche, l'espressione della matrice $M_N$ in funzione della sola matrice *M*, del numero di periodi *N* e della fase *β*:

$$M_N = M\frac{\sin N\beta}{\sin\beta} - I\frac{\sin(N-1)\beta}{\sin\beta} \qquad (1.4.1.9)$$

in cui *I* è la matrice identità.

Ora è opportuno osservare come la (1.4.1.9) può essere fonte di confusione e ambiguità.

Nell'ipotesi di struttura periodica infinita vale la (1.4.1.9) per i primi *N* periodi; solo sotto questa condizione la fase *β* assume il significato fisico suddetto.

Se la struttura è periodica finita, vale ancora la (1.4.1.9) ma la fase *β* non è più quella accumulata dal campo nella propagazione attraverso la cella elementare: difatti, quando la struttura è finita, non vale l'invarianza traslazionale ed è diverso il comportamento del campo elettromagnetico.

Infine si ricorda che nessuna specifica è stata data per l'indice di rifrazione all'interno della cella elementare e quindi che le relazioni ottenute sono del tutto generali.

## *1.4.1.a. Coefficiente di trasmissione.*

Ora si determina il coefficiente di trasmissione, la trasmittanza e si svolgono alcune considerazioni di carattere generale [7].

Inserendo le (1.4.1.3), (1.4.1.4) nella (1.4.1.9), si ricavano le espressioni dei coefficienti complessi di trasmissione e riflessione:

$$\frac{1}{t_N} = \frac{1}{t}\frac{sinN\beta}{sin\beta} - \frac{sin(N-1)\beta}{sin\beta} \qquad (1.4.1.10)$$

$$\frac{r_N}{t_N} = \frac{r}{t}\frac{sinN\beta}{sin\beta} \qquad (1.4.1.11)$$

Applicando il principio di conservazione dell'energia nella forma $|r|^2 = 1 - |t|^2$, dalla (1.4.1.11), si può ricavare l'espressione della trasmittanza:

$$\frac{1}{T_N} = 1 + \left(\frac{sinN\beta}{sin\beta}\right)^2 \left(\frac{1}{T} - 1\right) \qquad (1.4.1.12)$$

Il coefficiente di trasmissione può essere espresso come:

$$t_N = \sqrt{T_N}e^{i\phi_N} \qquad (1.4.1.13)$$

Torna che la fase di Bloch non ha significato fisico per una struttura finita; infatti, differisce da quella del coefficiente di trasmissione per la struttura intera e per la singola cella.

Tuttavia, la fase *β* assume un ruolo importante nel determinare le caratteristiche di trasmissione di una struttura 1D-PBG.

In particolare le bande passanti e quelle proibite di una struttura finita sono le stesse di una struttura infinita: riferendosi alla (1.4.1.6), bande passanti per una struttura infinita corrispondono a valori reali della fase di Bloch; anche per la (1.4.1.12) vale un discorso analogo.

Si osserva la (1.4.1.12). Per valori complessi della fase di Bloch, indipendentemente da *T*, il termine quadrato diventa rapidamente maggiore di uno e quindi la trasmissione $T_N$ diventa trascurabile; una banda è proibita per la propagazione del campo elettromagnetico quando, nella banda, la trasmissione è nulla. La trasmittanza $T_N$ è una funzione periodica rispetto alla fase *β* con periodo pari a $\pi/N$ e quindi, all'interno di una banda di trasmissione, in cui *β* è incrementata di *π*, la trasmittanza presenta *N* massimi; la rottura dell'invarianza traslazionale ad opera delle dimensioni finite della geometria porta alla divisione della banda di trasmissione in $N$ picchi di risonanza.

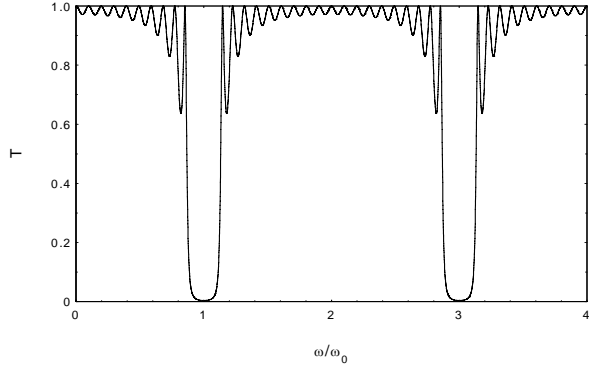

Figura 1.6. Spettro di trasmissione $T_N(\omega/\omega_{rif})$ di una struttura periodica la cui cella è costituita di due strati per i quali $n_1 d_1 = n_2 d_2 = \lambda_{rif}/4$ dove $\lambda_{rif}$ è una lunghezza d'onda di riferimento, $\lambda_{rif} = 1 m\mu$. Gli indici di rifrazione sono pari a $n_1 = 1$, $n_2 = 1.41$ e il numero di periodi è $N = 10$.

### *1.4.1.b. Densità dei modi (DOM).*

La DOM di un PBG unidimensionale con numero dei periodi $N$ è definita come la derivata del numero d'onda $k_N$ rispetto alla pulsazione $\omega$, quindi [7]:

$$\rho_N(\omega) = \frac{dk_N}{d\omega} \quad (1.4.1.14)$$

Se il coefficiente di trasmissione della struttura è $t_N = x_N + i y_N$ e quello della singola cella è $t = x + iy$, partendo dalle equazioni (1.4.1.7) e (1.4.1.10), si possono esprimere la parte reale e immaginaria di $t_N$ in funzione delle rispettive di $t$, del numero dei periodi $N$ e della fase $\beta$, vale a dire:

$$x_N = \frac{x \sin\beta \sin N\beta - (x^2 + y^2)\sin\beta \sin(N-1)\beta}{\sin^2 N\beta - 2x\sin N\beta \sin(N-1)\beta + (x^2 + y^2)\sin^2(N-1)\beta} \quad (1.4.1.15)$$

$$y_N = \frac{y \sin\beta \sin N\beta}{\sin^2 N\beta - 2x\sin N\beta \sin(N-1)\beta + (x^2 + y^2)\sin^2(N-1)\beta} \quad (1.4.1.16)$$

Si pone la sostituzione:

$$z_N = \frac{y_N}{x_N} \quad (1.4.1.17)$$

Inserendo le (1.4.1.15) e (1.4.1.16) nella (1.4.1.17), si ottiene:

$$z_N = \frac{y \sin N\beta}{x \sin N\beta - (x^2 + y^2) \sin(N-1)\beta} \qquad (1.4.1.18)$$

Sostituendo la trasmittanza della singola cella $T = x^2 + y^2$, le quantità scalate $\xi = x/T$ e $\eta = y/T$, introducendo la nuova posizione $z = y/x$ ed utilizzando che $\xi = \cos\beta$, la (1.4.1.18) diviene ancora:

$$z_N = z \cot\beta \tan N\beta \qquad (1.4.1.19)$$

Posto il coefficiente di trasmissione della struttura nella forma $t_N = \sqrt{T_N} e^{i\varphi_N}$, la sua fase ha espressione $\varphi_N = k_N N\Lambda$, dove $\Lambda$ è lo spessore della cella, quindi:

$$z_N = \tan\varphi_N = \tan(k_N N\Lambda) \qquad (1.4.1.20)$$

Inserendo la (1.4.1.19) nella (1.4.1.20), si ottiene la relazione di dispersione della struttura $k_N = k_N(\omega)$ nella forma:

$$\tan(k_N N\Lambda) = z \cot\beta \tan N\beta \qquad (1.4.1.21)$$

Infine, la DOM (1.4.1.14), tenendo conto della (1.4.1.20), diviene:

$$\rho_N = \frac{dk_N}{d\omega} = \frac{1}{N\Lambda}\frac{d}{d\omega}\tan^{-1}(z_N) = \frac{1}{N\Lambda}\frac{z_N'}{1+z_N^2} \qquad (1.4.1.22)$$

e, tenendo conto della (1.4.1.19), dopo alcune manipolazioni, assume la forma:

$$\rho_N = \frac{1}{N\Lambda} \frac{(1/2)[\sin(2N\beta)/\sin\beta][\eta' + \eta\xi\xi'/(1-\xi^2)] - N\eta\xi'/(1-\xi^2)}{\cos^2 N\beta + \eta^2[\sin(N\beta)/\sin\beta]^2} \qquad (1.4.1.23)$$

dove l'apice indica l'operatore di derivata rispetto ad $\omega$.

Quindi la DOM della struttura è funzione del numero dei periodi $N$, della fase di Bloch $\beta$ e del coefficiente di trasmissione della cella base $t$.

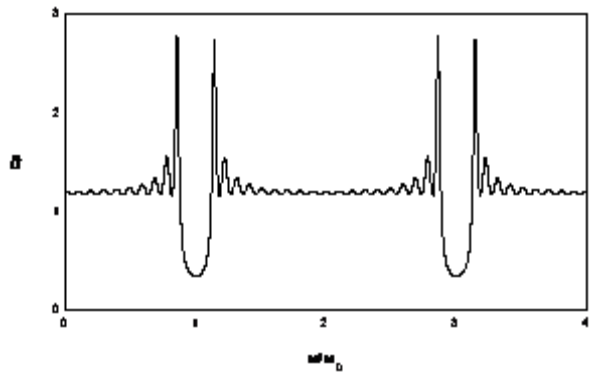

Figura 1.7. Densità dei modi per la stessa struttura di figura 1.6.

## 1.4.2. Proprietà dei PBG unidimensionali.

In questo paragrafo si illustrano le proprietà prima di un PBG generico e poi, in specifico, di un PBG a quarto d'onda [7].

## 1.4.2.a. Generalità sullo spettro di trasmissione e sulla DOM.

Dapprima si considera lo spettro di trasmissione $T_N$.

Se $N\beta = m\pi$ e $m \in \{0,1,...,N-1\}$, allora $sin^2 N\beta = 0$ e $T_N$ diviene pari ad uno.

I picchi di risonanza della struttura finale sono indipendenti dalla trasmittanza della cella unitaria $T$; per tali frequenze, supposto che $T$ è prossimo a zero, cioè che la cella elementare è completamente riflettente, $T_N$ vale uno, quindi l'intera struttura è completamente trasparente.

Quando $N\beta$ è uguale a multiplo dispari di $\pi/2$, allora la curva di trasmissione $T_N$ assume un valore tanto vicino al minimo relativo quanto più grande è il numero dei periodi $N$.

Difatti, per tali valori, $sin^2(N\beta)$ è massimo, vale a dire è il termine che varia più velocemente e l'intera funzione sarà vicina al suo minimo.

Il valore di $T_N$ nei punti di minimo è pari circa a:

$$\left.\frac{1}{T_N}\right|_{\beta=[(2m+1)/N]\pi/2} \cong 1 + \frac{1}{sin^2[(2m+1)\pi/2N]}\left[\frac{1}{T}-1\right] \qquad (1.4.2.1)$$

Un'altra regione d'interesse per la funzione $T_N$ è nel band-gap.

All'interno del primo band-gap, la fase di Bloch diventa complessa ed assume la forma $\beta = \pi + i\theta$ dove $\theta = cosh^{-1}(-\xi)$.

Applicando le $\cos(i\theta) = \cosh\theta$ ed $sen(i\theta) = senh\theta$, si ottiene un'espressione per calcolare la trasmissione nel gap:

$$\frac{1}{T_N^{gap}} = 1 + \frac{sinh^2 N\theta}{sinh^2 \theta}\left(\frac{1}{T}-1\right) \qquad (1.4.2.2)$$

Le stesse grandezze si possono ricavare per la densità dei modi $\rho_N$.

Quando $\beta = m\pi/N$ con $m \in \{0,1,...,N-1\}$, la DOM ha approssimativamente i suoi massimi.

I picchi di risonanza di $T_N$ e $\rho_N$ sono tanto più coincidenti, quanto più la struttura tende ad essere infinita ($N \to \infty$).

Nella (1.4.1.23), se $N = m\pi$, il termine che varia più rapidamente nel denominatore è vicino al suo minimo, mentre il numeratore è proporzionale a *N*, che è già abbastanza grande. Quindi:

$$\rho_N^{mac}\big|_{\beta=m\pi/N} \cong -\frac{1}{N\Lambda}\frac{N\eta\xi'}{1-\xi^2}\bigg|_{\beta=m\pi/N} \tag{1.4.2.3}$$

Per le frequenze all'interno del primo band-gap, accettando la posizione fatta per la $T_N$, cioè $\beta = \pi + i\theta$, si ottiene:

$$\rho_N^{gap}\big|_{\beta=\pi+i\theta} = \frac{1}{N\Lambda}\frac{-(1/2)[\sinh(2N\theta)/\sinh\theta]\left[\eta' + \xi\eta\xi'/(1-\xi^2)\right] - N\eta\xi'/(1-\xi^2)}{\cosh^2 N\theta + \eta^2[\sinh(N\theta)/\sinh\theta]^2}. \tag{1.4.2.4}$$

La trasmittanza è minima al centro della banda proibita: si ha un suo massimo o minimo per $\cos\beta = \xi(\omega)$, quindi $\xi' = 0$.

La densità dei modi al centro del gap diventa:

$$\rho_N^{MG}\big|_{\xi'=0} = \frac{1}{N\Lambda}\frac{-(1/2)\eta'[\sinh(2N\theta^{MG})/\sinh\theta^{MG}]}{\cosh^2(N\theta^{MG}) + \eta^2[\sinh(N\theta^{MG})/\sinh\theta^{MG}]^2} \tag{1.4.2.5}$$

dove $\theta^{MG} = \cosh^{-1}(-\xi^{MG})$.

Per grandi valori di *N* le funzioni iperboliche possono essere approssimate con esponenziali e si ottiene:

$$\rho_N^{MG}\big|_{\xi'=0} = -\frac{1}{N\Lambda}\frac{\eta'\sinh\theta^{MG}}{\eta^2 + \sinh^2\theta^{MG}} \tag{1.4.2.6}$$

La densità dei modi al centro del gap non dipende dalla trasmissione della cella unitaria e varia asintoticamente come $1/N$; quindi la velocità di gruppo al centro della banda proibita cresce linearmente con *N*, diventando eventualmente superluminale [9].

### *1.4.3.b. PBG a quarto d'onda.*

Il calcolo della matrice di scattering di tale struttura assume una forma molto semplice: il processo di trasmissione attraverso la cella elementare può essere suddiviso in due processi di riflessione, alle interfacce tra i mezzi diversi, e due processi di propagazione, attraverso gli strati dielettrici uniformi [6].

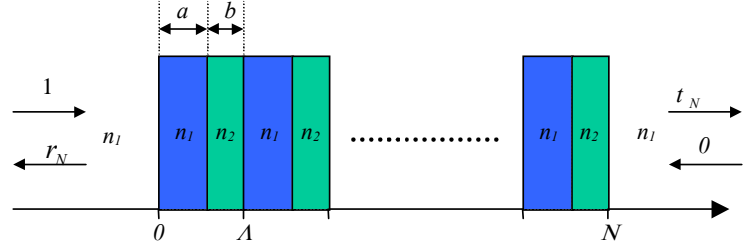

Figura 1.8. Struttura $\lambda/4 - \lambda/4$: gli spessori sono $a = \lambda_{rif}/4n_1$ e $b = \lambda_{rif}/4n_2$, dove $\lambda_{rif}$ è la lunghezza d'onda di riferimento. La prima banda proibita è centrata intorno a $\lambda_{rif}$.

Il coefficiente di trasmissione che si ottiene per la cella elementare è:

$$t = \frac{T_{12} e^{i(k_1 a + k_2 b)}}{1 - R_{12} e^{2ik_2 b}} \tag{1.4.2.7}$$

dove:

$$T_{12} = t_{12} t_{21} = \frac{4 n_1 n_2}{(n_1 + n_2)^2} \tag{1.4.2.8}$$

$$R_{12} = r_{12}^2 = \left(\frac{n_1 - n_2}{n_1 + n_2}\right)^2 \tag{1.4.2.9}$$

con $k_1 = n_1 \omega/c$ e $k_2 = n_2 \omega/c$.

La trasmittanza per l'intera struttura è:

$$T_N^{\lambda/4} = \frac{1 + \cos\beta}{1 + \cos\beta + 2(R_{12}/T_{12})\sin^2 N\beta} \tag{1.4.2.10}$$

I suoi punti di minimo sono:

$$T_N^{\lambda/4-\min} = \frac{1 + \cos[(2m+1)\pi/2N]}{1 + 2R_{12}/T_{12} + \cos[(2m+1)\pi/2N]} \tag{1.4.2.11}$$

dove $m \in \{0, 1, \ldots, N-1\}$.

Al bordo banda, dove $m = N-1$, la trasmissività (1.4.2.11), per $N$ molto grande, diviene, dopo uno sviluppo in serie di Taylor del termine in coseno:

$$T_N^{\lambda/4}\big|_{m=N-1} \underset{N \to \infty}{\approx} \frac{\pi^2}{16} \frac{T_{12}}{R_{12}} \frac{1}{N^2} \tag{1.4.2.12}$$

Se si tiene conto delle (1.4.2.8) e (1.4.2.9), allora:

$$T_N^{\lambda/4}\big|_{m=N-1} \underset{N\to\infty}{\approx} \frac{\pi^2}{16}\frac{4n_1 n_2}{(n_1-n_2)^2}\frac{1}{N^2} \tag{1.4.2.13}$$

Si definisce velocità di gruppo nel bulk la quantità:

$$\mathrm{v}_{bulk} = \frac{\Lambda}{t_1+t_2} = \frac{c}{2}\left(\frac{1}{n_1}+\frac{1}{n_2}\right) \tag{1.4.2.14}$$

dove $t_1$ e $t_2$ sono i tempi di transito del campo nei due strati.

Normalizzando la DOM alla velocità di gruppo si ottiene una grandezza che tende ad uno nelle bande passanti.

Nella (1.4.2.14) non sono state prese in considerazione le riflessioni alle varie interfacce: è proprio questo il motivo per cui la grandezza normalizzata, $\mathrm{v}_{bulk}\cdot\rho_N$, evidenzia l'effetto della struttura periodica sulla DOM e sulla velocità di gruppo, effetto prodotto alterando le proprietà di riflessione e trasmissione della struttura rispetto allo spazio libero.

La lunghezza della cella elementare $\Lambda$ può essere espressa in funzione di $\upsilon_{bulk}$:

$$\Lambda = a+b = \frac{\lambda}{4}\left(\frac{1}{n_1}+\frac{1}{n_2}\right) = \frac{\pi c}{2\omega_{rif}}\left(\frac{1}{n_1}+\frac{1}{n_2}\right) = \frac{\pi}{\omega_{rif}}\mathrm{v}_{bulk} \tag{1.4.2.15}$$

Se si tiene conto di questa posizione, $\Lambda$ non compare nella DOM.

I massimi della DOM sono:

$$\rho_N^{\lambda/4}\big|_{\beta=m\pi/N} \cong \frac{1}{\mathrm{v}_{bulk}}\frac{1-T_{12}\sin^2\left(\frac{m\pi}{2N}\right)}{T_{12}\cos^2\left(\frac{m\pi}{2N}\right)} \tag{1.4.2.16}$$

Il massimo di maggior interesse è il Band Edge Resonance (BER), che corrisponde a $m=N-1$. La DOM presenta un massimo assoluto nel BER che vale:

$$\rho_N^{\lambda/4}\big|_{BER} \cong \frac{1}{\mathrm{v}_{bulk}}\frac{1-T_{12}\cos^2\left(\frac{\pi}{2N}\right)}{T_{12}\sin^2\left(\frac{\pi}{2N}\right)} \tag{1.4.2.17}$$

Per $N$ molto grande, si ottiene, dopo aver sviluppato le funzioni trigonometriche in serie di Taylor:

$$\rho_N^{\lambda/4}\big|_{BER} \underset{N\to\infty}{\approx} \frac{4}{\pi^2 \mathrm{v}_{bulk}}\frac{R_{12}}{T_{12}}N^2 \tag{1.4.2.18}$$

Se si tiene conto delle (1.4.2.8) e (1.4.2.9), allora:

$$\rho_N^{\lambda/4}\big|_{BER} \underset{N\to\infty}{\approx} \frac{4}{\pi^2 v_{bulk}} \frac{(n_1-n_2)^2}{4n_1 n_2} N^2 \qquad (1.4.2.19)$$

Le (1.4.2.13) e (1.4.2.19) permettono di arricchire l'elenco delle caratteristiche di un PBG unidimensionale.

Le grandezze fondamentali sono il numero dei periodi e il contrasto d'indice: lo spettro di trasmissione è inversamente proporzionale ai loro quadrati mentre la DOM ha una dipendenza quadratica rispetto ad entrambe.

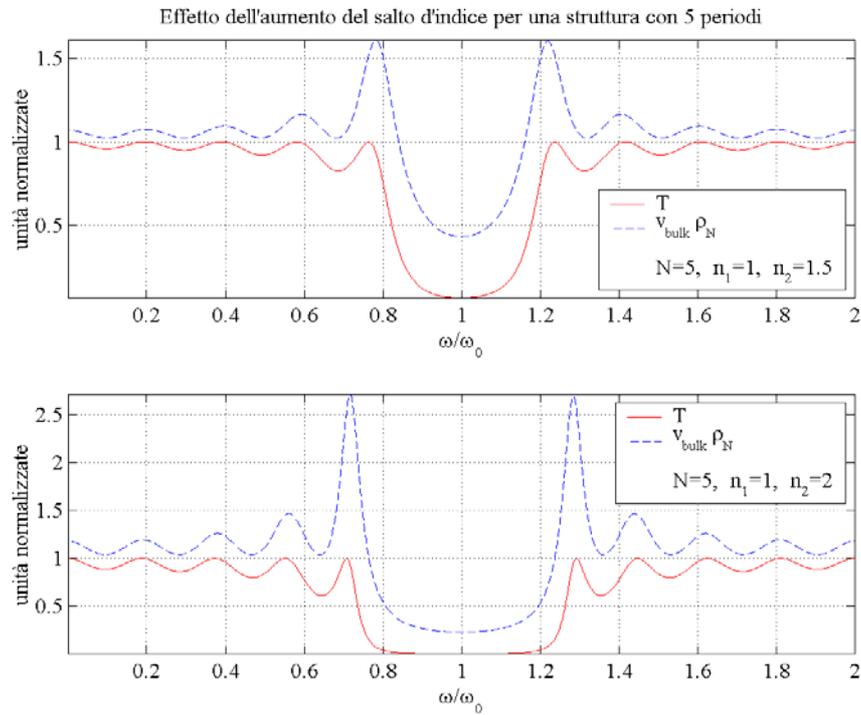

Figura 1.9. Trasmissione e DOM normalizzata per una struttura 1D-PBG: si osservano le risonanze della *T*, in corrispondenza delle quali la DOM ha un massimo relativo. I massimi e i minimi della *T* e della DOM sono leggermente spostati gli uni rispetto agli altri: la differenza diminuisce rapidamente all'aumentare di *N*.

Il numero dei periodi e il salto d'indice intervengono anche nel fattore di merito per la risonanza al BER, in quanto [10]:

$$\frac{\Delta\omega_N^{BER}}{\omega_{rif}} \underset{N\to\infty}{\approx} \frac{\pi}{2N^2}\sqrt{\frac{T_{12}}{R_{12}}} = \frac{\pi}{2N^2}\sqrt{\frac{4n_1 n_2}{(n_1-n_2)^2}} \qquad (1.4.2.20)$$

Il solo salto d'indice interviene nella larghezza di banda del gap, in quanto [2]:

$$\Delta\omega_{gap} = \omega_{rif}\frac{4}{\pi}\sin^{-1}\frac{|(n_2-n_1)|}{n_2+n_1} \cong \omega_{rif}\frac{2}{\pi}\frac{\Delta n}{n} \qquad (1.4.2.21)$$

dove $\Delta n = |n_2 - n_1|$ e $n = (1/2)(n_1 + n_2)$.

Si osservi che la larghezza di banda del gap è data dall'armonica di Fourier del salto della costante dielettrica:

$$\frac{\Delta \omega_{gap}}{\omega_{rif}} \underset{N \to \infty}{\approx} \frac{|\varepsilon_1|}{\varepsilon_0} \tag{1.4.2.22}$$

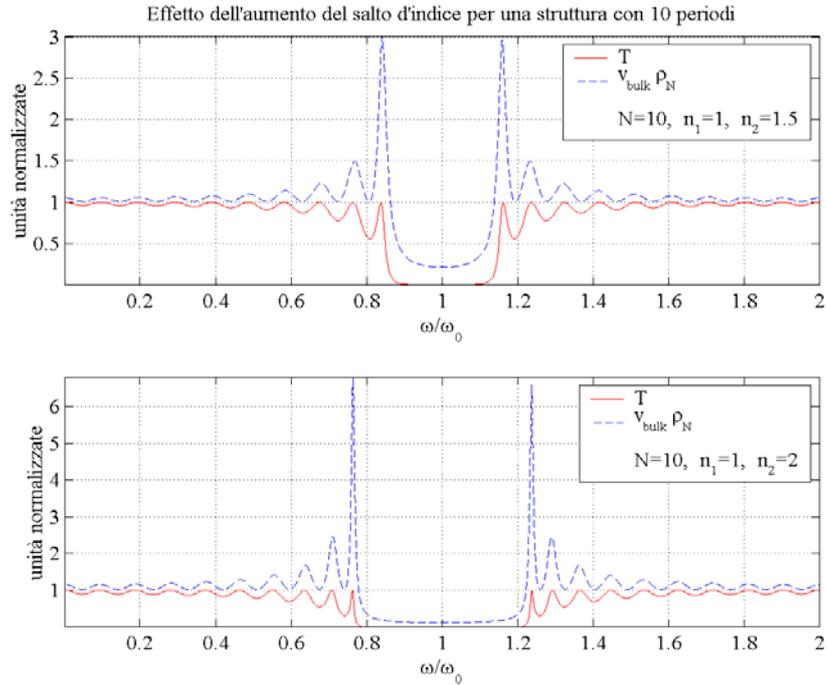

Figura 1.10. Trasmissione e DOM normalizzata per una struttura 1D PBG: ad un aumento della DOM corrisponde una risonanza dello spettro di trasmissione più definita in corrispondenza del BER; infatti, il sistema è lineare e per un aumento della lunghezza della struttura si ha una maggiore selettività in frequenza.

# Capitolo 2. Teoria dei QNMs per cavità semiaperte.

## Paragrafo 2.1. Introduzione.

Se un sistema è conservativo, può essere descritto da operatori hermitiani e ogni suo stato può essere espanso in modi normali. Questo avviene nella descrizione classica del campo e.m., con lo sviluppo in onde piane, o quando il campo e.m. è confinato in regioni limitate, come le guide d'onda.

Qui interessano i sistemi non conservativi: gli operatori associati non sono hermitiani, nel senso usuale, e non si possono introdurre i modi normali, almeno in senso stretto. Si considerano le cavità ottiche risonanti che consentono un confinamento particolare del campo e.m.: le cavità sono aperte, quindi vi sono perdite di radiazione.

La trattazione più conveniente è dapprima includere i gradi di libertà dell'esterno e poi eliminarli dalle equazioni del moto.

Si pone la cavità in un universo con dimensione lineare $\Lambda \to \infty$ e si impongono condizioni al contorno tali che l'intero sistema sia conservativo. L'elettrodinamica può essere formulata in termini dei modi dell'universo [1] ma il prezzo che si paga è triplice:

a) i modi dell'universo sono continui, con numeri d'onda separati da $\Delta k \approx \pi/\Lambda$;

b) non è chiaro se l'elettrodinamica della cavità è indipendente dalle assunzioni fatte per l'universo;

c) non c'è la riduzione intuitiva al limite di perdite nulle.

Una trattazione più appropriata è di eliminare da subito i gradi di libertà dell'esterno.

In presenza di perdite, i modi della cavità sono quasi normali (QNMs), caratterizzati da frequenze complesse. Inoltre, i QNMs sono discreti, con numeri d'onda separati da $\Delta k \approx \pi/a$, dove $a$ è la dimensione lineare della cavità.

In generale, poiché la cavità non è un sistema hermitiano, i QNMs non formano una base ortonormale completa: tuttavia, si possono introdurre delle condizioni sotto di cui sussiste la completezza [2] e si può stabilire un parallelismo formale opportuno con il caso conservativo, da cui si deduce l'ortogonalità [3].

## *Paragrafo 2.2. Modello fisico-matematico.*

Si considera lo spazio semi infinito $0 \leq x < \infty$: il sistema $S$ sia la cavità $I = [0, a]$ ed il bagno $B$ sia il resto dell'universo $I' = (a, \infty)$.

L'equazione delle onde nella cavità è la seguente:

$$\left[\rho(x)\frac{\partial^2}{\partial t^2} - \frac{\partial^2}{\partial x^2}\right]\phi(x,t) = 0 \qquad (2.2.1)$$

dove:

$$\rho(x) = \left(\frac{n(x)}{c}\right)^2 \qquad (2.2.2)$$

con $n(x)$ indice di rifrazione e $c$ velocità della luce.

Il singolo modo evolve nel tempo secondo la legge:

$$\phi(x,t) = f_n(x)e^{-i\omega_n t} \qquad (2.2.3)$$

che, inserita nella (2.2.1), dà luogo all'equazione per gli autovalori:

$$\left[\partial_x^2 + \rho(x)\omega_n^2\right]f_n(x) = 0 \qquad (2.2.4)$$

Definito l'operatore:

$$H = -\rho^{-1}(x)\partial_x^2 \qquad (2.2.5)$$

l'equazione (2.2.4) assume la forma:

$$Hf_n(x) = \omega_n^2 f_n(x) \qquad (2.2.6)$$

Vi sia uno specchio totalmente riflettente in $x = 0$; vale la condizione nodale [4]:

$$\phi(x = 0, t) = 0 \qquad (2.2.7)$$

che, tenendo conto della (2.2.3), si traduce nella:

$$f_n(0) = 0 \qquad (2.2.8)$$

Dapprima si considera il caso conservativo [5]. La cavità è chiusa; vi è uno specchio totalmente riflettente anche in $x = a$; vale l'altra condizione nodale:

$$\phi(x = a, t) = 0 \qquad (2.2.9)$$

che, tenendo conto della (2.2.3), si traduce nella:

$$f_n(a) = 0 \qquad (2.2.10)$$

L'operatore $H$ è Hermitiano, definito positivo, illimitato superiormente.

Gli autovalori $\omega_n^2$ sono reali e positivi; se si osserva la (2.2.3), si deduce che i modi sono stazionari, non decadono nel tempo.

Posto, per convenzione, che $0 < \omega_1 \leq \omega_2 \leq \ldots$, i modi sono tali che:

$$\omega_{-n} = -\omega_n \qquad (2.2.11)$$

$$f_{-n}(x) = f_n(x) \qquad (2.2.12)$$

Segue che sono significativi i soli modi con $n > 0$.

Le autofunzioni $f_n(x)$ costituiscono un sistema completo, sia in senso debole:

$$\phi(x) = \sum_{n>0} a_n f_n(x) \qquad (2.2.13)$$

che in senso stretto:

$$\phi(x,t) = \sum_{n>0} a_n f_n(x) e^{-i\omega_n t} \qquad (2.2.14)$$

Definendo il prodotto interno come:

$$\langle \phi | \psi \rangle = \int_I dx \, \phi^*(x) \rho(x) \psi(x) \qquad (2.2.15)$$

le autofunzioni $f_n(x)$ costituiscono una base ortogonale: quindi, si parla di modi normali (NMs).

Ora si considera il caso non conservativo [5]. Vi è uno scambio di energia tra $S$ e $B$; è stato posto uno specchio parzialmente riflettente in $x = a$. Il campo $\phi(x,t)$ è un'onda uscente $\phi(x - ct)$ fuori della cavità, quindi, posto $\rho(x) = \rho_0$ per $x > a$, vale la condizione di *"outgoing wave"*:

$$\partial_t \phi(x,t) = -\rho_0^{-1/2} \partial_x \phi(x,t) \quad \text{per} \quad x > a \qquad (2.2.16)$$

che, tenendo conto della (2.2.3), dà luogo alla:

$$f_n'(a^+) = i\omega_n \rho_0^{1/2} f_n(a^+) \qquad (2.2.17)$$

La condizione di *"outgoing wave"* rende l'operatore $H$ non più Hermitiano nella cavità $I$.

Le frequenze $\omega_n$ divengono complesse, con $\operatorname{Im}\omega_n < 0$: se si osserva la (2.2.3), si deduce che, ora, i modi non sono più stazionari; nel dominio della frequenza, sono visti come risonanze con ampiezze $\gamma_n = |\operatorname{Im}\omega_n|$, mentre, in quello del tempo, come oscillazioni smorzate.

Ad eccezione dei modi zero, indicati con $\omega_0$ e tali che $\operatorname{Re}\omega_0 = 0$, gli altri modi sono a coppie; vale a dire, posto $0 < \operatorname{Re}\omega_1 \leq \operatorname{Re}\omega_2 \leq \ldots$, valgono le:

$$\omega_{-n} = -\omega_n^* \tag{2.2.18}$$

$$f_{-n}(x) = f_n^*(x) \tag{2.2.19}$$

Poiché $\omega_{-n}^2 \neq \omega_n^2$, le autofunzioni $f_n(x)$ e $f_{-n}(x)$ sono linearmente indipendenti e sono significativi tutti i modi, non solo quelli con $n > 0$.

Come si dirà in seguito, si può recuperare la completezza, sotto opportune condizioni, tra le quali quella di porsi solo all'interno della cavità: quindi, si parlerà di modi quasi normali (QNMs).

### 2.2.1. Funzione di Green.

Si consideri la funzione causale di Green, definita dall'equazione:

$$\left[\rho(x)\frac{\partial^2}{\partial t^2} - \frac{\partial^2}{\partial x^2}\right]G(x,y;t) = \delta(t)\delta(x-y) \tag{2.2.1.1}$$

assieme alle condizioni iniziali:

$$G(x,y;t) = 0 \quad \text{per} \quad t \leq 0 \tag{2.2.1.2}$$

$$\rho(x)\frac{\partial G(x,y;t)}{\partial t}\bigg|_{t=0^+} = \delta(x-y) \tag{2.2.1.3}$$

La trasformata secondo Fourier della funzione di Green $\tilde{G}(x,y;\omega)$ soddisfa l'equazione:

$$\left[\frac{\partial^2}{\partial x^2} + \omega^2 \rho(x)\right]\tilde{G}(x,y;\omega) = -\delta(x-y) \tag{2.2.1.4}$$

ed è analitica per $\operatorname{Im}\omega > 0$ e va come $\exp(i\omega x/c)$ per $x \to \infty$.

Si introducono le due funzioni ausiliarie $g_\pm(x,\omega)$, soluzioni dell'equazione omogenea, indipendente dal tempo:

$$\left[\frac{\partial^2}{\partial x^2} + \omega^2 \rho(x)\right] g_\pm(x,\omega) = 0 \qquad (2.2.1.5)$$

con:

$$g_-(x=0,\omega) = 0 \qquad (2.2.1.6)$$

$$g_+(x,\omega) = \exp(i\omega x/c) \quad \text{per} \quad x \to \infty \qquad (2.2.1.7)$$

Il Wronskiano è indipendente da $x$, quindi [6]:

$$W(\omega) = g_+(x,\omega)g'_-(x,\omega) - g_-(x,\omega)g'_+(x,\omega) \qquad (2.2.1.8)$$

Si dimostra che [6]:

$$\tilde{G}(x,y;\omega) = \begin{cases} -\dfrac{g_-(x,\omega)g_+(y,\omega)}{W(\omega)} & , \quad \text{per} \quad 0 < x < y \\ -\dfrac{g_+(x,\omega)g_-(y,\omega)}{W(\omega)} & , \quad \text{per} \quad 0 < y < x \end{cases} \qquad (2.2.1.9)$$

La normalizzazione delle $g_\pm(x,\omega)$ è arbitraria e non influisce sulla (2.2.1.9).

I QNMs corrispondono ai poli di $\tilde{G}(x,y;\omega)$.

Alle frequenze $\omega = \omega_n$ dei QNMs, le funzioni ausiliarie $g_\pm(x,\omega)$ soddisfano sia la condizione di regolarità in $x = 0$, che quella di asintoticità per $x \to \infty$, quindi [6]:

$$g_-(x,\omega_n) \propto g_+(x,\omega_n) \qquad (2.2.1.10)$$

Segue che:

$$W(\omega_n) = 0 \qquad (2.2.1.11)$$

I residui della $\tilde{G}(x,y;\omega)$ nei poli $\omega = \omega_n$ sono dati da [6]:

$$R_n = -\frac{g_-(x,\omega_n)g_+(x,\omega_n)}{\left.\dfrac{dW}{d\omega}\right|_{\omega=\omega_n}} \qquad (2.2.1.12)$$

## 2.2.1.a. *Discussione sulla completezza dei QNMs.*

La funzione di Green $G(x,y;t)$ per $t \geq 0$ è calcolata come la trasformata inversa di Fourier della funzione $\tilde{G}(x,y;\omega)$, con il dominio di integrazione chiuso da un semicerchio all'infinito nel semipiano inferiore delle $\omega$.

In generale, vi sono tre contributi diversi [7]:

a) *Risposta "istantanea":* questa risposta è quella del transitorio iniziale ed é dovuta al valore di $\tilde{G}(x,y;\omega)$ lungo il semicerchio infinito.

b) *Risposta "ritardata":* questa risposta è quella per tempi lontani.
   E' dovuta alle singolarità della funzione $g_+(x,\omega)$. La $g_-(x,\omega)$ è ottenuta integrando l'equazione delle onde indipendente dal tempo, nella quale appare analiticamente $\omega$, lungo una distanza finita da $0$ a $x$; quindi, è una combinazione lineare di funzioni analitiche in $\omega$ e, come tale, è analitica in $\omega$. Invece, la $g_+(x,\omega)$ è ottenuta integrando da $x \to \infty$ e potrebbe avere un taglio sul semiasse negativo $\text{Im}\,\omega$.

c) *Risposta "intermedia":* questa risposta dà luogo all'oscillazione dei QNMs per tempi intermedi ed è dovuta agli zeri del Wronskiano.

Ora, si suppone che la cavità ottica confini con l'esterno tramite un'interfaccia costituita da un materiale sottile. Allora, valgono le condizioni [7]:

$\alpha$) *di discontinuità*: la funzione $\rho(x)$ ha una discontinuità in $x = a$, almeno a gradino, che delimita una regione di cavità ben definita;

$\beta$) *"no tail"*: la funzione $\rho(x)$ non ha alcuna coda fuori la cavità, cioè $\rho(x) = \rho_0$ per $x > a$, quindi l'esterno non riflette l'onda uscente indietro nella cavità.

Poiché vale la condizione di discontinuità, si trova che svanisce il contributo della risposta "istantanea"; mentre, poiché vale la condizione "no tail", si può imporre la condizione di "outgoing wave" per $g_+(x,\omega)$ in $x = a^+$ e integrare solo lungo una distanza finita: la funzione $g_+(x,\omega)$ è analitica ed è rimosso il contributo della risposta "ritardata".

Sotto le condizioni di discontinuità e di "no tail", vale la completezza dei QNMs: rimane solo il contributo della risposta "intermedia", dei QNMs.

Si vuole applicare l'approssimazione WKB [6] per studiare l'andamento di $\tilde{G}(x,y;\omega)$ quando $|\omega| \to \infty$.

Si suppone che il potenziale $\rho(x)$ abbia una discontinuità in $x = a$, quindi si introduce il coefficiente di riflessione:

$$R = \frac{n(a^-) - n(a^+)}{n(a^-) + n(a^+)} \qquad (2.2.1.13)$$

Si applica l'approssimazione WKB, separatamente per $0 < x < a$ e $x > a$, e si uniscono le due soluzioni tramite il coefficiente di riflessione $R$.
Si ottiene:

$$\tilde{G}(x,y;\omega) \cong \frac{sin[\omega I(0,y)]\left[e^{-i\omega I(x,a)} + \text{Re}^{i\omega I(x,a)}\right]}{\omega\sqrt{n(x)n(y)}\left[e^{-i\omega I(0,a)} + \text{Re}^{i\omega I(0,a)}\right]} \quad \text{per} \quad 0 < y \leq x < a \quad (2.2.1.14)$$

dove è stata definita la quantità:

$$I(u,v) = \int_u^v \rho(x)dx \quad (2.2.1.15)$$

che è positiva per $v > u$.

Si studia l'andamento per $\omega = \omega_R + i\omega_I \to \infty$, con $\omega_I < 0$. Sia il numeratore che il denominatore sono dominati dal termine proporzionale a $R$, e:

$$\tilde{G}(x,y;\omega) \cong \frac{e^{|\omega_I|I(0,y)}e^{|\omega_I|I(x,a)}}{\omega e^{|\omega_I|I(0,a)}} \quad (2.2.1.16)$$

Se $y \leq x$, poiché $I(0,y) + I(x,a) \leq I(0,a)$, la funzione $\tilde{G}(x,y;\omega)$ va a zero per $|\omega| \to \infty$ nel semipiano inferiore; mentre, se $x, y > a$, la funzione $\tilde{G}(x,y;\omega)$ non può soddisfare la stessa proprietà.
Solo all'interno della cavità, può sussistere la completezza dei QNMs.

## 2.2.2. Rappresentazione per la funzione di Green.

Alle frequenze $\omega = \omega_n$, le funzioni $g_-(x,\omega_n)$ e $g_+(x,\omega_n)$ sono proporzionali, quindi si adotta la notazione:

$$g_-(x,\omega_n) \propto g_+(x,\omega_n) = f_n(x) \quad (2.2.2.1)$$

La norma della funzione $f_n(x)$ viene cosi' definita [6]:

$$\langle f_n | f_n \rangle = \frac{dW}{d\omega}\bigg|_{\omega=\omega_n} \quad (2.2.2.2)$$

La funzione $f_n(x)$ viene cosi' normalizzata [7]:

$$f_n^N(x) = f_n(x)\sqrt{\frac{2\omega_n}{\langle f_n | f_n \rangle}} \quad (2.2.2.3)$$

Ciò posto, i residui (2.2.1.12) assumono la forma:

$$R_n = -\frac{f_n(x)f_n(y)}{\langle f_n | f_n \rangle} = -\frac{f_n^N(x)f_n^N(y)}{2\omega_n} \quad (2.2.2.4)$$

e viene determinata la funzione di Green, nell'ipotesi in cui sussiste la completezza, applicando il teorema dei residui [6][8]:

$$G(x,y;t) = \frac{i}{2}\sum_n \frac{f_n^N(x)f_n^N(y)}{\omega_n}e^{-i\omega_n t} \qquad (2.2.2.5)$$

Le condizioni iniziali sulla funzione di Green e la sua derivata conducono alle:

$$\sum_n \frac{f_n^N(x)f_n^N(y)}{\omega_n} = 0 \qquad (2.2.2.6)$$

$$\frac{\rho(x)}{2}\sum_n f_n^N(x)f_n^N(y) = \delta(x-y) \qquad (2.2.2.7)$$

L'equazione delle onde (2.2.1) è del secondo ordine nel tempo, quindi i QNMs hanno una ricorrenza a coppie $[\omega_n, f_n(x)]$, $[-\omega_n^*, f_n^*(x)]$ e la (2.2.2.7) può essere riscritta come:

$$\text{Re}\left[\rho(x)\sum_{\text{Re}\,\omega_n>0} f_n^N(x)f_n^N(y)\right] = \delta(x-y) \qquad (2.2.2.8)$$

dove si è supposto che non vi sono QNMs con $\text{Re}\,\omega_n = 0$.

Le (2.2.2.5)-(2.2.2.7) sono le relazioni di completezza dei QNMs e, nel caso di perdite nulle, si riducono a quelle di un sistema conservativo. La (2.2.2.5) rende conto della completezza in senso stretto, mentre la (2.2.2.7) di quella in senso debole. La (2.2.2.6) non ha un equivalente nel caso conservativo, mentre la (2.2.2.7) ha un fattore *1/2*, dovuto alla ricorrenza per coppie dei QNMs.

### *2.2.2.a. Osservazioni intorno alla norma dei QNMs.*

Si può dimostrare che [6]:

$$\langle f_n | f_n \rangle = 2\omega_n \int_0^R \rho(x)f_n^2(x)dx + i\rho_0^{1/2}f_n^2(R) \qquad (2.2.2.9)$$

dove si integra su una distanza $R > a$ tale che $\rho(R) = \rho_0$.

Si comprende la normalizzazione adottata (2.2.2.3), dalla quale:

$$\langle f_n^N | f_n^N \rangle = 2\omega_n \qquad (2.2.2.10)$$

La norma per i QNMs differisce da quella nel caso conservativo, in quanto:

a) è complessa, difatti compare nell'integrale il termine $f_n^2(x)$ invece di $|f_n(x)|^2$;

b) oltre all'integrale, compare un termine superficiale $i\rho_0^{1/2} f_n^2(R)$.

Inoltre la norma non dipende da *R*, seppure il termine integrale e quello superficiale, separatamente, vi dipendano.

## *Paragrafo 2.3. Parallelo formale con i sistemi conservativi.*

Ora, si imposta il problema dei QNMs in una cavità aperta in un modo più appropriato [7].

Si introduce la coppia di funzioni:

$$\phi(x) = \phi(x, t=0) \tag{2.3.1}$$

$$\hat{\phi}(x) = \rho(x) \frac{\partial \phi(x,t)}{\partial t}\bigg|_{t=0} \tag{2.3.2}$$

Poiché le funzioni $\phi(x)$ e $\hat{\phi}(x)$ sono arbitrarie e possono essere determinate indipendentemente, si può considerare lo sviluppo simultaneo della coppia di funzioni $(\phi, \hat{\phi})$, usando gli stessi coefficienti per entrambe le componenti [5]:

$$\begin{pmatrix} \phi(x) \\ \hat{\phi}(x) \end{pmatrix} = \sum_n a_n \begin{pmatrix} 1 \\ -i\omega_n \rho(x) \end{pmatrix} f_n(x) \tag{2.3.3}$$

La dissipazione distrugge la simmetria per cui $t \to -t$ e $\omega \to -\omega$: diventano misti i gruppi delle autofunzioni, quelli con $\text{Re}\,\omega > 0$ e $\text{Re}\,\omega < 0$.

Nel caso di sistema conservativo, gli autovalori sono reali e $\omega_{-n} = -\omega_n$, $f_{-n}(x) = f_n(x)$; si possono raggruppare i termini come segue [5]:

$$\phi(x) = \sum_{n>0} (a_n + a_{-n}) f_n(x) = \sum_{n>0} \beta_n f_n(x) \tag{2.3.4}$$

$$\hat{\phi}(x) = \rho(x) \sum_{n>0} -i\omega_n (a_n - a_{-n}) f_n(x) = \rho(x) \sum_{n>0} \gamma_n f_n(x) \tag{2.3.5}$$

In questo caso, lo sviluppo simultaneo della coppia di funzioni $(\phi, \hat{\phi})$, usando l'insieme dei modi normali $n = \pm 1, \pm 2, \ldots$, è equivalente allo sviluppo separato di ciascuna funzione, $\phi$ e $\hat{\phi}/\rho$, in modo indipendente (nel senso che $\beta_n$ e $\gamma_n$ sono indipendenti), usando i soli modi normali del semipiano destro delle frequenze $n = 1, 2, \ldots$.

### *2.3.1. Deduzione dello sviluppo a due componenti.*

Si consideri la classe $\Gamma$, costituita dalle coppie di funzioni $(\phi, \hat{\phi})$ che sono definite nella cavità $I = [0, a]$ e rispettano la condizioni nodale e di *"outgoing wave"*:

$$\phi(x = 0) = 0 \qquad (2.3.1.1)$$

$$\hat{\phi}(x = 0) = 0 \qquad (2.3.1.2)$$

$$\hat{\phi}(x = a^+) = -\rho_0^{1/2} \phi'(x = a^+) \qquad (2.3.1.3)$$

dove l'apice indica l'operatore di derivata rispetto a *x*.

Si osserva un altro motivo per cui è opportuno trattare con coppie di funzioni: la condizione di *"outgoing wave"* non può essere definita in termini di una sola funzione.

Si estende la funzione $\phi(x)$ all'intervallo $[0, \infty)$, con le condizioni [5]:

a) $\rho(x) = \rho_0$ per $x > a$;

b) $\phi(x) \to 0$ per $x \to \infty$;

c) $\hat{\phi}(x) = -\rho_0^{1/2} \phi'(x)$ per $x > a$;

Queste estensioni sono puramente matematiche, quindi la $\phi(x)$ non coincide con la funzione d'onda effettiva fuori della cavità.

Si consideri il campo $\phi(x, t)$ che soddisfi l'equazione delle onde:

$$\left[ \rho(x) \frac{\partial^2}{\partial t^2} - \frac{\partial^2}{\partial x^2} \right] \phi(x, t) = 0 \qquad (2.3.1.4)$$

con le condizioni al contorno:

$$\phi(x = 0, t) = 0 \qquad (2.3.1.5)$$

$$\phi(x, t) \to 0 \quad \text{per} \quad x \to \infty \qquad (2.3.1.6)$$

e le condizioni iniziali:

$$\phi(x, t = 0) = \phi(x) \qquad (2.3.1.7)$$

$$\rho(x) \frac{\partial \phi}{\partial t} \bigg|_{t=0} = \hat{\phi}(x) \qquad (2.3.1.8)$$

Se la funzione di Green è $G(x, y; t)$, il problema ha soluzione [5]:

$$\phi(x,t) = \int_0^\infty \left[ G(x,y;t)\hat{\phi}(y) + \partial_t G(x,y;t)\rho(y)\phi(y) \right] dy \qquad (2.3.1.9)$$

In condizioni di completezza, utilizzando la rappresentazione di Green, si dimostra che vale lo sviluppo [5]:

$$\phi(x,t) = \sum_n a_n f_n^N(x) e^{-i\omega_n t} \qquad (2.3.1.10)$$

dove i coefficienti sono forniti da:

$$a_n = \frac{i}{2\omega_n} \left\{ \int_0^{a^+} \left[ f_n^N(y)\hat{\phi}(y) + \hat{f}_n^N(y)\phi(y) \right] dy + \rho_0^{1/2} f_n^N(a)\phi(a) \right\} \qquad (2.3.1.11)$$

avendo posto:

$$\hat{f}_n^N(x) = -i\omega_n \rho(x) f_n^N(x) \qquad (2.3.1.12)$$

Si osservi che sono state utilizzate le autofunzioni normalizzate.

Tenendo conto della (2.3.1.10), si deduce lo sviluppo a due componenti (2.3.3).

Si evidenziano due caratteristiche:

a) il formalismo a due componenti deriva dalla struttura delle condizioni iniziali;
b) i coefficienti dello sviluppo non dipendono dall'estensione arbitraria dei dati iniziali fuori dalla cavità.

### 2.3.2. Costruzione dello spazio non hermitiano dei QNMs.

Si consideri uno spazio $W$ sul quale sono definiti una hamiltoniana non hermitiana $H$ e una trasformazione lineare di dualità $D$ [9]:

$$D(\alpha|\Phi\rangle + \beta|\Psi\rangle) = \alpha^* D|\Phi\rangle + \beta^* D|\Psi\rangle \qquad (2.3.2.1)$$

tale:

$$DH = H^+ D \qquad (2.3.2.2)$$

Una base biortogonale è costituita da due insiemi di autovettori [9]:

$$|F_n\rangle \in W$$

$$|G_n\rangle = D|F_n\rangle \in \widetilde{W} = D(W) \qquad (2.3.2.3)$$

che soddisfano le equazioni:

$$H|F_n\rangle = \omega_n |F_n\rangle \qquad (2.3.2.4)$$

$$H^+|G_n\rangle = \omega_n^* |G_n\rangle \qquad (2.3.2.5)$$

Proiettando le equazioni agli autovalori su $\langle G_n|$ e $|F_n\rangle$, si ottiene che:

$$\langle G_n|F_m\rangle = 0 \quad \text{per} \quad n \neq m \qquad (2.3.2.6)$$

Posto che questi autostati sono completi, ogni vettore può essere sviluppato come:

$$|\Phi\rangle = \sum_n a_n |F_n\rangle \qquad (2.3.2.7)$$

con:

$$a_n = \frac{\langle G_n|\Phi\rangle}{\langle G_n|F_n\rangle} \qquad (2.3.2.8)$$

Segue che l'operatore identità e quello di evoluzione nel tempo possono essere sviluppati come [4]:

$$I = \sum_n \frac{|F_n\rangle\langle G_n|}{\langle G_n|F_n\rangle} \qquad (2.3.2.9)$$

$$e^{-iHt} = \sum_n \frac{|F_n\rangle e^{-i\omega_n t}\langle G_n|}{\langle G_n|F_n\rangle} \qquad (2.3.2.10)$$

### *2.3.2.a. Primo passo: spazio lineare dei QNMs.*

La classe $\Gamma$ delle coppie di funzioni $(\phi, \hat{\phi})$, che sono definite nella cavità $I = [0, a]$ e soddisfano la condizione nodale e di *"outgoing wave"* (2.3.1.1)-(2.3.1.3), costituisce uno spazio lineare nel campo degli scalari complessi.

Si usa il ket per denotare il vettore colonna:

$$|\Phi\rangle = \begin{pmatrix} \phi(x) \\ \hat{\phi}(x) \end{pmatrix} \qquad (2.3.2.11)$$

I QNMs sono le coppie di funzioni $(f_n^N, \hat{f}_n^N) \in \Gamma$, dove la prima componente è l'autofunzione normalizzata $f_n^N(x)$ e la seconda è il suo momento coniugato dato da (2.3.1.12); gli autovalori sono le frequenze complesse $\omega_n$, con $\text{Im}\,\omega_n < 0$, che descrivono la velocità di smorzamento dell'oscillazione.

Con la notazione dei ket:

$$|F_n^N\rangle = \begin{pmatrix} f_n^N(x) \\ \hat{f}_n^N(x) \end{pmatrix} \qquad (2.3.2.12)$$

### 2.3.2.b. Secondo passo: Hamiltoniana H e trasformazione di dualità D.

Definito il vettore $|\Phi(t)\rangle = (\phi, \hat{\phi})$, dove la prima componente è il campo $\phi(x,t)$ e la seconda è il suo momento coniugato $\hat{\phi}(x,t) = \rho(x)\partial_t \phi$, il problema della dinamica è ben descritto dall'Hamiltoniana [10]:

$$H = i \begin{pmatrix} 0 & \rho^{-1}(x) \\ \partial_x^2 & 0 \end{pmatrix} \qquad (2.3.2.13)$$

ed è risolto tramite un'eq. formalmente analoga a quella di Schroedinger [10]:

$$\frac{\partial}{\partial t}|\Phi(t)\rangle = -iH|\Phi(t)\rangle \qquad (2.3.2.14)$$

Definita la trasformazione [10]:

$$|\Psi\rangle = H|\Phi\rangle \qquad (2.3.2.15)$$

dove, le singole componenti sono:

$$\psi(x) = i\rho^{-1}(x)\hat{\phi}(x) \qquad (2.3.2.16)$$

$$\hat{\psi}(x) = i\partial_x^2 \phi(x) \qquad (2.3.2.17)$$

l'operatore $H$ è valido, in quanto mantiene le funzioni d'onda nello spazio $\Gamma$ esteso; le (2.3.2.16) e (2.3.2.17) soddisfano la condizione di "*outgoing wave*":

$$\hat{\psi}(x) = -\rho_0^{1/2}\psi'(x) \quad \text{per} \quad x > a \qquad (2.3.2.18)$$

La trasformazione di dualità è la seguente [10]:

$$D\begin{pmatrix}\phi_1 \\ \phi_2\end{pmatrix} = -i\begin{pmatrix}\phi_2^* \\ \phi_1^*\end{pmatrix} \qquad (2.3.2.19)$$

### 2.3.2.c. Terzo passo: prodotto interno.

La definizione naturale di prodotto interno tra $|\Psi\rangle$ e $|\Phi\rangle$ su $[0,\infty)$ è la seguente:

$$\langle \Psi | \Phi \rangle = \int_0^\infty \left[ \psi^*(x)\phi(x) + \hat{\psi}^*(x)\hat{\phi}(x) \right] dx \qquad (2.3.2.20)$$

Tenendo conto del comportamento asintotico assunto, l'integrale è convergente.

Per le cavità aperte, un concetto importante è il prodotto interno tra il duale di un vettore $|\Psi\rangle$ e un altro $|\Phi\rangle$, a cui si dà la notazione compatta [4]:

$$\langle \Psi, \Phi \rangle = \langle D\Psi | \Phi \rangle = i\int_0^\infty \left[ \hat{\psi}(x)\phi(x) + \psi(x)\hat{\phi}(x) \right] dx \qquad (2.3.2.21)$$

Questa mappa è bilineare, nel senso che è simmetrica e lineare rispetto ad entrambi i vettori bra e ket (piuttosto che coniugata lineare rispetto al bra), e gioca il ruolo di un prodotto interno generalizzato.

La notazione adottata non distingue tra le funzioni definite nella cavità $I = [0, a]$, che appartengono allo spazio $W$, e quelle estese all'intervallo $[0, \infty)$, che appartengono ad uno spazio $U$. Il prodotto interno include funzioni d'onda definite fuori della cavità $I$, quindi sembra che sia definito su $U$ piuttosto che su $W$. Tuttavia, si possono eliminare del tutto i gradi di libertà dell'esterno; utilizzando le condizioni di *"outgoing wave"*, la (2.3.2.21) può essere riscritta solo in termini delle variabili interne [10]:

$$\langle \Psi, \Phi \rangle = i\int_0^{a^+} \left[ \hat{\psi}(x)\phi(x) + \psi(x)\hat{\phi}(x) \right] dx + i\rho_0^{1/2}\psi(a)\phi(a) \qquad (2.3.2.22)$$

Cosi', il prodotto interno è definito su $W$. Inoltre, si osserva che:

a) la (2.3.2.22) ben si accorda con la norma di un QNM (2.2.2.9); il vantaggio è che non vi è riferimento ad alcun autovalore poiché è stata utilizzata una seconda componente.

b) I coefficienti (2.3.1.11) possono essere riscritti come:

$$a_n = \frac{1}{2\omega_n} \langle F_n, \Phi \rangle \qquad (2.3.2.23)$$

### 2.3.2.d. Quarto passo: simmetria di H ed ortogonalità dei QNMs.

Definito il prodotto interno generalizzato, o meglio la mappa bilineare, (2.3.2.22), dato un operatore $A$, si definisce il suo aggiunto $A^\dagger$ come segue:

$$\langle \psi | \{ A^\dagger | \phi \rangle \} = \langle \phi | \{ A | \psi \rangle \} \qquad (2.3.2.24)$$

L'operatore $A$ è autoaggiunto, o meglio simmetrico ($A^\dagger = A$), quando:

$$\langle \psi | \{ A | \phi \rangle \} = \langle \phi | \{ A | \psi \rangle \} = \langle \psi | A | \phi \rangle \qquad (2.3.2.25)$$

Se l'operatore *A* è simmetrico, i suoi autovettori destri:

$$A|\Phi_n\rangle = a_n|\Phi_n\rangle \qquad (2.3.2.26)$$

formano una base ortogonale [11]:

$$\langle \Phi_n, \Phi_m \rangle = 0 \quad \text{per} \quad n \neq m \qquad (2.3.2.27)$$

Si dimostra che l'Hamiltoniana *H* è un operatore simmetrico [10].

I QNMs $|F_n\rangle$, che sono gli autovettori destri dell'Hamiltoniana, $H|F_n\rangle = \omega_n|F_n\rangle$, formano una base ortogonale, $\langle F_n, F_m \rangle = 0$ per $n \neq m$; quindi, lo sviluppo (2.3.3) è unico.

In definitiva, solo con il formalismo a due componenti, si può recuperare, oltre alla completezza, anche l'ortogonalità dei QNMs.

## *Paragrafo 2.4. Applicazione dei QNMs:*
## *seconda quantizzazione delle cavità aperte.*

Si vuole introdurre la meccanica quantistica nei sistemi costituiti da cavità [12]. Dapprima si consideri una cavità chiusa di lunghezza *a*. Il sistema è conservativo, con modi normali (NMs) ordinati secondo un indice discreto $n = 1, 2, \ldots$, con numeri d'onda $p_n \cong n\pi/a$ spaziati di $\Delta p \cong \pi/a$. I campi quantistici sono distrutti e creati dagli operatori $a_n$ ed $a_n^+$, mentre i processi di ordine più alto comportano somme discrete $\Sigma_n$.

Ora si consideri una cavità aperta, sempre di lunghezza *a*: il sistema è dissipativo, vale a dire l'energia si disperde verso l'esterno, e non può essere normalmente quantizzato da solo [13].

Come primo approccio, si considera anche il bagno esterno ove l'energia fluisce, cosicché l'intero sistema è conservativo. La cavità è immersa in un universo con dimensione $\Lambda \to \infty$: la quantizzazione è sui modi dell'universo, con numero d'onda p, spaziati di $\Delta p \cong \pi/\Lambda \to 0$; i campi quantistici sono distrutti e creati dagli operatori $a(p)$ e $a^\dagger(p)$, mentre processi di ordine più alto comportano integrali circuitali $\oint dp$.

In questo paragrafo, la quantizzazione del campo è implementata tramite i QNMs: in particolare, si esprimeranno la funzione di correlazione per il campo e la densità dei modi in termini dei QNMs.

### 2.4.1. Prima metodologia: funzione di Green.

Questa metodologia è molto semplice ed ha il vantaggio che si applica a situazioni dove il formalismo a due componenti è più complicato [12].

La meccanica quantistica del sistema costituito dalla cavità e dal bagno esterno è determinata dall'Hamiltoniana:

$$H = \int_0^\infty h(x)dx = \int_0^\infty \left[ \frac{\hat{\phi}^2}{2\rho} + \frac{1}{2}\left(\frac{\partial \phi}{\partial x}\right)^2 \right]dx \qquad (2.4.1.1)$$

assieme alla regola canonica di commutazione:

$$[\phi(x), \hat{\phi}(y)] = i\delta(x-y) \qquad (2.4.1.2)$$

Si concentra l'attenzione sul propagatore ritardato:

$$G^R(x,y;t) = -i\theta(t)\langle [\phi(x,t), \phi(y)] \rangle \qquad (2.4.1.3)$$

dove $\phi$ è considerato come operatore, $\langle ... \rangle$ denota il valore atteso alla temperatura $T$, $\theta(t)$ è la funzione gradino unitario.

Tenendo conto della (2.4.1.2), si dimostra che il propagatore $G^R(x,y;t)$ corrisponde alla funzione di Green $G(x,y;t)$ [14], che, se vale la completezza, è stata sviluppata in termini dei QNMs secondo la (2.2.2.5); quindi, il propagatore può essere valutato senza introdurre esplicitamente un'espansione per l'operatore di campo.

Si definisce la funzione di correlazione per il campo all'equilibrio [14]:

$$F(x,y;t) = \langle \phi(x,t), \phi(y) \rangle \qquad (2.4.1.4)$$

Si dimostra che la funzione di correlazione si ottiene dal propagatore ritardato, nel dominio delle frequenze, tramite la [15]:

$$\tilde{F}(x,y;\omega) = -\frac{2}{1-e^{-\beta\hbar\omega}} \operatorname{Im} \tilde{G}^R(x,y;\omega) \qquad (2.4.1.5)$$

dove $\beta = 1/k_B T$, con $k_B$ costante di Boltzman, e $\hbar = h/2\pi$, con h costante di Planck.

Trasformando secondo Fourier la (2.2.2.5) ed inserendola nella (2.4.1.5), si ottiene:

$$\widetilde{F}(x,y;\omega) = \frac{i\omega}{1-e^{-\beta\hbar\omega}} \sum_n \frac{f_n^N(x)f_n^N(y)}{\omega_n(\omega^2-\omega_n^2)} \qquad (2.4.1.6)$$

La funzione di correlazione $\widetilde{F}(x,y;\omega)$ presenta due classi di poli:

a) quelli di Matsubara, alle frequenze $\mu_m = 2m\pi k_B T/\hbar$;

b) quelli dovuti ai QNMs, alle frequenze $\omega_n$.

Si è interessati alla densità degli stati, della quale si parla abbondantemente nella trattazione euristica di Purcell [16].

La densità degli stati locale $d(x,\omega)$ è definita operativamente come [15]:

$$d(x,\omega) = -\frac{2\omega}{\pi} \operatorname{Im} \widetilde{G}^R(x,y;\omega) \qquad (2.4.1.7)$$

Si dimostra che la densità locale degli stati $d(x,\omega)$ si ottiene dalla funzione di correlazione $\widetilde{F}(x,y;\omega)$, per frequenze $\omega$ reali e positive, secondo la [12]:

$$d(x,\omega) = \frac{\omega}{\pi}\left(1-e^{-\beta\hbar\omega}\right)\widetilde{F}(x,x;\omega) \qquad (2.4.1.8)$$

Inserendo la (2.4.1.6) nella (2.4.1.8), si ottiene l'espressione diagonale della densità locale degli stati $d(x,\omega)$ in termini dei QNMs:

$$d(x,\omega) = \frac{\omega}{\pi} \sum_n \operatorname{Im} \frac{\left[f_n^N(x)\right]^2}{\omega_n(\omega_n-\omega)} \qquad (2.4.1.9)$$

Nell'approssimazione di singola risonanza, con un solo termine in *n*, si ottiene:

$$d(x,\omega) \cong \frac{\omega}{\pi} \operatorname{Im} \frac{\left[f_n^N(x)\right]^2}{\omega_n(\omega_n-\omega)} \qquad (2.4.1.10)$$

Emerge un limite della metodologia: la forma diagonale (2.4.1.10), seppure sia semplice, non è definita positiva.

### 2.4.2. Seconda metodologia: sviluppo a due componenti.

Poiché le fluttuazioni quantistiche di punto zero contengono inevitabilmente onde provenienti dall'esterno, si vuole estendere la teoria, affinché sia applicabile a situazioni dove esiste un campo di pompaggio esterno [12].

Non si considera più la classe $\Gamma$, ma un'altra classe $\Gamma'$ di coppie di funzioni $(\phi,\hat{\phi})$ che, se definite nella cavità $I = [0,a]$, rispettino sempre la condizione nodale (2.3.1.1)-(2.3.1.2) e, se estese all'intervallo $[0,\infty)$, posto $\rho(x) = \rho_0$ per $x > a$, sono sempre tali che $\phi(x) \to 0$ per $x \to \infty$ e non verificano più la semplice condizione di *"outgoing wave"*,

ma la seguente [12]:

$$\phi(x) = \phi_{in}(x) + \phi_{out}(x) \qquad (2.4.2.1)$$

dove la $\phi_{in}$ verifica la condizione di *"incoming wave"*:

$$\hat{\phi}_{in}(x) = +\rho_0^{1/2}\phi'_{in}(x) \qquad (2.4.2.2)$$

mentre la $\phi_{out}$ verifica la condizione di "*outgoing wave*":

$$\hat{\phi}_{out}(x) = -\rho_0^{1/2}\phi'_{out}(x) \qquad (2.4.2.3)$$

Il campo $\phi(x,t)$, che soddisfa l'equazione delle onde (2.3.1.4), con le condizioni al contorno (2.3.1.5)-(2.3.1.6) e le condizioni iniziali (2.3.1.7)-(2.3.1.8), non verifica più la semplice condizione di "*outgoing wave*" ma la seguente condizione al contorno [12]:

$$\hat{\phi}(a^+,t) + \rho_0^{1/2}\phi'(a^+,t) = b(t) \qquad (2.4.2.4)$$

dove il termine forzante $b(t)$, determinato dalle condizioni iniziali, è una funzione nota, almeno in senso statistico, che caratterizza le onde provenienti dall'esterno:

$$b(t) = 2\rho_0^{1/2}\hat{\phi}_{in}(a^+,t) \qquad (2.4.2.5)$$

Si mantiene che il campo $\phi(x,t)$ possa essere sviluppato in termini dei QNMs, all'interno della cavità [12]:

$$\phi(x,t) = \sum_n a_n(t) f_n^N(x) \qquad (2.4.2.6)$$

dove i coefficienti dello sviluppo sono forniti dalla:

$$a_n(t) = \frac{1}{2\omega_n}\left\langle F_n^N, \Phi(t)\right\rangle =$$
$$= \frac{i}{2\omega_n}\int_0^{a^+}[f_n^N(x)\hat{\phi}(x,t) + \hat{f}_n^N(x)\phi(x,t)]dx + \frac{i}{2\omega_n}\rho_0^{1/2}f_n^N(a)\phi(a,t) \qquad (2.4.2.7)$$

Anche se non si è nello spazio $\Gamma$, sono mantenute valide la formula di espansione, la notazione e la definizione di prodotto interno. Ciò nonostante, la somma della prima componente converge, ovunque, a $\phi(x,t)$, mentre la somma della seconda componente converge a $\hat{\phi}(x,t)$, eccetto che nel punto ove è imposta la condizione al contorno, cioè $x = a$. Comunque, questa pecca non induce problemi su un insieme di misura zero.

## *2.4.2.a. Regole di commutazione.*

L'intero universo è un sistema conservativo al quale si può applicare la quantizzazione canonica [12]: si promuovono ad operatori le funzioni $\phi(x,t)$ e $\hat{\phi}(x,t)$; la formula di proiezione (2.4.2.7) definisce gli operatori $a_n(t)$.

La formula di proiezione (2.4.2.7), la regola di commutazione (2.4.1.2) e la proprietà di ortogonalità dei QNMs portano a determinare i commutatori degli operatori $a_n(t)$, cioè [12]:

$$[a_n, a_m] = \frac{\omega_m - \omega_n}{4\omega_m \omega_n} \int_0^{a^+} f_n^N(x) \rho(x) f_m^N(x) dx$$

$$= \frac{i\rho_0^{1/2}(\omega_n - \omega_m)}{4\omega_n \omega_m (\omega_n + \omega_m)} f_n^N(a) f_m^N(a) \qquad (2.4.2.8)$$

Per avere un confronto più trasparente con il caso conservativo, è opportuno riscrivere queste espressioni ponendo la sostituzione $n \to -n$, per cui $a_{-n} = a_n^\dagger$, $\omega_{-n} = -\omega_n^*$ e $f_{-n}^N(x) = [f_n^N(x)]^*$. Si ottiene [12]:

$$[a_n^\dagger, a_m] = -\frac{\omega_m + \omega_n^*}{4\omega_m \omega_n^*} \int_0^{a^+} [f_n^N(x)]^* \rho(x) f_m^N(x) dx =$$

$$= -\frac{i\rho_0^{1/2}(\omega_n^* + \omega_m)}{4\omega_n^* \omega_m (\omega_n^* - \omega_m)} [f_n^N(a)]^* f_m^N(a) \qquad (2.4.2.9)$$

Per uno smorzamento finito, gli operatori $a_n(t)$ hanno carattere misto di distruzione e creazione.

Definiti i nuovi operatori:

$$\alpha_n = \sqrt{2\omega_n} a_n \qquad (2.4.2.10)$$

$$\alpha_n^\dagger = \sqrt{2\omega_n^*} a_n^\dagger \qquad (2.4.2.11)$$

lo sviluppo del campo nei QNMs assume la forma [12]:

$$\begin{pmatrix} \phi(x) \\ \hat{\phi}(x) \end{pmatrix} = \sum_{n>0} \begin{pmatrix} (\alpha_n^\dagger + \alpha_n)/\sqrt{2\omega_n} \\ i\rho(x)\sqrt{\omega_n/2}(\alpha_n^\dagger - \alpha_n) \end{pmatrix} f_n^N(x) \qquad (2.4.2.12)$$

Nel limite conservativo, gli operatori (2.4.2.10) e (2.4.2.11) si riducono rispettivamente agli operatori di distruzione e creazione.

## 2.4.2.b. Equazione dinamica di evoluzione.

Partendo dalla (2.4.2.7), si ricava l'equazione del moto per gli operatori $a_n(t)$, cioè [12]:

$$\dot{a}_n(t) + i\omega_n a_n(t) = \frac{i\rho_0^{-1/2}}{2\omega_n} f_n^N(a)b(t) \tag{2.4.2.13}$$

Rispetto al caso di soli onde uscenti, l'equazione non è omogenea: l'accoppiamento con l'esterno è determinato dal valore del QNM nel punto $x = a$; ciascun QNM è guidato dal termine forzante $b(t)$ e assieme decade per la $\mathrm{Im}\,\omega_n$.

Si è supposto che la cavità sia in equilibrio, quindi le condizioni iniziali sono irrilevanti e la dinamica è completamente determinata dalla forza guidante $b(t)$, cioè [12]:

$$a_n(t) = \frac{i\rho_0^{-1/2} f_n^N(a)}{2\omega_n} \int_{-\infty}^{t} b(\tau) e^{i\omega_n(\tau-t)} d\tau \tag{2.4.2.14}$$

La $\mathrm{Im}\,\omega_n$ rende l'integrale rapidamente convergente.

## 2.4.2.c. Funzione di autocorrelazione per il campo e.m.

Dapprima si esprime la densità spettrale degli operatori $a_n(t)$; dopo aver trasformato secondo Fourier la (2.4.2.14), si ottiene [12]:

$$\langle \tilde{a}_n(\omega) a_m \rangle = \frac{f_n^N(a) f_m^N(a)}{4\rho_0 \omega_n \omega_m (\omega_n - \omega)(\omega_m + \omega)} \langle \tilde{b}(\omega) b \rangle \tag{2.4.2.15}$$

Ora si determina la densità spettrale della forza guidante $b(t)$.

Questa è completamente specificata dalle onde provenienti dalla fune infinita $a < x < \infty$ la cui funzione di correlazione è [12]:

$$\tilde{F}_{free}(x, y; \omega) = \langle \hat{\phi}(x,\omega)\phi(y) \rangle_{free} = K \frac{2\cos[\frac{\omega}{c}(x-y)]}{\omega(1-e^{-\beta\hbar\omega})} \tag{2.4.2.16}$$

dove $K$ è una costante opportuna che restituisce le dimensioni fisiche corrette alla correlazione.

Si parte dalla definizione (2.4.2.4), la si trasforma secondo Fourier, si usa la funzione di correlazione della fune infinita (2.4.2.16) e si ottiene [12]:

$$\langle \tilde{b}(\omega)b \rangle = -(\rho_0^{1/2}\partial_x - i\omega\rho_0)^2 \tilde{F}_{free}(x,x;\omega) = K\rho_0^2 \frac{2\omega}{1-e^{-\beta\hbar\omega}} \qquad (2.4.2.17)$$

Infine si determina la funzione di correlazione per il campo; tenendo conto della definizione (2.4.1.4), dello sviluppo (2.4.2.6), delle densità spettrali dei coefficienti (2.4.2.15) e della forza guidante (2.4.2.17), si ottiene [12]:

$$\tilde{F}(x,y;\omega) = K\frac{\rho_0}{2}\frac{\omega}{1-e^{-\beta\hbar\omega}}\sum_{n,m}\frac{f_n^N(a)f_m^N(a)}{\omega_n\omega_m(\omega_n-\omega)(\omega_m+\omega)}f_n^N(x)f_m^N(y)$$

(2.4.2.18)

Si deduce una rappresentazione fisica chiara dei poli nel piano complesso:
a) i poli di Matsubara, alle frequenze $\mu_m = 2m\pi k_B T/\hbar$, nascono dal carattere termico del rumore entrante;
b) i poli dei QNMs, alle frequenze $\omega_n$, corrispondono a risonanze della cavità. che sono eccitate da questo rumore.

## 2.4.2.d. Densità degli stati.

Si vuole determinare l'espressione non diagonale della densità locale dei modi in termini dei QNMs.

Se si tiene conto del legame tra la densità locale degli stati $d(x,\omega)$ e la funzione di correlazione $\tilde{F}(x,y;\omega)$ per frequenze $\omega$ reali e positive, fornito dalla (2.4.1.8), ed ancora dell'espressione di $\tilde{F}(x,y;\omega)$ in termini dei QNMs, fornito dalla (2.4.2.18), si ottiene [12]:

$$d(x,\omega) = K\rho_0\frac{\omega^2}{2\pi}\sum_{n,m}\frac{f_n^N(a)f_m^N(a)}{\omega_n\omega_m(\omega_n-\omega)(\omega_m+\omega)}f_n^N(x)f_m^N(x) \qquad (2.4.2.19)$$

Applicando la relazione di completezza debole (2.2.2.6), la densità locale dei modi diviene [12]:

$$d(x,\omega) = K\frac{\rho_0}{2\pi}\sum_{n,m}\frac{f_n^N(a)f_m^N(a)}{(\omega-\omega_n)(\omega+\omega_m)}f_n^N(x)f_m^N(x) \qquad (2.4.2.20)$$

Nell'approssimazione di singola risonanza, che, tenendo conto della ricorrenza a coppie dei QNMs $(-\omega_n^*, f_n^*) = (-\omega_{-n}, f_{-n})$, si esprime come $m = -n$, la densità locale dei modi diviene [12]:

$$d(x,\omega) \cong K \frac{\rho_0}{2\pi} \frac{\left|f_n^N(a)f_n^N(x)\right|^2}{[(\omega - \text{Re}\,\omega_n)^2 + (\text{Im}\,\omega_n)^2]} \qquad (2.4.2.21)$$

Questa forma è definita positiva ed, oltretutto, è lorenziana.

In condizioni di risonanze strette, quando:

$$\int_0^{a^+} \rho(x)\left|f_n^N(x)\right|^2 dx \cong 1 \qquad (2.4.2.22)$$

e nel limite conservativo, quando:

$$\left|f_n^N(a)\right|^2 = \frac{2|\text{Im}\,\omega_n|}{\rho_0^{1/2}} \qquad (2.4.2.23)$$

vale la normalizzazione [12]:

$$\int_{ris} d\omega \int_0^{a^+} dx\, \rho(x) d(x,\omega) \cong 1 \qquad (2.4.2.24)$$

La costante $K$ è determinata proprio da questa condizione di normalizzazione.

## *Bibliografia*

# *Capitolo 3. Teoria dei QNMs per i cristalli fotonici. Incidenza normale.*

## *Paragrafo 3.1. Introduzione.*

In questo capitolo, si evidenziano, in modo schematico e coinciso, i risultati finali del lavoro [1], dove si è estesa la teoria dei QNMs ai PC unidimensionali, quando un'onda si propaga in direzione normale.

Nei capitoli successivi, saranno dimostrate in modo rigoroso, del lavoro [1], le ipotesi euristiche, seppur corrette, quali le estensioni sulla norma e sul prodotto interno con due termini superficiali, qui non verificate a scapito di una comprensione più profonda del problema, ed anche la DOM come somma di lorenziane, qui confermata con simulazioni ma non accertata formalmente.

La cavità è completamente aperta quindi, in assenza di pompaggio, valgono le condizioni di *"outgoing wave"*. Queste rendono il sistema non hermitiano: le autofrequenze sono complesse, mentre i modi sono denominati quasinormali (QNMs) poiché, in generale, non é più verificata la loro completezza né l'ortogonalità.

Come nel capitolo precedente, la trattazione procede per tre passi logici.

a) Si recupera la completezza dei QNMs, all'interno della cavità, sotto le condizioni di discontinuità e di *"no tail"* per l'indice di rifrazione, stavolta ad entrambe le estremità; inoltre, si introduce la norma dei QNMs ma definita complessa e, stavolta, con due termini superficiali.

b) Si recupera l'ortogonalità dei QNMs, all'interno della cavità, svolgendo un parallelo con i sistemi conservativi, tramite un formalismo a due componenti; si costruisce uno spazio biortogonale dei QNMs, dove si introduce una definizione di prodotto interno tale che l'operatore del sistema è autoaggiunto. Si osserva che, stavolta, il prodotto interno coinvolge due termini superficiali.

c) Si applica la teoria dei QNMs per la seconda quantizzazione della cavità: stavolta, si suppone vi siano due pompe scorrelate che incidono ad entrambe le estremità; si introducono le regole di commutazione e si determinano la funzione di correlazione per il campo e la densità dei modi. Quindi, l'accoppiamento con l'esterno è determinato dai valori dei QNMs ad entrambe le estremità.

## *Paragrafo 3.2. Estensione della teoria QNM's dalle cavità semiaperte a quelle aperte.*

Si denotano come sistema *S* la cavità $I = [0, a]$ e come bagno *B* il resto dell'universo $I' = (-\infty, a) \cup (a, \infty)$.

Si suppone che l'onda si propaga normalmente alla cavità, quindi il campo dipende dalla sola coordinata spaziale *x*.

Se l'onda è monocromatica, nel dominio dei fasori [2], l'equazione delle onde è quella di Helmotz:

$$\frac{d^2 E}{dx^2} + q^2(x)E(x) = 0 \tag{3.2.1}$$

dove:

$$q(x) = k_0 n(x) \tag{3.2.2}$$

$$k_0 = \frac{\omega_0}{c} \tag{3.2.3}$$

con $n(x)$ profilo d'indice, $\omega_0$ frequenza dell'onda e *c* velocità della luce.

Se l'onda contiene tutte le componenti frequenziali, nel domino di Fourier (si tiene conto del legame fra trasformata di Fourier e fasore [3]), l'equazione delle onde è:

$$\frac{\partial^2 \tilde{E}}{\partial x^2} + \omega^2 \rho(x) \tilde{E}(x, \omega) = 0 \tag{3.2.4}$$

dove:

$$\rho(x) = \left(\frac{n(x)}{c}\right)^2 \tag{3.2.5}$$

Nel dominio del tempo (antitrasformando secondo Fourier [3]), l'equazione delle onde è:

$$\left[\rho(x)\frac{\partial^2}{\partial t^2} - \frac{\partial^2}{\partial x^2}\right] E(x,t) = 0 \tag{3.2.6}$$

Vi è uno scambio di energia tra il sistema $S$ ed il bagno $B$: fuoriescono delle onde dal sistema $S$.

Posto:

$$\rho(x) = \rho_0 \quad \text{per} \quad x < 0 \quad \text{ed} \quad x > a \tag{3.2.7}$$

valgono le condizioni di *"outgoing wave"* [4][1]:

$$\partial_x E(x,t) = +\rho_0^{1/2} \partial_t E(x,t) \quad \text{per} \quad x < 0 \tag{3.2.8}$$

$$\partial_x E(x,t) = -\rho_0^{1/2} \partial_t E(x,t) \quad \text{per} \quad x > a \tag{3.2.9}$$

Ora, si illustra schematicamente l'estensione della teoria QNM's per le cavità doppiamente aperte; in particolare, si evidenziando solo i risultati finali. L'estensione procede secondo i tre passi logici noti, corrispondenti, il primo, al formalismo con la funzione di Green, il secondo, a quello degli spazi biortogonali, il terzo, alla seconda quantizzazione.

I risultati finali sono, per il primo passo, la norma dei QNMs, per il secondo, il prodotto interno dei QNMs, per il terzo, la densità dei modi.

*Primo passo*

<u>Norma dei QNMs</u>

I QNMs sono rappresentati dalle coppie $[\omega_n, f_n(x)]$ che soddisfano la (2.2.4).

Si estende la definizione di norma per i QNMs (2.2.2.9) nella forma [1]:

$$\langle f_n | f_n \rangle = 2\omega_n \int_{-R}^{R} \rho(x) f_n^2(x) dx + i\rho_0^{1/2} [f_n^2(R) + f_n^2(-R)] \tag{3.2.10}$$

dove si è introdotto un $R > a$ tale che $\rho(\pm R) = \rho_0$.

La norma non dipende da $R$, seppure il termine integrale e quelli superficiali, separatamente, vi dipendono.

### *Secondo passo*

*Sviluppo a due componenti*

Si consideri la coppia di funzioni:

$$E(x, t = 0) = E(x) \tag{3.2.11}$$

$$\rho(x)\frac{\partial E}{\partial t}\bigg|_{t=0} = \hat{E}(x) \tag{3.2.12}$$

che soddisfano le condizioni [5][1]:

a) $E(x) \to 0$ \qquad per $x \to \pm\infty$;

b) $\hat{E}(x) = +\rho_0^{1/2} E'(x)$ \quad per $x < 0$;

c) $\hat{E}(x) = -\rho_0^{1/2} E'(x)$ \quad per $x > a$;

In condizioni di completezza, utilizzando la rappresentazione di Green [6], vale lo sviluppo:

$$E(x,t) = \sum_n a_n f_n^N(x) e^{-i\omega_n t} \tag{3.2.13}$$

dove i coefficienti sono forniti da [1]:

$$a_n = \frac{i}{2\omega_n}\left\{\int_0^{a^+}\left[f_n^N(y)\hat{E}(y) + \hat{f}_n^N(y)E(y)\right]dy + \rho_0^{1/2}[f_n^N(a)E(a) + f_n^N(0)E(0)]\right\} \tag{3.2.14}$$

avendo introdotto le autofunzioni normalizzate $f_n^N(x)$, secondo le (2.2.2.2) - (2.2.2.3), e i momenti coniugati $\hat{f}_n^N(x)$, secondo la (2.3.1.12).

*Prodotto interno*

Presa la coppia di elementi:

$$|\Psi\rangle = \begin{pmatrix} \psi(x) \\ \hat{\psi}(x) \end{pmatrix} \quad , \quad |\Phi\rangle = \begin{pmatrix} \phi(x) \\ \hat{\phi}(x) \end{pmatrix} \tag{3.2.15}$$

si estende la definizione di prodotto interno (2.3.2.22) nella forma [1]:

$$\langle\Psi,\Phi\rangle = i\int_0^{a^+}\left[\hat{\psi}(x)\phi(x) + \psi(x)\hat{\phi}(x)\right]dx + i\rho_0^{1/2}[\psi(a)\phi(a) + \psi(0)\phi(0)] \tag{3.2.16}$$

### *Terzo passo*

#### Coppia di pompe esterne

Vi siano due pompe provenienti da destra e da sinistra, caratterizzate dai termini forzanti $b(t)$ e $b^\dagger(t)$, funzioni note, almeno in senso statistico, e determinate dalle condizioni iniziali [7][1]:

$$b^\dagger(t) = 2\rho_0^{1/2} E'_{in}(a^+, t) \tag{3.2.17}$$

$$b(t) = -2\rho_0^{1/2} E'_{in}(0^-, t) \tag{3.2.18}$$

Il campo $E(x,t)$ non verifica più le semplici condizioni di *"outgoing wave"* ma la seguenti condizioni al contorno [7][1]:

$$\hat{E}(a^+, t) + \rho_0^{1/2} E'(a^+, t) = b^\dagger(t) \tag{3.2.19}$$

$$\hat{E}(0^-, t) - \rho_0^{1/2} \tilde{E}'(0^-, t) = b(t) \tag{3.2.20}$$

#### Equazione dinamica

Se i coefficienti dello sviluppo per il campo, in presenza delle due pompe, sono indicati con $a_n(t)$, si può scrivere la loro equazione dinamica.
Si estende la (2.4.2.13) nella forma [1]:

$$\dot{a}_n(t) + i\omega_n a_n(t) = \frac{i\rho_0^{-1/2}}{2\omega_n}[f_n^N(a)b^\dagger(t) + f_n^N(0)b(t)] \tag{3.2.21}$$

#### Regole di commutazione

Se si promuovono i coefficienti $a_n(t)$ ad operatori con carattere di distruzione e creazione, si possono scrivere le loro regole di commutazione.
Si estende la (2.4.2.9) nella forma [1]:

$$[a_n^\dagger, a_m] = -\frac{\omega_m + \omega_n^*}{4\omega_m \omega_n^*}\int_0^{a^+}[f_n^N(x)]^* \rho(x) f_m^N(x)dx$$

$$= -\frac{i\rho_0^{1/2}(\omega_n^* + \omega_m)}{4\omega_n^* \omega_m (\omega_n^* - \omega_m)}\{[f_n^N(a)]^* f_m^N(a) + [f_n^N(0)]^* f_m^N(0)\} \tag{3.2.22}$$

#### Funzione di correlazione per il campo

Si suppone che le forze guidanti $b(t)$ e $b^\dagger(t)$ siano scorrelate:

$$\langle \tilde{b}^\dagger(\omega)b \rangle = \langle \tilde{b}(\omega)b^\dagger \rangle = 0 \tag{3.2.23}$$

Allora, le loro densità spettrali $\langle \tilde{b}(\omega)b \rangle$ e $\langle \tilde{b}^{\dagger}(\omega)b^{\dagger} \rangle$ sono specificate solo dalle onde provenienti rispettivamente dalle funi infinite $a < x < \infty$ e $-\infty < x < 0$.

Per la simmetria del problema, è immediato che entrambe le pompe seguono la stessa distribuzione [7][1]:

$$\langle \tilde{b}^{\dagger}(\omega)b^{\dagger} \rangle = \langle \tilde{b}(\omega)b \rangle = K\rho_0^2 \frac{2\omega}{1-e^{-\beta\hbar\omega}} \qquad (3.2.24)$$

dove $K$ è una costante opportuna che restituisce le dimensioni fisiche corrette alla correlazione.

Ciò posto, si può estendere la densità spettrale degli operatori $a_n(t)$, fornita dalla (2.4.2.15), nella forma [7][1]:

$$\langle \tilde{a}_n(\omega)a_m \rangle = \frac{f_n^N(a)f_m^N(a)}{4\rho_0 \omega_n \omega_m (\omega_n - \omega)(\omega_m + \omega)} \langle \tilde{b}^{\dagger}(\omega)b^{\dagger} \rangle + $$
$$+ \frac{f_n^N(0)f_m^N(0)}{4\rho_0 \omega_n \omega_m (\omega_n - \omega)(\omega_m + \omega)} \langle \tilde{b}(\omega)b \rangle \qquad (3.2.25)$$

Infine, si può estendere la funzione di correlazione per il campo, fornita dalla (2.4.2.18), nella forma [1]:

$$\tilde{F}(x,y;\omega) = K\frac{\rho_0}{2}\frac{\omega}{1-e^{-\beta\hbar\omega}} \sum_{n,m} \frac{f_n^N(a)f_m^N(a) + f_n^N(a)f_m^N(a)}{\omega_n \omega_m (\omega_n - \omega)(\omega_m + \omega)} f_n^N(x)f_m^N(y)$$
$$(3.2.26)$$

*Densità locale dei modi*

Solo se le forze guidanti $b(t)$ e $b^{\dagger}(t)$ sono scorrelate, si può estendere la densità locale dei modi, fornita dalla (2.4.2.20), nella forma [1]:

$$d(x,\omega) = K\frac{\rho_0}{2\pi} \sum_{n,m} \frac{f_n^N(a)f_m^N(a) + f_n^N(0)f_m^N(0)}{(\omega - \omega_n)(\omega + \omega_m)} f_n^N(x)f_m^N(x) \qquad (3.2.27)$$

Nell'approssimazione di singola risonanza, si estende la lorenziana (2.4.2.21) nella forma [1]:

$$d(x,\omega) \cong K\frac{\rho_0}{2\pi} \frac{\left|f_n^N(a)\right|^2 + \left|f_n^N(0)\right|^2}{[(\omega - \text{Re}\,\omega_n)^2 + (\text{Im}\,\omega_n)^2]} \left|f_n^N(x)\right|^2 \qquad (3.2.28)$$

La costante $K$ è determinata dalla condizione di normalizzazione (2.4.2.24).

## *Paragrafo 3.3. Deduzioni della teoria QNM's per i PC.*

In questo paragrafo, si riprende la teoria QNM's delle cavità doppiamente aperte per applicarla ai cristalli fotonici: dapprima si discute intorno alla completezza e alla norma dei QNMs, poi si riportano i risultati per le autofrequenze e per la densità dei modi, in particolare per i PBG simmetrici a quarto d'onda.

### *3.3.1. Completezza e norma dei QNMs.*

Si discute intorno alle funzioni ausiliarie $g_\pm(x,\omega)$ che soddisfano le equazioni:

$$\frac{\partial^2 g_\pm}{\partial x^2} + \frac{n^2(x)}{c^2}\omega^2 g_\pm(x,\omega) = 0 \qquad (3.3.1.1)$$

con le condizioni al contorno [1]:

$$g_\pm(x,\omega) = e^{\pm i n_0 \frac{\omega}{c} x} \quad \text{per} \quad x \to \pm\infty \qquad (3.3.1.2)$$

### *3.3.1.a. Verifica sulla condizione di completezza.*

L'indice di rifrazione è una funzione continua a tratti, con *N* discontinuità:

$$n(x) = \begin{cases} n_0(x) & , \text{ per } x < x_0 \\ n_j(x) & , \text{ per } x_{j-1} < x < x_j \quad , \text{ ove } j \in [1,N] \\ n_{N+1}(x) & , \text{ per } x > x_N \end{cases} \qquad (3.3.1.3)$$

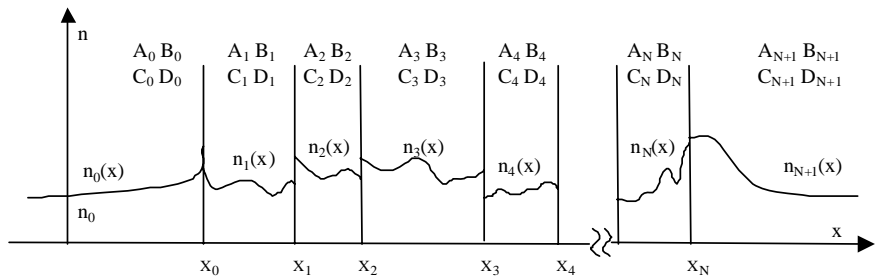

Figura 3.1. Andamento dell'indice di rifrazione n(x),
continuo a tratti, con N discontinuità,
tendente ad un valore costante per x infinito.

Si applica l'approssimazione WKB per frequenze elevate $|\omega| \to \infty$.

Per la funzione $g_-(x,\omega)$, si ottiene [1]:

$$\begin{cases} g_-(x,\omega) = A_j(\omega) e^{i\frac{\omega}{c}\int_x^{x_j} n(\xi)d\xi} + B_j(\omega) e^{-i\frac{\omega}{c}\int_x^{x_j} n(\xi)d\xi} & \text{per } x_{j-1} < x < x_j \\ g_-(x,\omega) = e^{-i\frac{\omega}{c}\int_x^{x_0} n(\xi)d\xi} & \text{per } x < x_0 \end{cases} \quad (3.3.1.4)$$

dove $j \in [1, N+1]$ ed $x_{N+1} = +\infty$.

Per la funzione $g_+(x,\omega)$, si ottiene [1]:

$$\begin{cases} g_+(x,\omega) = C_j(\omega) e^{i\frac{\omega}{c}\int_{x_{j-1}}^{x} n(\xi)d\xi} + D_j(\omega) e^{-i\frac{\omega}{c}\int_{x_{j-1}}^{x} n(\xi)d\xi} & \text{per } x_{j-1} < x < x_j \\ g_+(x,\omega) = e^{i\frac{\omega}{c}\int_{x_N}^{x} n(\xi)d\xi} & \text{per } x > x_N \end{cases} \quad (3.3.1.5)$$

dove $j \in [0, N]$ ed $x_{-1} = -\infty$.

Quindi $\forall (x,y)/x_0 < y \leq x < x_N$ si ha che $\exists k / x_0 < y \leq x_k$ e $x_{k-1} \leq x < x_N$; per $k \leq m \leq N$ e $1 \leq n \leq k$ si ottengono le:

$$g_-(y,\omega) = A_n(\omega) e^{i\frac{\omega}{c}\int_y^{x_n} n(\xi)d\xi} + B_n(\omega) e^{-i\frac{\omega}{c}\int_y^{x_n} n(\xi)d\xi} \quad (3.3.1.6)$$

$$g_+(x,\omega) = C_m(\omega) e^{i\frac{\omega}{c}\int_{x_{m-1}}^{x} n(\xi)d\xi} + D_m(\omega) e^{-i\frac{\omega}{c}\int_{x_{m-1}}^{x} n(\xi)d\xi} \quad (3.3.1.7)$$

Allora, la funzione di Green per $y \leq x$ ha espressione:

$$\tilde{G}(x,y;\omega) = -\frac{\left[A_n(\omega) e^{i\frac{\omega}{c}\int_y^{x_n} n(\xi)d\xi} + B_n(\omega) e^{-i\frac{\omega}{c}\int_y^{x_n} n(\xi)d\xi}\right]\left[C_m(\omega) e^{i\frac{\omega}{c}\int_{x_{m-1}}^{x} n(\xi)d\xi} + D_m(\omega) e^{-i\frac{\omega}{c}\int_{x_{m-1}}^{x} n(\xi)d\xi}\right]}{2i\frac{\omega}{c}n(x)\left[D_m(\omega)B_n(\omega) e^{-i\frac{\omega}{c}\int_{x_{m-1}}^{x_n} n(\xi)d\xi} - C_m(\omega)A_n(\omega) e^{i\frac{\omega}{c}\int_{x_{m-1}}^{x_n} n(\xi)d\xi}\right]}$$

(3.3.1.8)

Si determinano i coefficienti che compaiono nella funzione di Green.

Si impongono le condizioni di raccordo in $x = x_j$. Si ottengono le:

$$\begin{pmatrix} A_{j+1} \\ B_{j+1} \end{pmatrix} = S_j \begin{pmatrix} A_j e^{-i\vartheta_j} + R_j B_j e^{-i\vartheta_j} \\ R_j A_j e^{i\vartheta_j} + B_j e^{i\vartheta_j} \end{pmatrix} \quad (3.3.1.9)$$

$$\begin{pmatrix} C_j \\ D_j \end{pmatrix} = S'_j \begin{pmatrix} C_{j+1} e^{-i\vartheta_{j-1}} - R_j D_{j+1} e^{-i\vartheta_{j-1}} \\ -R_j C_{j+1} e^{i\vartheta_{j-1}} + D_{j+1} e^{i\vartheta_{j-1}} \end{pmatrix} \quad (3.3.1.10)$$

dove:

$$\begin{cases} \vartheta_j = \dfrac{\omega}{c} \int\limits_{x_j}^{x_{j-1}} n(x)dx \\ R_j = \dfrac{[n(x_j^+) - n(x_j^-)]}{[n(x_j^+) + n(x_j^-)]} \\ S_j = \dfrac{[n(x_j^+) + n(x_j^-)]}{2n(x_j^+)} \\ S'_j = \dfrac{[n(x_j^+) + n(x_j^-)]}{2n(x_j^-)} \end{cases} \qquad (3.3.1.11)$$

Se si tiene conto solo dei termini dominanti, si ottengono le:

$$\begin{pmatrix} A_n \\ B_n \end{pmatrix} \cong \prod_{k=0}^{n-1} S_k \begin{pmatrix} R_n e^{i\sum_{k=0}^{n-2}\vartheta_k - \vartheta_{n-1}} \\ e^{i\sum_{k=0}^{n-1}\vartheta_k} \end{pmatrix} \qquad (3.3.1.12)$$

$$\begin{pmatrix} C_m \\ D_m \end{pmatrix} \cong \prod_{k=m}^{N} S'_k \begin{pmatrix} R_N R_m e^{i\sum_{k=m}^{N-1}\vartheta_k - \vartheta_{m-1}} \\ -R_N e^{i\sum_{k=m-1}^{N-1}\vartheta_k} \end{pmatrix} \qquad (3.3.1.13)$$

che valgono $\forall n \in [0, N]$ e $\forall m \in [-1, N-1]$.

Poiché $B_n$ domina su $A_n$, mentre $D_m$ domina su $C_m$, l'espressione della funzione di Green (3.3.1.8) si semplifica,

cioè:

$$\widetilde{G}(x,y,\omega) \cong - \dfrac{e^{-i\frac{\omega}{c}\left[\int_y^{x_0} n(\xi)d\xi + \int_{x_{m-1}}^x n(\xi)d\xi\right]}}{2i\dfrac{\omega}{c}n(x)e^{-i\frac{\omega}{c}\int_{x_{m-1}}^{x_0} n(\xi)d\xi}} \qquad (3.3.1.14)$$

Poiché per $y \leq x$ vale la:

$$\int_y^{x_n} n(\xi)d\xi + \int_{x_{m-1}}^x n(\xi)d\xi = \int_{x_{m-1}}^{x_n} n(\xi)d\xi + \int_y^x n(\xi)d\xi \leq \int_{x_{m-1}}^{x_n} n(\xi)d\xi \qquad (3.3.1.15)$$

la funzione di Green soddisfa la:

$$\widetilde{G}(x,y,\omega) \to 0 \quad \text{per} \quad |\omega| \to \infty \qquad (3.3.1.16)$$

Quindi sussiste la condizione per cui, all'interno del PBG, vale la completezza dei QNMs.

### *3.3.1.b. Calcolo della norma.*

Si suppone che l'indice di rifrazione $n(x)$ sia una funzione costante a tratti, con $N$ discontinuità, definita secondo la:

$$n(x) = \begin{cases} n_0 & , \text{ per } x < x_0 \\ n_j & , \text{ per } x_{j-1} < x < x_j \\ n_{N+1} & , \text{ per } x > x_{N+1} \end{cases}, \text{ ove } j \in [1, N] \qquad (3.3.1.17)$$

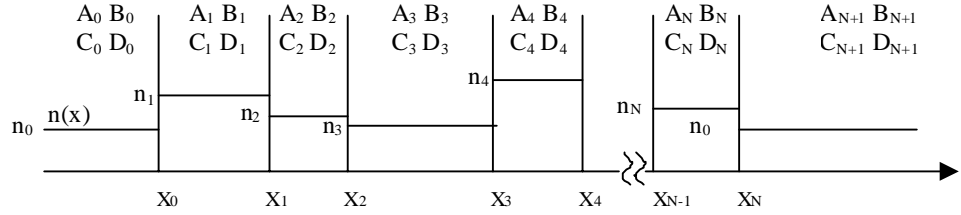

Figura 3.2. Indice di rifrazione n(x) definito dalla (3.3.1.17).

Si risolvono le equazioni (3.3.1.1), tenendo conto dell'indice (3.3.1.17).
Per la funzione $g_-(x, \omega)$ si ottiene [1]:

$$g_-(x, \omega) = \begin{cases} A_0(\omega)e^{in_0\frac{\omega}{c}x} + B_0(\omega)e^{-in_0\frac{\omega}{c}x} & , \text{ per } x < x_0 \\ A_j(\omega)e^{in_j\frac{\omega}{c}x} + B_j(\omega)e^{-in_j\frac{\omega}{c}x} & , \text{ per } x_{j-1} < x < x_j \\ A_{N+1}(\omega)e^{in_{N+1}\frac{\omega}{c}x} + B_{N+1}(\omega)e^{-in_{N+1}\frac{\omega}{c}x} & , \text{ per } x > x_N \end{cases} \quad (3.3.1.18)$$

dove $j \in [1, N]$.

Per la funzione $g_+(x, \omega)$ si ottiene un'espressione analoga, dove vi sono dei coefficienti $C_j(\omega)$ e $D_j(\omega)$ al posto dei $A_j(\omega)$ e $B_j(\omega)$.

S'impongono le condizioni di continuità per le funzioni $g_\pm(x, \omega)$ e le loro derivate ad ogni interfaccia $x_j$, con $j \in [0, N]$.

Quelle per la funzione $g_-(x,\omega)$ sono:

$$\begin{cases} (A_j e^{i\frac{\omega}{c}n_j x} + B_j e^{-i\frac{\omega}{c}n_j x})_{x=x_j} = (A_{j+1} e^{i\frac{\omega}{c}n_{j+1} x} + B_{j+1} e^{-i\frac{\omega}{c}n_{j+1} x})_{x=x_j} \\ \left(\frac{\partial}{\partial x}(A_j e^{i\frac{\omega}{c}n_j x} + B_j e^{-i\frac{\omega}{c}n_j x})\right)_{x=x_j} = \left(\frac{\partial}{\partial x}(A_{j+1} e^{i\frac{\omega}{c}n_{j+1} x} + B_{j+1} e^{-i\frac{\omega}{c}n_{j+1} x})\right)_{x=x_j} \end{cases}$$

(3.3.1.19)

Quelle per la funzione $g_+(x,\omega)$ sono analoghe ma vi sono i coefficienti $C_j$ e $D_j$ al posto dei $A_j$ e $B_j$.

La funzione $g_-(x,\omega)$ è un'onda contropropagante per $x < x_0$, quindi $A_0(\omega) = 0$, mentre la $g_+(x,\omega)$ è propagante per $x > x_N$, quindi $D_{N+1}(\omega) = 0$.

Alle frequenze $\omega_n$ dei QNMs, le funzioni $g_\pm(x,\omega)$ sono proporzionali, quindi $g_-(x,\omega_n) \propto g_+(x,\omega_n) = f_n(x)$; entrambe sono contropropaganti per $x < x_0$, quindi $C_0(\omega_n) = 0$, e propaganti per $x > x_N$, quindi:

$$\begin{cases} B_{N+1}(\omega_n) = 0 \\ A_{N+1}(\omega_n) = C_{N+1}(\omega_n) \end{cases} \quad (3.3.1.20)$$

Per quanto riguarda la funzione $g_-(x,\omega)$, scelta la condizione di normalizzazione, viene fissato il coefficiente $B_0$ e, applicando la (3.3.1.19), sono determinate tutte le coppie di coefficienti sino alla $(A_{N+1}, B_{N+1})$.

Anche per la funzione $g_+(x,\omega)$ si procede analogamente.

Il Wronskiano $W = g_- g'_+ - g_+ g'_-$ é fornito dalla [1]:

$$W(\omega) = \begin{cases} 2in_0 \frac{\omega}{c} B_0(\omega) C_0(\omega) & , \text{ per } x < x_0 \\ 2in_j \frac{\omega}{c}[-A_j(\omega) D_j(\omega) + B_j(\omega) C_j(\omega)] & , \text{ per } x_{j-1} < x < x_j \\ 2in_{N+1} \frac{\omega}{c} B_{N+1}(\omega) C_{N+1}(\omega) & , \text{ per } x > x_{N+1} \end{cases} \quad (3.3.1.21)$$

dove $j \in [0, N]$.

Applicando la (3.3.1.19) per la $g_-(x,\omega)$ e l'analoga per la $g_+(x,\omega)$, si trova che il Wronskiano $W$ non dipende dalla coordinata $x$ poiché le espressioni (3.3.1.21) sono tutte equivalenti. Conviene porre:

$$W(\omega) = 2in_{N+1} \frac{\omega}{c} B_{N+1}(\omega) C_{N+1}(\omega) \quad , \quad \forall x \in \Re \quad (3.3.1.22)$$

La norma dei QNMs, definita come $\langle f_n | f_n \rangle = (dW/d\omega)_{\omega=\omega_n}$, viene determinata utilizzando la (3.3.1.22) e poi la (3.3.1.20). Si ottiene:

$$\langle f_n | f_n \rangle = 2in_{N+1}\frac{\omega_n}{c} A_{N+1}(\omega_n)\left(\frac{dB_{N+1}}{d\omega}\right)_{\omega=\omega_n} \qquad (3.3.1.23)$$

### 3.3.2. Autofrequenze dei QNMs.

Si consideri un 1D PBG tale che ogni cella sia costituita da due materiali, rispettivamente con lunghezza ed indice di rifrazione $h, n_h$ e $l, n_l$.

Il calcolo per le autofrequenze dei QNMs si basa sul metodo delle matrici [8] che utilizza i polinomi di Chebyshev [1].

Con riferimento alla (3.3.1.18), posto $A_0(\omega) = 0$, si determina $B_{N+1}(\omega)$, quindi si risolve l'equazione $B_{N+1}(\omega_n) = 0$.

### 3.3.2.a. Calcolo per un PBG simmetrico a quarto d'onda.

Si divide lo spazio x in un numero di regioni $2N+3$ dove l'indice di rifrazione $n(x)$ è costante. Nell'intervallo $(x_{j-1}, x_j)$, con $j = 0, 1, \ldots, 2N+1, 2N+2$, dove $x_{-1} = -\infty$ e $x_{2N+1} = \infty$, l'indice di rifrazione assume il valore costante $n_j$ definito da:

$$n_j = \begin{cases} 1 & , \text{ per } j = 0, 2N+2 \\ n_h & , \text{ per } j = 1, 3, \ldots, 2N-1, 2N+1 \\ n_l & , \text{ per } j = 2, 4, \ldots, 2N \end{cases} \qquad (3.3.2.1)$$

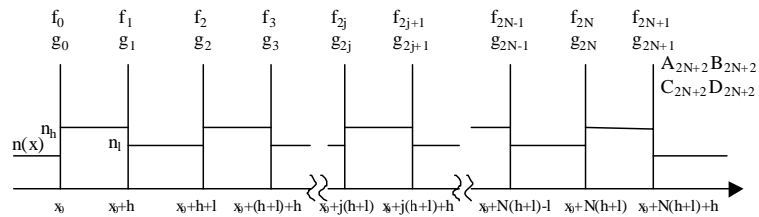

Figura 3.3. Indice di rifrazione n(x) per un 1D-PBG simmetrico definito dalla (3.3.2.1).

Si introducono le fasi:

$$\begin{cases} \delta_l = q_l l = n_l l \dfrac{\omega}{c} \\ \delta_h = q_h h = n_h h \dfrac{\omega}{c} \end{cases} \quad (3.3.2.2)$$

dove si è denotata la frequenza con $\omega$ e la velocità della luce con $c$.

Si trova che le autofrequenze dei QNMs soddisfano l'eq. trascendente [1]:

$$\alpha \sum_{j=0}^{\left[\frac{N-1}{2}\right]} \frac{(-1)^j}{j!} \frac{(N-1-j)!}{(N-1-2j)!} (\gamma)^{N-1-2j} + \beta \sum_{j=0}^{\left[\frac{N-2}{2}\right]} \frac{(-1)^j}{j!} \frac{(N-2-j)!}{(N-2-2j)!} (\gamma)^{N-2-2j} = 0$$

(3.3.2.3)

a coefficienti reali [1]:

$$\begin{cases} \alpha = \dfrac{1}{4}\left\{\left[\left(n_l + \dfrac{1}{n_l}\right) + 2\left(n_h + \dfrac{1}{n_h}\right) - 4 - 2\left(\dfrac{n_h}{n_l} + \dfrac{n_l}{n_h}\right) + \left(\dfrac{n_h^2}{n_l} + \dfrac{n_l}{n_h^2}\right)\right] e^{2i\delta_h + i\delta_l} + \right. \\ \qquad + \left[\left(n_l + \dfrac{1}{n_l}\right) - 2\left(n_h + \dfrac{1}{n_h}\right) - 4 + 2\left(\dfrac{n_h}{n_l} + \dfrac{n_l}{n_h}\right) + \left(\dfrac{n_h^2}{n_l} + \dfrac{n_l}{n_h^2}\right)\right] e^{-2i\delta_h + i\delta_l} + \\ \qquad + \left[2\left(n_l + \dfrac{1}{n_l}\right) - 2\left(\dfrac{n_h^2}{n_l} + \dfrac{n_l}{n_h^2}\right)\right] e^{i\delta_l} + \\ \qquad + \left[-\left(n_l + \dfrac{1}{n_l}\right) + 2\left(n_h + \dfrac{1}{n_h}\right) - 4 + 2\left(\dfrac{n_h}{n_l} + \dfrac{n_l}{n_h}\right) - \left(\dfrac{n_h^2}{n_l} + \dfrac{n_l}{n_h^2}\right)\right] e^{2i\delta_h - i\delta_l} + \\ \qquad + \left[-\left(n_l + \dfrac{1}{n_l}\right) - 2\left(n_h + \dfrac{1}{n_h}\right) - 4 - 2\left(\dfrac{n_h}{n_l} + \dfrac{n_l}{n_h}\right) - \left(\dfrac{n_h^2}{n_l} + \dfrac{n_l}{n_h^2}\right)\right] e^{-2i\delta_h - i\delta_l} + \\ \qquad \left. + \left[-2\left(n_l + \dfrac{1}{n_l}\right) + 2\left(\dfrac{n_h^2}{n_l} + \dfrac{n_l}{n_h^2}\right)\right] e^{-i\delta_l} \right\} \\ \beta = \left\{\left[2 - \left(n_h + \dfrac{1}{n_h}\right)\right] e^{i\delta_h} + \left[2 + \left(n_h + \dfrac{1}{n_h}\right)\right] e^{-i\delta_h}\right\} \\ \gamma = \dfrac{1}{4 n_h n_l}\left\{(n_h + n_l)^2 e^{i(\delta_h + \delta_l)} - (n_h - n_l)^2 e^{i(\delta_h - \delta_l)} - (n_h - n_l)^2 e^{i(\delta_l - \delta_h)} + (n_h + n_l)^2 e^{-i(\delta_h + \delta_l)}\right\} \end{cases}$$

(3.3.2.4)

Si osservi che l'equazione non dipende dal sistema di riferimento, in altre parole da $x_0$, ma solo dai parametri del PBG.

Si suppone che il PBG sia a quarto d'onda, allora:

$$\begin{cases} n_l l = m_l \dfrac{\lambda_{rif}}{4} \\ n_h h = m_h \dfrac{\lambda_{rif}}{4} \end{cases} \quad (3.3.2.5)$$

dove $\lambda_{rif}$ è una lunghezza d'onda di riferimento e $m_l, m_h \in \mathbb{Z}$.

Ciò posto, le fasi (3.3.2.2) vengono riscritte come:

$$\begin{cases} \delta_l = m_l \delta \\ \delta_h = m_h \delta \end{cases} \qquad (3.3.2.6)$$

dove si è introdotto $\delta = (\lambda_{rif}/4)(\omega/c)$.

La (3.3.2.3) diviene un'equazione polinomiale a coefficienti reali di grado $N(m_l + m_h) + m_h$ nella variabile composta $e^{2i\delta}$. Vi sono $N(m_l + m_h) + m_h$ famiglie di QNMs; presa una generica famiglia, i QNMs della stessa hanno la parte immaginaria uguale e si ripetono periodicamente in direzione dell'asse reale con passo $\Delta = \pi m_l c / n_l l$.

Le autofunzioni dei QNMs si costruiscono secondo la [1]:

$$f(x) = \left( A_0 e^{i\frac{\omega}{c}x} + B_0 e^{-i\frac{\omega}{c}x} \right) \vartheta(-x) +$$
$$+ \sum_{j=0}^{N} \left( A_{2j+1} e^{i\frac{\omega}{c}n_h x} + B_{2j+1} e^{-i\frac{\omega}{c}n_h x} \right) \vartheta(x - j(h+l)) \vartheta(j(h+l) + h - x) +$$
$$+ \sum_{j=0}^{N-1} \left( A_{2j+2} e^{i\frac{\omega}{c}n_l x} + B_{2j+2} e^{-i\frac{\omega}{c}n_l x} \right) \vartheta(x - j(h+l) - h) \vartheta((j+1)(h+l) - x) +$$
$$+ \left( A_{2N+2} e^{i\frac{\omega}{c}x} + B_{2N+2} e^{-i\frac{\omega}{c}x} \right) \vartheta(x - N(h+l) - h)$$

(3.3.2.7)

dove si è introdotta la funzione gradino unitario $\vartheta(x)$.

I coefficienti della (3.3.2.7) si costruiscono secondo le [1]:

$$\begin{cases}
\begin{pmatrix} A_1 \\ B_1 \end{pmatrix} = \frac{1}{2} \begin{pmatrix} e^{-in_h \frac{\omega}{c} x_0} & \frac{1}{n_h} e^{-in_h \frac{\omega}{c} x_0} \\ e^{in_h \frac{\omega}{c} x_0} & \frac{-1}{n_h} e^{in_h \frac{\omega}{c} x_0} \end{pmatrix} \begin{pmatrix} e^{in_0 \frac{\omega}{c} x_0} & e^{-in_h \frac{\omega}{c} x_0} \\ n_0 e^{in_h \frac{\omega}{c} x_0} & -n_0 e^{-in_h \frac{\omega}{c} x_0} \end{pmatrix} \begin{pmatrix} A_0 \\ B_0 \end{pmatrix} \\
\begin{pmatrix} A_{2j} \\ B_{2j} \end{pmatrix} = \frac{1}{2} \begin{pmatrix} e^{-in_l \frac{\omega}{c}(x_0+(j-1)(h+l)+h)} & \frac{1}{n_l} e^{-in_l \frac{\omega}{c}(x_0+(j-1)(h+l)+h)} \\ e^{in_l \frac{\omega}{c}(x_0+(j-1)(h+l)+h)} & \frac{-1}{n_l} e^{in_l \frac{\omega}{c}(x_0+(j-1)(h+l)+h)} \end{pmatrix} \begin{pmatrix} e^{in_h \frac{\omega}{c}(x_0+(j-1)(h+l)+h)} & e^{-in_h \frac{\omega}{c}(x_0+(j-1)(h+l)+h)} \\ n_h e^{in_h \frac{\omega}{c}(x_0+(j-1)(h+l)+h)} & -n_h e^{-in_h \frac{\omega}{c}(x_0+(j-1)(h+l)+h)} \end{pmatrix} \begin{pmatrix} A_{2j-1} \\ B_{2j-1} \end{pmatrix} \\
\begin{pmatrix} A_{2j+1} \\ B_{2j+1} \end{pmatrix} = \frac{1}{2} \begin{pmatrix} e^{-in_h \frac{\omega}{c}(x_0+j(h+l))} & \frac{1}{n_h} e^{-in_h \frac{\omega}{c}(x_0+j(h+l))} \\ e^{in_h \frac{\omega}{c}(x_0+j(h+l))} & \frac{-1}{n_h} e^{in_h \frac{\omega}{c}(x_0+j(h+l))} \end{pmatrix} \begin{pmatrix} e^{in_l \frac{\omega}{c}(x_0+j(h+l))} & e^{-in_l \frac{\omega}{c}(x_0+j(h+l))} \\ n_l e^{in_l \frac{\omega}{c}(x_0+j(h+l))} & -n_l e^{-in_l \frac{\omega}{c}(x_0+j(h+l))} \end{pmatrix} \begin{pmatrix} A_{2j} \\ B_{2j} \end{pmatrix} \\
\begin{pmatrix} A_{2N+2} \\ B_{2N+2} \end{pmatrix} = \frac{1}{2} \begin{pmatrix} e^{-in_0 \frac{\omega}{c}(x_0+N(h+l)+h)} & \frac{1}{n_0} e^{-in_0 \frac{\omega}{c}(x_0+N(h+l)+h)} \\ e^{in_0 \frac{\omega}{c}(x_0+N(h+l)+h)} & \frac{-1}{n_0} e^{in_0 \frac{\omega}{c}(x_0+N(h+l)+h)} \end{pmatrix} \begin{pmatrix} e^{in_h \frac{\omega}{c}(x_0+N(h+l)+h)} & e^{-in_h \frac{\omega}{c}(x_0+N(h+l)+h)} \\ n_h e^{in_h \frac{\omega}{c}(x_0+N(h+l)+h)} & -n_h e^{-in_h \frac{\omega}{c}(x_0+N(h+l)+h)} \end{pmatrix} \begin{pmatrix} A_{2N+1} \\ B_{2N+1} \end{pmatrix}
\end{cases}$$

(3.3.2.8)

dove $A_0(\omega) = 0$ e $B_{2N+1}(\omega_n) = 0$.

Dalla (3.3.1.23), la norma per la famiglia *n*-esima di QNMs vale [1]:

$$\langle f_n | f_n \rangle = 2\omega_n cost \tag{3.3.2.9}$$

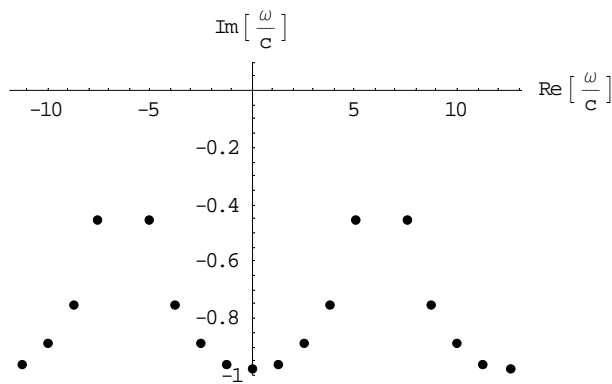

Figura 3.4. QNMs per un PBG simmetrico a quarto d'onda, ove $\lambda_{rif} = 1\mu m$, con numero dei periodi $N = 4$ ed indici di rifrazione $n_h = 1.41$, $n_l = 1$.

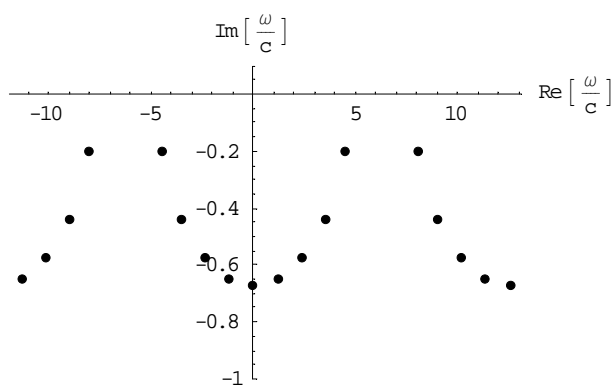

Figura 3.5. QNMs per un PBG simmetrico a quarto d'onda, ove $\lambda_{rif} = 1\mu m$, con numero dei periodi $N = 4$ ed indici di rifrazione $n_h = 2$, $n_l = 1$.

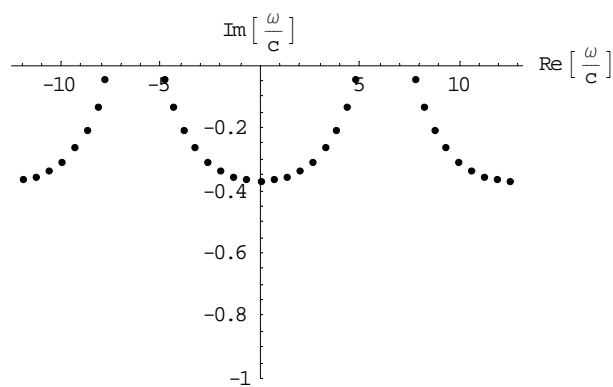

Figura 3.6. QNMs per un PBG simmetrico a quarto d'onda, ove $\lambda_{rif} = 1\mu m$, con numero dei periodi $N = 8$ ed indici di rifrazione $n_h = 2$, $n_l = 1$.

I QNMs non si distribuiscono uniformemente nello spazio ma tendono ad addensarsi, generando dei veri e propri gap. I QNMs hanno come parte reale la frequenza di risonanza e come parte immaginaria la larghezza di riga.

Dalla figura 3.4, i QNMs più vicini al gap hanno parte immaginaria più piccola in modulo, quindi sono quelli con picco di risonanza più stretto.

Dalle figure 3.4 e 3.5, se il salto d'indice aumenta, il gap si allarga e la parte immaginaria dei QNMs tende a diminuire in modulo, quindi i picchi di risonanza si stringono.

Dalle figure 3.5 e 3.6, se il numero dei periodi aumenta, la posizione del gap rimane invariata: come prima, la parte immaginaria dei QNMs diminuisce in modulo e i picchi di risonanza si stringono.

### *3.3.3. Densità dei modi.*

La densità dei modi $d(\omega)$ si ottiene da quella locale $d(x,\omega)$ saturando sulla variabile spaziale $x \in [0, a^+]$ con peso $\rho(x)$. Quindi [7][1]:

$$d(\omega) = \int_0^{a^+} dx \rho(x) d(x,\omega) \qquad (3.3.3.1)$$

Dalla (3.3.3.1), se si inserisce la densità locale (3.2.27), si ottiene:

$$d(\omega) = K \frac{\rho_0}{2\pi} \sum_{n,m} \frac{f_n^N(a) f_m^N(a) + f_n^N(0) f_m^N(0)}{(\omega - \omega_n)(\omega + \omega_m)} \int_0^{a^+} f_n^N(x) \rho(x) f_m^N(x) dx \quad (3.3.3.2)$$

La densità dei modi (3.3.3.2) è reale.

I QNMs ricorrono a coppie: per ogni modo $[\omega_n, f_n(x)]$ esiste il corrispondente complesso coniugato $[\omega_{-n}, f_{-n}(x)] = [-\omega_n^*, f_n^*(x)]$.

Quindi, nella (3.3.3.2), per ogni termine:

$$I_{n,m} = K \frac{\rho_0}{2\pi} \frac{f_n^N(a) f_m^N(a) + f_n^N(0) f_m^N(0)}{(\omega - \omega_n)(\omega + \omega_m)} \int_0^{a^+} f_n^N(x) \rho(x) f_m^N(x) dx \qquad (3.3.3.3)$$

esiste il corrispondente complesso coniugato $I_{-n,-m} = I_{n,m}^*$.

Dalla normalizzazione (2.2.2.2) e dalla definizione di prodotto interno (3.2.16), si ottiene che [1]:

$$\int_0^{a^+} f_n^N(x) \rho(x) f_m^N(x) dx = \delta_{n,m} - i \frac{\rho_0^{1/2}}{\omega_n + \omega_m} [f_n^N(a) f_m^N(a) + f_n^N(0) f_m^N(0)] \quad (3.3.3.4)$$

La densità dei modi (3.3.3.2) diviene:

$$\begin{aligned} d(\omega) = & K \frac{\rho_0}{2\pi} \sum_n \frac{[f_n^N(a)]^2 + [f_n^N(0)]^2}{(\omega^2 - \omega_n^2)} + \\ & -iK \frac{\rho_0^{3/2}}{2\pi} \sum_{n,m} \frac{[f_n^N(a) f_m^N(a) + f_n^N(0) f_m^N(0)]^2}{(\omega - \omega_n)(\omega + \omega_m)(\omega_n + \omega_m)} \end{aligned} \qquad (3.3.3.5)$$

### 3.3.3.a. Calcolo per strutture multistrato simmetriche.

Per una struttura simmetrica, le autofunzioni $f_n^N(x)$ sono a parità definita. Come per i sistemi hermitiani, se il modo *n*-esimo ha una certa parità, il modo *(n+1)*-esimo ha parità opposta. In particolare [1]:

$$f_n^N(a) = (-1)^n f_n^N(0) \tag{3.3.3.6}$$

La densità (3.3.3.5) diviene:

$$d(\omega) = K \frac{\rho_0}{\pi} \sum_n \frac{[f_n^N(0)]^2}{(\omega^2 - \omega_n^2)} - iK \frac{\rho_0^{3/2}}{2\pi} \sum_{n,m} \frac{[1+(-1)^{n+m}]^2 [f_n^N(0) f_m^N(0)]^2}{(\omega - \omega_n)(\omega + \omega_m)(\omega_n + \omega_m)} \tag{3.3.3.7}$$

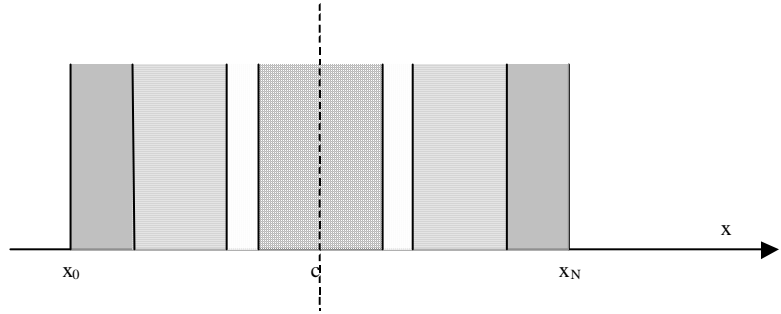

Figura 3.7: Struttura multistrato simmetrica rispetto al suo centro.
L'indice di rifrazione è funzione di $|x - c|$.

I QNMs possono essere interpretati in termini di frequenze di risonanza; poiché, nell'approssimazione di singola risonanza, la densità locale dei modi ha la forma di una lorenziana, si propone, nel caso generale, una densità dei modi ottenuta come somma di lorenziane, con parametri forniti dalla parte reale e immaginaria dei QNMs [1].

Definita la densità dei modi adimensionata:

$$d_a(\omega) = c d(\omega) \tag{3.3.3.8}$$

dove *c* è la velocità della luce, si propone [1]:

$$d_a(\omega/\omega_{rif}) = \frac{1}{\pi^2} \sum_n \frac{|\text{Im}(\omega_n/\omega_{rif})|}{[(\omega/\omega_{rif}) - \text{Re}(\omega_n/\omega_{rif})]^2 + \text{Im}^2(\omega_n/\omega_{rif})} \tag{3.3.3.9}$$

dove $\omega_{rif}$ è una frequenza di riferimento (per i PBG, quella di centro banda).

Si ricorda che, nel lavoro [1], la densità dei modi si riferisce a due pompe scorrelate.

Il riscontro della plausibilità dell'ipotesi sta nel fatto che le simulazioni della (3.3.3.9) per i PBG simmetrici mostrano una densità dei modi con le caratteristiche illustrate nell'articolo [9].

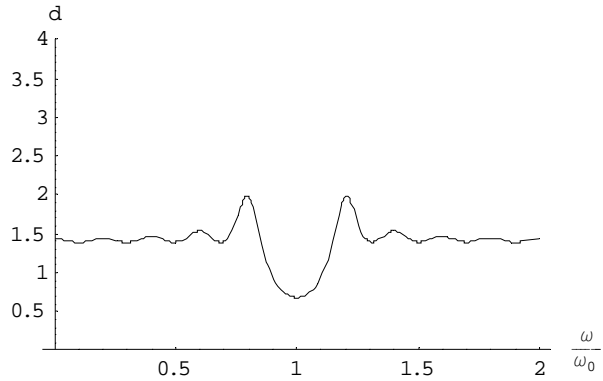

Figura 3.8. DOM definita dalla (3.3.3.9), per un PBG simmetrico a quarto d'onda, ove $\lambda_{rif} = 1\mu m$, con numero dei periodi $N = 4$ ed indici di rifrazione $n_h = 1.41$, $n_l = 1$.

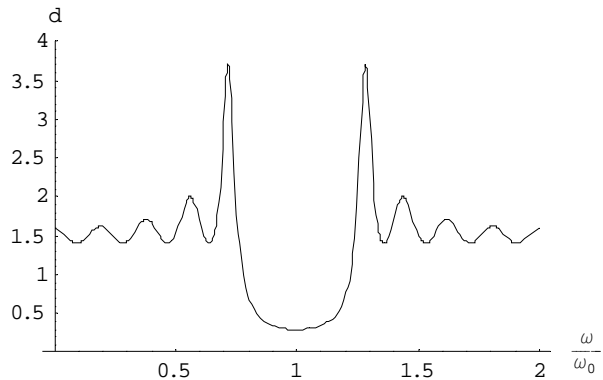

Figura 3.9. DOM definita dalla (3.3.3.9), per un PBG simmetrico a quarto d'onda, ove $\lambda_{rif} = 1\mu m$, con numero dei periodi $N = 4$ ed indici di rifrazione $n_h = 2$, $n_l = 1$.

Normalizzando la DOM (3.3.3.9) alla velocità di gruppo (1.4.3.14) si ottiene una grandezza che tende ad uno nelle bande passanti.

### 3.3.3.b. Calcolo per un PBG simmetrico a quarto d'onda.

Qui si specializza la densità dei modi (3.3.3.7) per un PBG simmetrico a *λ/4*.
Dapprima si tiene conto:
a) della normalizzazione sui QNMs, introdotta dalle (2.2.2.2) e (2.2.2.3);
b) che le autofunzioni, costruite tramite la (3.3.2.7), hanno $A_0 = 0$ e, per convenienza, $B_0 = 1$, da cui $f_n(0) = 1$;
c) che il termine $(1+(-1)^{n+m})^2$ vale *0* per $n+m$ dispari e *4* per $n+m$ pari.

La densità dei modi (3.3.3.7) diviene [1]:

$$d(\omega) = K\frac{\rho_0}{\pi} \sum_n \frac{2\omega_n}{\langle f_n | f_n \rangle} \frac{1}{(\omega^2 - \omega_n^2)} +$$
$$-iK\frac{\rho_0^{3/2}}{\pi} \sum_{n,m} \frac{2\omega_n}{\langle f_n | f_n \rangle} \frac{2\omega_m}{\langle f_m | f_m \rangle} \frac{(1+(-1)^{n+m})}{(\omega - \omega_n)(\omega + \omega_m)(\omega_n + \omega_m)}$$
(3.3.3.10)

Poi si tiene conto che, per un PBG simmetrico a $\lambda/4$:

a) i QNMs sono suddivisi in famiglie indipendenti di numero: $N_F = N(m_l + m_h) + m_h$;

b) le autofrequenze della n-esima famiglia sono infinite, con parte immaginaria identica e parte reale periodica, e sono indicate con:

$\omega_{n+j\Delta}$, $j \in \mathbb{Z}$;

c) le autofunzioni della famiglia n-esima hanno la stessa norma, vale a dire

$\langle f_{n+j\Delta} | f_{n+j\Delta} \rangle = \langle f_n | f_n \rangle = 2\omega_n \cdot cost$, $j \in \mathbb{Z}$.

La densità dei modi (3.3.3.10) diviene [1]:

$$d(\omega) = K\frac{\rho_0}{\pi} \sum_{n=1}^{N_F} \sum_{j=-\infty}^{\infty} \frac{2\omega_{n+j\Delta}}{\langle f_{n+j\Delta} | f_{n+j\Delta} \rangle} \frac{1}{(\omega^2 - \omega_{n+j\Delta}^2)} +$$
$$-iK\frac{\rho_0^{3/2}}{\pi} \sum_{n,m=1}^{N_F} \sum_{j,k=-\infty}^{\infty} \frac{2\omega_{n+j\Delta}}{\langle f_{n+j\Delta} | f_{n+j\Delta} \rangle} \frac{2\omega_{m+k\Delta}}{\langle f_{m+k\Delta} | f_{m+k\Delta} \rangle} \cdot$$
$$\cdot \frac{[1+(-1)^{n+m+(j+k)\Delta}]}{(\omega - \omega_{n+j\Delta})(\omega + \omega_{m+k\Delta})(\omega_{n+j\Delta} + \omega_{m+k\Delta})}$$
(3.3.3.11)

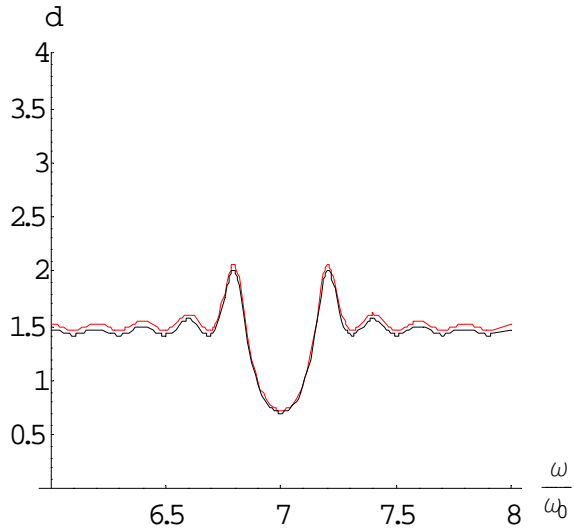

Figura 3.10. Sovrapposizione delle (3.3.3.11) (nero) e (3.3.3.9) (rosso), per un PBG simmetrico a quarto d'onda, ove $\lambda_{rif} = 1\mu m$, con numero dei periodi $N = 4$, ed indici di rifrazione $n_h = 1.41$, $n_l = 1$.

## *Bibliografia.*

# *Capitolo 4. QNMs TE e TM per cavità aperte.*
# *Norma e prodotto interno.*

## *Paragrafo 4.1. Introduzione.*

In questo capitolo, si utilizza il modello teorico dei QNMs per studiare il problema elettromagnetico di un'onda che si propaga in una cavità aperta lungo una direzione obliqua; nello specifico, si discute intorno alla norma ed il prodotto interno per i quasi modi TE e TM.

Si prosegue un filo logico: nel capitolo 2, è stata introdotta la teoria dei QNMs per le cavità semiaperte [1] che prevedeva un prodotto interno per i QNMs complesso e con un solo termine superficiale [2]; all'inizio del capitolo 3, è stata estesa la teoria per le cavità aperte ma l'estensione era limitata al caso d'incidenza normale e, solo euristicamente, suggeriva per il prodotto interno due termini superficiali [3].

Questo capitolo prevede un corpo principale ed un'appendice: nel corpo principale, si studiano prima i quasi modi TE e poi quelli TM; nell'appendice, si riportano, solo per il caso TE, le dimostrazioni, in quanto, per il caso TM, queste seguono la stessa logica.

Si prosegue il lavoro [3]. Introdotto un angolo d'incidenza $\theta_0$, per i quasi modi TE, nell'ambito del metodo della funzione di Green [4]:

a) si introducono le funzioni ausiliarie $g_\pm(z,\omega)$, con le note condizioni di asintoticità (e risonanza) [4] e si osserva che non sono delle onde e.m., poiché non rispettano le condizioni radiative [5]; tuttavia, si impone che soddisfino delle condizioni formali di *"outgoing waves"* e si evidenzia l'analogia tra le condizioni di asintoticità e quelle di *"outgoing waves"*;

b) tramite le condizioni di asintoticità (e risonanza), utilizzando una procedura già nota [4], si deduce una prima formula per la norma, definita su tutto lo spazio, che è coerente con quella del lavoro [3]; invece, tramite le condizioni formali di *"outgoing waves"*, si perviene ad una seconda formula per la norma, definita solo dentro la cavità; infine, dopo aver dimostrato l'equivalenza tra i due metodi di calcolo, si giustifica l'introduzione delle condizioni formali di *"outgoing waves"*, poiché si sceglie la seconda formula come definizione operativa per la norma;

e, sempre per i quasi modi TE, nell'ambito del formalismo a due componenti [6][2]:

c) qualora sussistano le condizioni di completezza per i quasi modi [1], si evidenzia il ruolo delle condizioni radiative e di *"outgoing waves"* per il campo elettrico, affinché questo possa essere sviluppato nei quasi modi all'interno della cavità;

d) si osserva come, proprio da queste condizioni, nasce la definizione di prodotto interno, che è coerente con quella del lavoro [3]; inoltre, giustificata l'ortogonalità dei quasi modi, si intuisce, per cavità simmetriche con perdite minime, una formula approssimata per la norma in termini delle sole quasi autofrequenze.

Infine, si trattano i quasi modi TM: si definisce una trasformazione di dilatazione [7], che è necessaria per l'analisi formale; si estendono il metodo della funzione di Green e il formalismo a due componenti, da cui si deducono, sempre per i quasi modi TM, la definizione di norma, di prodotto interno e l'ortogonalità.

## *Paragrafo 4.2. Modi quasi normali TE.*

Come primo passo, base degli sviluppi successivi, si vuole definire il problema della propagazione elettromagnetica, in un mezzo isotropo e non magnetico, di un'onda TE, piana e omogenea.

Inizialmente, si suppone che il campo sia monocromatico, poi, dopo aver introdotto per semplicità le ipotesi di parassialità e la legge generalizzata di Snell [8], si estende con immediatezza il problema al caso non monocromatico.

Con riferimento alla figura 1.1.a, il campo elettrico è del tipo [7]:

$$E_x = E(z)e^{ik_0 \alpha y} \qquad (4.2.1)$$

dove $\alpha = n_0 \cdot \sin\theta_0 = cost$, con $\theta_0$ angolo di propagazione e $n_0$ indice di rifrazione dell'esterno, e $k_0 = \dfrac{\omega_0}{c}$, con $\omega_0$ frequenza dell'onda e $c$ velocità della luce nel vuoto.

L'ampiezza complessa $E(z)$ soddisfa l'equazione di Helmholtz [7]:

$$\frac{d^2 E}{dz^2} + q^2(z) E(z) = 0 \qquad (4.2.2)$$

dove la componente del vettore d'onda $q(z)$ lungo la direzione $z$ è tale che:

$$q^2(z) = k_0^2 (n^2(z) - \alpha^2) \qquad (4.2.3)$$

con $n(z)$ profilo per l'indice di rifrazione.

Si vuole trascurare l'effetto trasversale per studiare il problema con riferimento alla sola coordinata spaziale $z$.

Si introduce l'ipotesi di parassialità; la dimensione trasversale $L_T$ della cavità sia tale che:

$$|k_0 n_0 L_T| \ll 2\pi \qquad (4.2.4)$$

Il campo è ben rappresentato dalla sola ampiezza complessa $E(z)$ in quanto $|k_0 \alpha y| < |k_0 n_0 L_T| \ll 2\pi$, da cui $e^{ik_0 \alpha y} \cong 1$.

Vale la legge generalizzata di Snell:

$$n(z)\,\text{sen}\,\theta(z) = n_0\,\text{sen}\,\theta_0 = \alpha = cost \qquad (4.2.5)$$

dove $\theta(z)$ è il profilo d'angolo.

La componente $q(z)$ diviene:

$$q(z) = k_0 n(z) \cos\theta(z) \qquad (4.2.6)$$

Si suppone che l'onda TE non sia più monocromatica ma contenga tutte le componenti frequenziali.

Tenendo conto del legame tra fasore e trasformata di Fourier [5], si passa al dominio di Fourier; per l'ipotesi di parassialità, l'onda è ben rappresentata dalla trasformata $\widetilde{E}(z,\omega)$ che, per la legge generalizzata di Snell, soddisfa l'equazione:

$$\frac{\partial^2 \widetilde{E}}{\partial z^2} + \omega^2 \rho(z) \widetilde{E}(z,\omega) = 0 \qquad (4.2.7)$$

dove:

$$\rho(z) = \left(\frac{n(z)\cos\theta(z)}{c}\right)^2 \qquad (4.2.8)$$

Antitrasformando secondo Fourier [5], si passa al dominio del tempo; l'onda, rappresentata da $E(z,t)$, soddisfa l'equazione:

$$\left[\frac{\partial^2}{\partial z^2} - \rho(z)\frac{\partial^2}{\partial t^2}\right]E(z,t) = 0 \qquad (4.2.9)$$

Se la cavità è $C = [0, L]$, posto $\rho(z) = \rho_0$ per $z \leq 0$ e per $z \geq L$, l'onda soddisfa le condizioni di *"outgoing waves"*:

$$\begin{cases} \partial_z E = \sqrt{\rho_0}\,\partial_t E & \text{per } z \leq 0 \\ \partial_z E = -\sqrt{\rho_0}\,\partial_t E & \text{per } z \geq L \end{cases} \qquad (4.2.10)$$

### *4.2.1. Metodo della funzione di Green.*

In questo paragrafo, si studiano i quasi modi TE, utilizzando il metodo della funzione di Green.

Il paragrafo si suddivide in tre parti. Con riferimento al caso TE:

I) si definisce il problema e.m. per la funzione di Green e si introducono le funzioni ausiliarie associate, che sono necessarie per ogni sviluppo formale successivo;

II) si risolve il problema e.m., ottenendo un'espressione per la funzione di Green in termini delle funzioni ausilarie;

III) sotto le condizioni di completezza per i QNMs, si esprime la funzione di Green in termini dei QNMs all'interno della cavità.

Il lavoro [3], nel caso di incidenza normale, dava per acquisiti questi risultati; nel presente lavoro, per una migliore comprensione dell'argomento, si sono rese necessarie le dimostrazioni.

La funzione di Green TE, nel dominio di Fourier, si indica con $\tilde{G}(z, z', \omega)$ e risolve l'equazione non omogenea [4]:

$$\frac{\partial_z^2 \tilde{G}}{\partial z^2} + \omega^2 \rho(z)\tilde{G}(z, z', \omega) = -\delta(z - z') \qquad (4.2.1.1)$$

mentre, nel dominio del tempo, si indica con $G(z,z',t)$, risolve l'equazione delle onde:

$$\frac{\partial_z^2 G}{\partial z^2} - \rho(z)\frac{\partial_t^2 G}{\partial t^2} = -\delta(z-z')\delta(t) \qquad (4.2.1.2)$$

e rispetta le condizioni di *"outgoing waves"*:

$$\begin{cases} \partial_z G = \sqrt{\rho_0}\partial_t G & \text{per } z' \leq 0 \\ \partial_z G = -\sqrt{\rho_0}\partial_t G & \text{per } z' \geq L \end{cases} \qquad (4.2.1.3)$$

Nel dominio di Fourier, si possono definire delle funzioni ausiliarie TE $g_\pm(z,\omega)$ che risolvono le equazioni omogenee associate [4]:

$$\frac{\partial^2 g_-}{\partial z^2} + \omega^2 \rho(z) g_-(z,\omega) = 0 \qquad (4.2.1.4)$$

$$\frac{\partial^2 g_+}{\partial z^2} + \omega^2 \rho(z) g_+(z,\omega) = 0 \qquad (4.2.1.5)$$

e rispettano le condizioni asintotiche:

$$g_-(z,\omega) = e^{-iq_0(\omega)z} \quad \text{per } z \to -\infty \qquad (4.2.1.6)$$

$$g_+(z,\omega) = e^{iq_0(\omega)z} \quad \text{per } z \to +\infty \qquad (4.2.1.7)$$

dove:

$$q_0(\omega) = n_0 \frac{\omega}{c} \cos\theta_0 \qquad (4.2.1.8)$$

Le autofrequenze dei quasi modi TE, $\omega_n \in \mathbb{C}$ con $\text{Im}\,\omega_n < 0$, soddisfano la condizione di risonanza [4]:

$$g_-(z,\omega_n) \propto g_+(z,\omega_n) = f_n(z) \cong e^{\pm i n_0 \frac{\omega_n}{c}\cos\theta_0 z} \quad \text{per } z \to \pm\infty \qquad (4.2.1.9)$$

In generale, le funzioni ausiliarie TE $g_\pm(z,\omega)$ non possono rappresentare delle onde e.m., poiché non rispettano le condizioni di radiazione; infatti, calcolate alle frequenze di risonanza, divergono per grandi distanze.

In ogni caso, si suppone che le funzioni $g_\pm(z,\omega)$ soddisfino delle condizioni formali di *"outgoing waves"*; tenendo conto dell'equivalenza $\partial_t \Leftrightarrow -i\omega$, queste si scrivono come:

$$\left[\partial_z g_\pm(z,\omega)\right]_{z=0^+} = -i\omega\sqrt{\rho_0}\, g_\pm(0,\omega) \qquad (4.2.1.10)$$

$$\left[\partial_z g_\pm(z,\omega)\right]_{z=L^-} = i\omega\sqrt{\rho_0}\, g_\pm(L,\omega) \qquad (4.2.1.11)$$

Si osserva l'analogia formale tra le condizioni asintotiche (4.2.1.6-7) per $z \to \pm\infty$ e le condizioni formali di *"outgoing waves"* (4.2.1.10-11) agli estremi della cavità $z = 0$ e $z = L$.

Si dimostra che:

a) il wronskiano associato all'equazione (4.2.1.4-5) dipende dalla sola frequenza ed è diverso da zero su tutto lo spazio [4]:

$$W(z,\omega) = g_-(z,\omega)\partial_z g_+(z,\omega) - g_+(z,\omega)\partial_z g_-(z,\omega) = \\ = W(\omega) \neq 0 \quad , \quad \forall z \in \Re \tag{4.2.1.12}$$

b) la funzione di Green TE può essere espressa come [4]:

$$G(z,z',\omega) = \begin{cases} -\dfrac{g_-(z,\omega)g_+(z',\omega)}{W(\omega)} & , \quad z < z' \\ -\dfrac{g_-(z',\omega)g_+(z,\omega)}{W(\omega)} & , \quad z > z' \end{cases} \tag{4.2.1.13}$$

Si suppone che valgano le condizioni di completezza per i QNMs, in particolare quella di discontinuità [1]; la funzione $\rho(z)$ ne abbia una in $z = L$, almeno a gradino, cosicché [4]:

$$\tilde{G}(z,z',\omega) \to 0 \quad \text{per} \quad |\omega| \to \infty \quad , \quad \forall (z,z') \in C = [0,L] \tag{4.2.1.14}$$

Definita la norma dei quasi modi TE come:

$$\langle f_n | f_n \rangle = \left(\dfrac{dW}{d\omega}\right)_{\omega=\omega_n} \tag{4.2.1.15}$$

si dimostra che la funzione di Green TE può essere così sviluppata [4]:

$$G(z,z',t) = i\sum_n \dfrac{f_n(z)f_n(z')}{\langle f_n | f_n \rangle} e^{-i\omega_n t} \quad , \quad \forall (z,z') \in C \tag{4.2.1.16}$$

Segue che, introdotta la normalizzazione:

$$f_n^N(z) = f_n(z)\sqrt{\dfrac{2\omega_n}{\langle f_n | f_n \rangle}} \tag{4.2.1.17}$$

la (4.2.1.16) diviene:

$$G(z,z';t) = \dfrac{i}{2}\sum_n \dfrac{f_n^N(z)f_n^N(z')}{\omega_n} e^{-i\omega_n t} \quad , \quad \forall (z,z') \in C \tag{4.2.1.18}$$

Vedi appendice per le dimostrazioni.

### *4.2.2. Costruzione della norma per i QNMs TE.*

In questo paragrafo, si utilizzano le funzioni ausiliarie secondo il metodo di Green per costruire la norma dei quasi modi TE.

Come anticipato in precedenza, viene seguita questa linea logica:

I) partendo dalle condizioni asintotiche e di risonanza, già note, si ottiene una prima definizione operativa per la norma dei quasi modi TE, definita su tutto lo spazio;

II) partendo dalle condizioni formali di *"outgoing waves"*, introdotte nel presente lavoro, si ottiene una seconda definizione operativa per la norma dei quasi modi TE, che è definita solo dentro la cavità;

III) la verifica che le due definizioni operative sono equivalenti giustifica l'introduzione delle condizioni formali di *"outgoing waves"* e consente di scegliere, come standard, la seconda definizione, più elegante; la norma è complessa e prevede, oltre al termine integrale sulla cavità, anche due termini superficiali, calcolati agli estremi.

Quindi, mentre nel lavoro [3] è stata proposta, solo euristicamente, la prima definizione, nel presente lavoro, seppure utilizzando tecniche di analisi già note [4], se ne verifica la validità e la si raffina.

In sintesi:

a) Utilizzando le condizioni asintotiche (4.2.1.6-7) e di risonanza (4.2.1.9), si perviene alla prima definizione operativa per la norma dei quasi modi TE [3]:

$$\langle f_n | f_n \rangle = 2\omega_n \int_{-R}^{R} \rho(z) f_n^2(z) dz + i\sqrt{\rho_0} [f_n^2(R) + f_n^2(-R)] \qquad (4.2.2.1)$$

dove $R \gg L$ e tale che $\rho(\pm R) = \rho_0$;

b) Utilizzando le condizioni formali di *"outgoing waves"* (4.2.1.10-11), si perviene alla seconda definizione operativa per la norma dei quasi modi TE:

$$\langle f_n | f_n \rangle = 2\omega_n \int_{0^+}^{L^-} \rho(z) f_n^2(z) dz + i\sqrt{\rho_0} [f_n^2(L) + f_n^2(0)] \qquad (4.2.2.2)$$

c) Vi è equivalenza tra le due definizioni operative di norma per i quasi modi TE.

Difatti, la (4.2.2.1) è equivalente alla:

$$\langle f_n | f_n \rangle = 2\omega_n \int_{0^+}^{R} \rho(z) f_n^2(z) dz + i\sqrt{\rho_0} [f_n^2(R) + f_n^2(0)] \qquad (4.2.2.3)$$

cioè, sottraendo membro a membro le (4.2.2.1) e (4.2.2.3), vale la:

$$2\omega_n \int_{-R}^{0^+} \rho(z) f_n^2(z) dz = i\sqrt{\rho_0} \left[ f_n^2(0) - f_n^2(-R) \right] \qquad (4.2.2.4)$$

Inoltre, la (4.2.2.3) è equivalente alla (4.2.2.2), cioè, sottraendo membro a membro le (4.2.2.3) e (4.2.2.2), vale la:

$$2\omega_n \int_{L^-}^{R} \rho(z) f_n^2(z) dz = i\sqrt{\rho_0} \left[ f_n^2(L) - f_n^2(R) \right] \qquad (4.2.2.5)$$

Vedi appendice per le dimostrazioni.

### *4.2.3. Sviluppo a due componenti.*

In questo paragrafo, si studia il campo elettrico con polarizzazione TE, tramite il formalismo a due componenti.

Il paragrafo si suddivide in due parti. Con riferimento al caso TE:

I) si impongono le condizioni radiative sulla distribuzione iniziale di campo elettrico, e, tramite le condizioni di *"outgoing waves"*, si introduce una posizione per determinare la distribuzione iniziale del momento coniugato fuori della cavità;

II) in condizioni di completezza per i QNMs, si sviluppa il campo elettrico come sovrapposizione dei QNMs all'interno della cavità.

Nel lavoro [3], per incidenza normale, non era chiara l'origine della posizione iniziale sul momento coniugato e si dava per acquisito lo sviluppo di campo; nel presente lavoro, si sciolgono i dubbi sulle posizioni iniziali e, seppure con tecniche già note [6], si individua meglio il loro ruolo nella dimostrazione per lo sviluppo di campo.

Il campo elettrico abbia una distribuzione iniziale:

$$E_0(z) = E(z, t=0) \qquad (4.2.3.1)$$

con momento coniugato [6]:

$$\hat{E}_0(z) = \rho(z) \left( \frac{\partial E}{\partial t} \right)_{t=0} \qquad (4.2.3.2)$$

Affinché valga la completezza dei QNMs, per la condizione di *"no tail"* [1]:

$$\rho(z) = \rho_0 \quad \text{per} \quad z \leq 0 \text{ e } z \geq L \tag{4.2.3.3}$$

Il campo soddisfa le condizioni radiative che, applicate alla (4.2.3.1), forniscono la:

$$E_0(z) \to 0 \quad \text{per} \quad z \to \pm\infty \tag{4.2.3.4}$$

Inoltre, il campo soddisfa le condizioni di *"outgoing waves"* (4.2.10); applicandole alla (4.2.3.2), si ottengono proprio le [3]:

$$\begin{cases} \hat{E}_0(z) = \sqrt{\rho_0}\dfrac{dE_0}{dz} & \text{per} \quad z \leq 0 \\ \hat{E}_0(z) = -\sqrt{\rho_0}\dfrac{dE_0}{dz} & \text{per} \quad z \geq L \end{cases} \tag{4.2.3.5}$$

Il campo elettrico può essere rappresentato in termini dei QNMs all'interno della cavità [1]:

$$E(z,t) = \sum_n a_n f_n^N(z) e^{-i\omega_n t} \quad, \quad \forall z \in [0,L] \tag{4.2.3.6}$$

dove i coefficienti dello sviluppo sono forniti proprio da [3]:

$$a_n = \frac{i}{2\omega_n}\left\{\int_{0^+}^{L^-}\left[f_n^N(z)\hat{E}_0(z) + \hat{f}_n^N(z)E_0(z)\right]dz + \sqrt{\rho_0}\left[f_n^N(L)E_0(L) + f_n^N(0)E_0(0)\right]\right\} \tag{4.2.3.7}$$

con:

$$\hat{f}_n^N(z) = -i\omega_n \rho(z) f_n^N(z) \tag{4.2.3.8}$$

Vedi appendice per la dimostrazione.

### *4.2.4. Prodotto interno dei QNMs TE.*

In questo paragrafo, si utilizza il formalismo a due componenti per discutere intorno al prodotto interno dei quasi modi TE.

Come anticipato in precedenza, viene seguita questa linea logica:

I) coerentemente con le condizioni di *"outgoing waves"*, viene definito il prodotto interno per i quasi modi TE, che è complesso e ben si accorda con la norma stessa;

II) giustificata l'ortogonalità dei quasi modi TE nella cavità, si calcolano gli integrali di normalizzazione e si propone, per cavità simmetriche con perdite minime, una formula approssimata per la norma, deducendone alcune caratteristiche.

Rispetto al lavoro [3], si inserisce il prodotto interno dei quasi modi TE, inquadrandolo meglio nell'ambito del formalismo a due componenti ed evidenziando la coerenza con le condizioni di *"outgoing waves"*; in più, tramite la condizione di ortogonalità per i quasi modi TE nella cavità, passando per gli integrali di normalizzazione, si perviene ad una formalizzazione per la norma, approssimata ma significativa.

Il prodotto interno dei quasi modi TE può essere definito sullo spazio biortogonale degli stessi [2]; oltre all'integrale, definito sulla cavità, si aggiungono i due termini superficiali, calcolati agli estremi, tenendo così conto delle condizioni di *"outgoig waves"* per il campo elettrico (4.2.10).

Si introduce la coppia di ket:

$$\left| f_n^N \right\rangle = \begin{pmatrix} f_n^N(z) \\ \hat{f}_n^N(z) \end{pmatrix} \quad , \quad \left| f_m^N \right\rangle = \begin{pmatrix} f_m^N(z) \\ \hat{f}_m^N(z) \end{pmatrix} \qquad (4.2.4.1)$$

dove i momenti coniugati $f_n^N(z)$ e $f_m^N(z)$ sono definiti dalla (4.2.3.8).

Si definisce il prodotto interno come [3]:

$$\left\langle f_n^N, f_m^N \right\rangle = i \int_{0^+}^{L^-} \left[ f_n^N(z) \hat{f}_m^N(z) + \hat{f}_n^N(z) f_m^N(z) \right] dz + i\sqrt{\rho_0} \left[ f_n^N(L) f_m^N(L) + f_n^N(0) f_m^N(0) \right]$$

(4.2.4.2)

Tenendo conto della definizione (4.2.3.8), si osserva che il prodotto interno (4.2.4.2) ben si accorda con la norma (4.2.2.2).

Introdotto il vettore $\left| E(t) \right\rangle = (E, \hat{E})$, dove la prima componente è il campo $E(z,t)$ e la seconda è il suo momento coniugato $\hat{E}(z,t) = \rho(z)\partial_t E(z,t)$, il problema e.m. della propagazione per un'onda piana TE è ben descritto dall'Hamiltoniana:

$$H = \begin{pmatrix} 0 & \rho^{-1}(z) \\ \partial_z^2 & 0 \end{pmatrix} \qquad (4.2.4.3)$$

ed è risolto tramite un'eq. formalmente analoga a quella di Schroedinger [6]:

$$\frac{\partial}{\partial t} \left| E(t) \right\rangle = -iH \left| E(t) \right\rangle \qquad (4.2.4.4)$$

Dalla definizione di prodotto interno (4.2.4.2), si dimostra che l'Hamiltoniana $H$ è un operatore simmetrico e, quindi, che i quasi modi TE $\left|f_n^N\right\rangle$, autovettori destri dell'Hamiltoniana, $H\left|f_n^N\right\rangle = \omega_n\left|f_n^N\right\rangle$, formano una base ortogonale [2]:

$$\left\langle f_n^N, f_m^N \right\rangle = \delta_{n,m}(\omega_n + \omega_m) \tag{4.2.4.5}$$

Segue che gli integrali di normalizzazione sono [3]:

$$\int_{0^+}^{L^-} f_n^N(z)\rho(z)f_m^N(z)dz = \delta_{n,m} - i\frac{\sqrt{\rho_0}}{\omega_n + \omega_m}[f_n^N(L)f_m^N(L) + f_n^N(0)f_m^N(0)]$$

(4.2.4.6)

e che, per cavità simmetriche [3] con perdite minime [9], si può proporre la:

$$\left\langle f_n, f_n \right\rangle \cong \sqrt{\rho_0}\frac{2\omega_n}{\operatorname{Im}\omega_n} \tag{4.2.4.7}$$

Si osserva che:

a) la teoria QNM's si applica alle cavità aperte, si basa sulle condizioni di *"outgoing waves"* che impongono delle perdite ($\operatorname{Im}\omega_n < 0$), quindi non può prevedere il caso conservativo, quando le cavità sono chiuse e non hanno perdite ($\operatorname{Im}\omega_n = 0$);

b) la norma è espressa in termini delle sole autofrequenze QNM's;

c) la norma decresce, in modulo, quando aumentano l'angolo di incidenza ($\rho_0 \to 0$ per $\theta_0 \to \pi/2$) e le perdite ($\left|\operatorname{Im}\omega_n\right| \gg 0$).

Vedi appendice per le dimostrazioni.

## *Paragrafo 4.3. Modi quasi normali TM.*

Nel lavoro [3], la radiazione si propagava normalmente alla cavità aperta, quindi non vi era distinzione tra i quasi modi TE e TM.

Nel presente lavoro, la propagazione è obliqua: finora, sono stati studiati i quasi modi TE, prima con il metodo di Green, determinandone la norma, poi con il formalismo a due componenti, determinandone il prodotto interno e deducendone l'ortogonalità; adesso, si studiano i quasi modi TM, con le stesse metodologie ed intenti.

Come primo passo, base degli sviluppi successivi, si vuole definire il problema della propagazione elettromagnetica, in un mezzo isotropo e non magnetico, di un'onda TM, piana e omogenea.

Come nel caso TE, si introducono le ipotesi di parassialità e la legge generalizzata di Snell. Inoltre, si definisce una trasformazione biunivoca di dilatazione che riduce formalmente il problema e.m. TM in quello TE, così da poter utilizzare risultati già noti per i modi TE; nella trasformazione inversa $TE \Rightarrow TM$, si ottengono i risultati corrispondenti per i modi TM.

Con riferimento alla figura 1.1.b, l'induzione magnetica ha la forma [7]:

$$B_x = B(z)e^{ik_0 \alpha y} \tag{4.3.1}$$

dove $\alpha = n_0 \cdot \sin\theta_0 = cost$, con $\theta_0$ angolo di propagazione e $n_0$ indice di rifrazione dell'esterno, e $k_0 = \dfrac{\omega_0}{c}$, con $\omega_0$ frequenza dell'onda e $c$ velocità della luce nel vuoto.

L'ampiezza complessa $B(z)$ soddisfa l'equazione [7]:

$$\frac{d}{dz}\left(\frac{1}{n^2}\frac{dB}{dz}\right) + \left(\frac{q(z)}{n(z)}\right)^2 B(z) = 0 \tag{4.3.2}$$

dove la componente del vettore d'onda $q(z)$ lungo la direzione $z$ è tale che:

$$q^2(z) = k_0^2 (n^2(z) - \alpha^2) \tag{4.3.3}$$

con $n(z)$ profilo per l'indice di rifrazione.

Si vuole trascurare l'effetto trasversale per studiare il problema con riferimento alla sola coordinata spaziale $z$.

Si introduce l'ipotesi di parassialità; la dimensione traversale $L_T$ della cavità sia tale che:

$$|k_0 n_0 L_T| \ll 2\pi \tag{4.3.4}$$

L'induzione è ben rappresentata dalla sola ampiezza complessa $B(z)$ in quanto $|k_0 \alpha y| < |k_0 n_0 L_T| \ll 2\pi$, da cui $e^{ik_0 \alpha y} \cong 1$.

Vale la legge generalizzata di Snell:

$$n(z)\operatorname{sen}\theta(z) = n_0 \operatorname{sen}\theta_0 = \alpha = cost \tag{4.3.5}$$

dove $\theta(z)$ è il profilo d'angolo.

La componente $q(z)$ diviene:

$$q(z) = k_0 n(z) \cos\theta(z) \qquad (4.3.6)$$

Si definisce la variabile indipendente dilatata $Z$, tale che [7]:

$$dZ = n^2(z)dz \qquad (4.3.7)$$

Vale la trasformazione tra dominio effettivo in $z$ e dominio dilatato in $Z$:

$$\begin{cases} z = 0 & \Leftrightarrow \quad Z = 0 \\ z = L & \Leftrightarrow \quad Z = L_d = \int_0^L n^2(z)dz \\ z = \pm\infty & \Leftrightarrow \quad Z = \pm\infty \end{cases} \qquad (4.3.8)$$

Inoltre, per le funzioni definite nel dominio:

$$\begin{cases} n(z) \Leftrightarrow n_d(Z) \\ \theta(z) \Leftrightarrow \theta_d(Z) \end{cases} \qquad (4.3.9)$$

Posto $B(z) \Leftrightarrow B_d(Z)$, l'equazione TM (4.3.2) con numero d'onda (4.3.6) si trasforma in un'equazione formalmente analoga a quella TE:

$$\frac{d^2 B_d}{dZ^2} + \omega_0^2 \rho_d(Z) B_d(Z) = 0 \qquad (4.3.10)$$

dove:

$$\rho_d(Z) = \left(\frac{1}{c}\frac{\cos\theta_d(Z)}{n_d(Z)}\right)^2 \qquad (4.3.11)$$

Dimostrazione. Primo punto:

$$\frac{dB}{dz} = n_d^2(Z)\frac{dB_d}{dZ} \qquad (4.3.12)$$

$$\frac{1}{n^2(z)}\frac{dB}{dz} = \frac{dB_d}{dZ} \qquad (4.3.13)$$

$$\frac{d}{dz}\left[\frac{1}{n^2(z)}\frac{dB}{dz}\right] = n_d^2(Z)\frac{d^2 B_d}{dZ^2} \qquad (4.3.14)$$

Secondo punto:

$$\frac{q(z)}{n(z)} = k_0 \cos\theta(z) = \frac{\omega_0}{c}\cos\theta_d(Z) \qquad (4.3.15)$$

Terzo punto; si inseriscono nella (4.3.2) le (4.3.14) e (4.3.15):

$$n_d^2(Z)\frac{d^2 B_d}{dZ^2} + \left(\frac{\omega_0}{c}\cos\theta_d(Z)\right)^2 B_d(Z) = 0 \qquad (4.3.16)$$

Dalla (4.3.16), si perviene alle (4.3.10-11).

### *4.3.1. Metodo della funzione di Green.*

Questo metodo, che è stato definito con riferimento ai quasi modi TE, qui viene applicato anche ai quasi modi TM.

Procedendo come nel caso TE, anche nel caso TM, si possono introdurre la funzione di Green $G(z,z',\omega)$ e le funzioni ausiliarie $g_\pm(z,\omega)$.

La funzione di Green $G(z,z',\omega)$ è soluzione dell'equazione non omogenea:

$$\frac{\partial}{\partial z}\left(\frac{1}{n^2}\frac{\partial G}{\partial z}\right)+\omega^2 n^2(z)\rho(z)G(z,z',\omega)=-\delta(z-z') \qquad (4.3.1.1)$$

dove:

$$\rho(z)=\left(\frac{1}{c}\frac{\cos\theta(z)}{n(z)}\right)^2 \qquad (4.3.1.2)$$

Le funzioni ausiliarie $g_\pm(z,\omega)$ sono soluzione dell'equazione omogenea:

$$\frac{\partial}{\partial z}\left(\frac{1}{n^2}\frac{\partial g_\pm}{\partial z}\right)+\omega^2 n^2(z)\rho(z)g_\pm(z,\omega)=0 \qquad (4.3.1.3)$$

Le stesse funzioni soddisfano ancora le condizioni asintotiche:

$$\begin{cases} g_-(z,\omega)=e^{-iq_0(\omega)z} & \text{per } z\to-\infty \\ g_+(z,\omega)=e^{+iq_0(\omega)z} & \text{per } z\to+\infty \end{cases} \qquad (4.3.1.4)$$

dove:

$$q_0(\omega)=n_0\frac{\omega}{c}\cos\theta_0 \qquad (4.3.1.5)$$

Le autofrequenze dei quasi modi TM, $\omega_n\in\mathbb{C}$ con $\text{Im}\,\omega_n<0$, sono determinate ancora dalla condizione di risonanza:

$$g_-(z,\omega_n)\propto g_+(z,\omega_n)=f_n(z)\cong e^{\pm in_0\frac{\omega_n}{c}\cos\theta_0 z} \quad \text{per } z\to\pm\infty \qquad (4.3.1.6)$$

Si suppone che i salti di indice siano sufficientemente piccoli $\Delta n\ll 1$, per cui il wronskiano é in pratica funzione della sola frequenza:

$$W(z,\omega)=g_-(z,\omega)\partial_z g_+(z,\omega)-g_+(z,\omega)\partial_z g_-(z,\omega)\cong W(\omega) \qquad (4.3.1.7)$$

Procedendo come nel caso TE, si può dimostrare, nel caso TM, che:
a) il wronskiano è diverso da zero su tutto lo spazio:

$$W(\omega)\neq 0 \quad , \quad \forall z\in\Re \qquad (4.3.1.8)$$

b) la funzione di Green è esprimibile come:

$$G(z,z',\omega) = \begin{cases} -n^2(z')\dfrac{g_-(z,\omega)g_+(z',\omega)}{W(\omega)} & , \quad z < z' \\ -n^2(z')\dfrac{g_-(z',\omega)g_+(z,\omega)}{W(\omega)} & , \quad z > z' \end{cases} \qquad (4.3.1.9)$$

c) se vale la completezza dei QNMs all'interno della cavità, per cui:

$$\tilde{G}(z,z',\omega) \to 0 \quad \text{per} \quad |\omega| \to \infty \, , \, \forall (z,z') \in C = [0,L] \qquad (4.3.1.10)$$

definita la norma dei QNMs come:

$$\langle f_n | f_n \rangle = \left( \frac{dW}{d\omega} \right)_{\omega = \omega_n} \qquad (4.3.1.11)$$

la funzione di Green è rappresentabile in termini dei QNMs:

$$G(z,z',t) = i \sum_n \frac{n^2(z') f_n(z) f_n(z')}{\langle f_n | f_n \rangle} e^{-i\omega_n t} \quad , \quad \forall (z,z') \in C \qquad (4.3.1.12)$$

Segue che, introdotta la normalizzazione:

$$f_n^N(z) = f_n(z) \sqrt{\frac{2\omega_n}{\langle f_n | f_n \rangle}} \qquad (4.3.1.13)$$

la (4.3.1.12) diviene:

$$G(z,z';t) = \frac{i}{2} \sum_n \frac{n^2(z') f_n^N(z) f_n^N(z')}{\omega_n} e^{-i\omega_n t} \quad , \quad \forall (z,z') \in C \qquad (4.3.1.14)$$

### *4.3.2. Norma dei QNMs TM.*

Procedendo come nel caso TE, si utilizzano le condizioni formali di *"outgoing waves"*:

$$\begin{cases} [\partial_z g_\pm(z,\omega)]_{z=0^+} = -i\omega \sqrt{\rho_0}\, g_\pm(0,\omega) \\ [\partial_z g_\pm(z,\omega)]_{z=L^-} = i\omega \sqrt{\rho_0}\, g_\pm(L,\omega) \end{cases} \qquad (4.3.2.1)$$

dove:

$$\rho_0 = \left( \frac{n_0 \cos\theta_0}{c} \right)^2 \qquad (4.3.2.2)$$

e si perviene alla definizione operativa per la norma dei quasi modi TM:

$$\langle f_n | f_n \rangle = 2\omega_n \int_{0^+}^{L^-} \rho(z) f_n^2(z) n^2(z) dz + i\sqrt{\rho_0'}\,[f_n^2(L) + f_n^2(0)] \qquad (4.3.2.3)$$

dove:

$$\rho'_0 = \left(\frac{1}{c}\frac{\cos\theta_0}{n_0}\right)^2 \tag{4.3.2.4}$$

### *4.3.3. Sviluppo a due componenti.*

Il formalismo a due componenti, che è stato definito con riferimento ai quasi modi TE, qui viene applicato ai quasi modi TM.

Tenendo conto della trasformazione di dilatazione (4.3.7), si passa al dominio dilatato in **Z**, dove il problema elettromagnetico TM si riduce in quello TE; in questo dominio, il formalismo a due componenti per il caso TM è come nel caso TE: di seguito, viene riportato con le ipotesi ed i risultati.

L'induzione magnetica abbia una distribuzione iniziale:

$$B_{d,0}(Z) = B_d(Z, t=0) \tag{4.3.3.1}$$

con momento coniugato [6]:

$$\hat{B}_{d,0}(Z) = \rho_d(Z)\left(\frac{\partial B_d}{\partial t}\right)_{t=0} \tag{4.3.3.2}$$

Affinché valga la completezza dei QNMs, per la condizione di *"no tail"* [1]:

$$\rho_d(Z) = \rho'_0 \quad \text{per} \quad Z \leq 0 \text{ e } Z \geq L_d \tag{4.3.3.3}$$

L'induzione soddisfa le condizioni radiative che, applicate alla (4.3.3.1), forniscono la:

$$B_{d,0}(Z) \to 0 \quad \text{per} \quad Z \to \pm\infty \tag{4.3.3.4}$$

Inoltre, l'induzione soddisfa le condizioni di *"outgoing waves"* che danno luogo alle:

$$\begin{cases} \hat{B}_{d,0}(Z) = \sqrt{\rho'_0}\dfrac{dB_{d,0}}{dZ} & \text{per} \quad Z \leq 0 \\ \hat{B}_{d,0}(Z) = -\sqrt{\rho'_0}\dfrac{dB_{d,0}}{dZ} & \text{per} \quad Z \geq L_d \end{cases} \tag{4.3.3.5}$$

L'induzione magnetica può essere rappresentata in termini dei QNMs all'interno della cavità dilatata:

$$B_d(Z,t) = \sum_n a_n f_{d,n}^N(Z) e^{-i\omega_n t} \quad , \quad \forall Z \in [0, L_d] \tag{4.3.3.6}$$

dove i coefficienti dello sviluppo sono forniti da:

$$a_n = \frac{i}{2\omega_n} \{ \int_{0^+}^{L_d^-} \left[ f_{d,n}^N(Z) \hat{B}_{d,0}(Z) + \hat{f}_{d,n}^N(Z) B_{d,0}(Z) \right] dZ + \\ + \sqrt{\rho_0'} [f_{d,n}^N(L_d) B_{d,0}(L_d) + f_{d,n}^N(0) B_{n,0}(0)] \}$$ (4.3.3.7)

con:

$$\hat{f}_{d,n}^N(Z) = -i\omega_n \rho_d(Z) f_{d,n}^N(Z)$$ (4.3.3.8)

Invertendo la trasformazione di dilatazione (4.3.7), si torna nel dominio *z*; di seguito, viene ridefinito il formalismo a due componenti per il caso TM, ma nel dominio spaziale effettivo.

L'induzione magnetica ha distribuzione iniziale:

$$B_0(z) = B(z, t=0)$$ (4.3.3.9)

con momento coniugato:

$$\hat{B}_0(z) = \rho(z) \left( \frac{\partial B}{\partial t} \right)_{t=0}$$ (4.3.3.10)

Vale la completezza dei QNMs quindi, per la condizione di *"no tail"*:

$$\rho(z) = \rho_0' \quad \text{per} \quad z \leq 0 \text{ e } z \geq L$$ (4.3.3.11)

L'induzione soddisfa le condizioni radiative che forniscono la:

$$B_0(z) \to 0 \quad \text{per} \quad z \to \pm\infty$$ (4.3.3.12)

Le condizioni di *"outgoing waves"* per l'induzione (4.3.3.5) si traducono nelle:

$$\begin{cases} \hat{B}_0(z) = \frac{\sqrt{\rho_0'}}{n_0^2} \frac{dB_0}{dz} & \text{per} \quad z \leq 0 \\ \hat{B}_0(z) = -\frac{\sqrt{\rho_0'}}{n_0^2} \frac{dB_0}{dz} & \text{per} \quad z \geq L \end{cases}$$ (4.3.3.13)

L'induzione può essere rappresentata in termini dei QNMs all'interno della cavità:

$$B(z,t) = \sum_n a_n f_n^N(z) e^{-i\omega_n t} \quad, \quad \forall z \in [0, L]$$ (4.3.3.14)

I coefficienti (4.3.3.7) possono essere riscritti come:

$$a_n = \frac{i}{2\omega_n} \{ \int_{0^+}^{L^-} \left[ f_n^N(z) \hat{B}_0(z) + \hat{f}_n^N(z) B_0(z) \right] n^2(z) dz + \\ + \sqrt{\rho_0'} [f_n^N(L) B_0(L) + f_n^N(0) B_0(0)] \}$$ (4.3.3.15)

con:

$$\hat{f}_n^N(z) = -i\omega_n \rho(z) f_n^N(z) \tag{4.3.3.16}$$

### *4.3.4. Prodotto interno dei QNMs TM.*

Procedendo come nel caso TE, si può definire il prodotto interno dei quasi modi TM sullo spazio biortogonale corrispondente [2]; oltre all'integrale, definito sulla cavità, si aggiungono i due termini superficiali, calcolati agli estremi, tenendo così conto delle condizioni di *"outgoig waves"* per l'induzione magnetica.

Si introduce la coppia di ket:

$$\left| f_n^N \right\rangle = \begin{pmatrix} f_n^N(z) \\ \hat{f}_n^N(z) \end{pmatrix} \quad , \quad \left| f_m^N \right\rangle = \begin{pmatrix} f_m^N(z) \\ \hat{f}_m^N(z) \end{pmatrix} \tag{4.3.4.1}$$

dove i momenti coniugati $f_n^N(z)$ e $f_m^N(z)$ sono definiti dalla (4.3.3.16).

Si definisce il prodotto interno come:

$$\left\langle f_n^N, f_m^N \right\rangle = i \int_{0^+}^{L^-} \left[ f_n^N(z) \hat{f}_m^N(z) + \hat{f}_n^N(z) f_m^N(z) \right] n^2(z) dz + \\ + i\sqrt{\rho_0'} \left[ f_n^N(L) f_m^N(L) + f_n^N(0) f_m^N(0) \right] \tag{4.3.4.2}$$

Tenendo conto della definizione (4.3.3.16), si osserva che il prodotto interno (4.3.4.2) ben si accorda con la norma (4.3.2.3).

*Ortogonalità dei quasi modi TM*

Definito il vettore $\left| B(t) \right\rangle = (B, \hat{B})$, dove la prima componente è l'induzione $B(z,t)$ e la seconda è il suo momento coniugato $\hat{B}(z,t) = \rho(z)\partial_t B(z,t)$, il problema e.m. della propagazione per un'onda piana TM è ben descritto dalla Hamiltoniana:

$$H = \begin{pmatrix} 0 & \rho^{-1}(z) \\ \dfrac{1}{n^2(z)} \dfrac{\partial}{\partial z} \left( \dfrac{1}{n^2(z)} \dfrac{\partial}{\partial z} \right) & 0 \end{pmatrix} \tag{4.3.4.3}$$

ed è risolto tramite un'eq. formalmente analoga a quella di Schroedinger:

$$\frac{\partial}{\partial t} \left| B(t) \right\rangle = -iH \left| B(t) \right\rangle \tag{4.3.4.4}$$

Come nel caso TE, dalla definizione di prodotto interno (4.3.4.2), si dimostra che l'Hamiltoniana $H$ è un operatore simmetrico e, quindi, che i quasi modi TM $\left| f_n^N \right\rangle$, autovettori destri dell'Hamiltoniana, $H\left| f_n^N \right\rangle = \omega_n \left| f_n^N \right\rangle$, formano una base ortogonale:

$$\left\langle f_n^N, f_m^N \right\rangle = \delta_{n,m}(\omega_n + \omega_m) \tag{4.3.4.5}$$

*Risultati utili*

Procedendo come nel caso TE, si dimostra, nel caso TM, che gli integrali di normalizzazione sono:

$$\int_{0^+}^{L^-} f_n^N(z)\rho(z)f_m^N(z)n^2(z)dz = \delta_{n,m} - i\frac{\sqrt{\rho_0'}}{\omega_n + \omega_m}[f_n^N(L)f_m^N(L) + f_n^N(0)f_m^N(0)]$$

$$\tag{4.3.4.6}$$

e si propone, sempre nel caso TM, che, per strutture simmetriche con perdite minime, la norma sia approssimata da:

$$\left\langle f_n, f_n \right\rangle \cong \sqrt{\rho_0'}\,\frac{2\omega_n}{\operatorname{Im}\omega_n} \tag{4.3.4.7}$$

# *Appendice A.*

In questa appendice, vengono riportate le dimostrazioni dei risultati principali per i quasi modi TE; si omettono le dimostrazioni corrispondenti per i quasi modi TM, poiché sono formalmente analoghe.

## *A.1. Dimostrazioni per il metodo di Green.*

*Prima dimostrazione*

Si vuole dimostrare che:

$$W(z,\omega) = g_-(z,\omega)\partial_z g_+(z,\omega) - g_+(z,\omega)\partial_z g_-(z,\omega) =$$
$$= W(\omega) \neq 0 \quad , \quad \forall z \in \Re$$

Dalle proprietà delle equazioni differenziali lineari [10], poiché le (4.2.1.4-5) mancano del termine in $\partial_z$, il wronskiano non dipende dalla variabile $z$, quindi:

$$W(z,\omega) = W(\omega) \tag{A.1.1}$$

Se si risolvono le equazioni (4.2.1.4-5) per $z \gg L$, si ottengono le funzioni ausiliarie a destra della cavità:

$$\begin{cases} g_-(z,\omega) = A(\omega)e^{in_0\frac{\omega}{c}\cos\theta_0 z} + B(\omega)e^{-in_0\frac{\omega}{c}\cos\theta_0 z} \\ g_+(z,\omega) = e^{in_0\frac{\omega}{c}\cos\theta_0 z} \end{cases} \tag{A.1.2}$$

dove, in particolare:

$$B(\omega) \neq 0 \tag{A.1.3}$$

Dalla (A.1.2), si può determinare il wronskiano:

$$W(\omega) = 2in_0\frac{\omega}{c}\cos\theta_0 B(\omega) \tag{A.1.4}$$

che, per la (A.1.3), è diverso da zero:

$$W(\omega) \neq 0 \tag{A.1.5}$$

*Seconda dimostrazione*

Si vuole dimostrare che:

$$G(z,z',\omega) = \begin{cases} -\dfrac{g_-(z,\omega)g_+(z',\omega)}{W(\omega)} &, \ z < z' \\ -\dfrac{g_-(z',\omega)g_+(z,\omega)}{W(\omega)} &, \ z > z' \end{cases}$$

La funzione di Green è la soluzione dell'equazione (4.2.1.1): in generale, è la sovrapposizione di un integrale particolare della stessa e dell'integrale generale dell'omogenea associata, quest'ultimo combinazione lineare delle funzioni ausiliarie. Tuttavia la funzione di Green, al contrario delle funzioni ausiliarie, deve rispettare la condizione di radiazione; quindi, la soluzione d'interesse fisico è la funzione di Green come integrale particolare della (4.2.1.1).

Si applica il metodo della variazione delle costanti arbitrarie [10] e si cerca una soluzione del tipo:

$$\tilde{G}(z,z',\omega) = u(z,z',\omega)g_-(z,\omega) + v(z,z',\omega)g_+(z,\omega) \tag{A.1.6}$$

dove, per la condizione di radiazione:

$$\begin{cases} u(z,z',\omega) \to 0 & \text{per} \ z \to +\infty \\ v(z,z',\omega) \to 0 & \text{per} \ z \to -\infty \end{cases} \tag{A.1.7}$$

Si impone che la derivata prima in z della (A.1.6) sia:

$$\partial_z \tilde{G}(z,z',\omega) = u(z,z',\omega)\partial_z g_-(z,\omega) + v(z,z',\omega)\partial_z g_+(z,\omega) \tag{A.1.8}$$

per cui:

$$g_-(z,\omega)\partial_z u(z,z',\omega) + g_+(z,\omega)\partial_z v(z,z',\omega) = 0 \tag{A.1.9}$$

Si calcola la derivata seconda in z della (A.1.6) dalla (A.1.8):

$$\begin{aligned}\partial_z^2 \tilde{G}(z,z',\omega) = {} & \partial_z u(z,z',\omega)\partial_z g_-(z,\omega) + u(z,z',\omega)\partial_z^2 g_-(z,\omega) + \\ & + \partial_z v(z,z',\omega)\partial_z g_+(z,\omega) + v(z,z',\omega)\partial_z^2 g_+(z,\omega)\end{aligned} \tag{A.1.10}$$

Se si inseriscono le (A.1.6) e (A.1.10) nella (4.2.1.1) e si tiene conto delle (4.2.1.4-5), si perviene alla:

$$\partial_z u(z,z',\omega)\partial_z g_-(z,\omega) + \partial_z v(z,z',\omega)\partial_z g_+(z,\omega) = -\delta(z-z') \tag{A.1.11}$$

Si risolve il sistema costituito dalle equazioni (A.1.9) e (A.1.11) nelle incognite $\partial_z u$ e $\partial_z v$:

$$\begin{cases} \partial_z u(z,z',\omega) = \delta(z-z')g_+(z,\omega)/W(\omega) \\ \partial_z v(z,z',\omega) = -\delta(z-z')g_-(z,\omega)/W(\omega) \end{cases} \tag{A.1.12}$$

Si integrano le (A.1.12), tenendo conto delle (A.1.7) e della definizione della delta di Dirac [11]:

$$u(z,z',\omega) = \int\limits_{+\infty}^{z} \frac{\delta(\xi-z')g_+(\xi,\omega)}{W(\omega)}d\xi = \begin{cases} -\dfrac{g_+(z',\omega)}{W(\omega)} & \text{per } z < z' \\ 0 & \text{per } z > z' \end{cases} \quad \text{(A.1.13)}$$

$$v(z,z',\omega) = -\int\limits_{-\infty}^{z} \frac{\delta(\xi-z')g_-(\xi,\omega)}{W(\omega)}d\xi = \begin{cases} 0 & \text{per } z < z' \\ -\dfrac{g_-(z',\omega)}{W(\omega)} & \text{per } z > z' \end{cases} \quad \text{(A.1.14)}$$

Infine, si perviene alla tesi inserendo nella (A.1.6) le (A.1.13) e (A.1.14).

*Terza dimostrazione*

Si vuole dimostrare che:

$$G(z,z',t) = i\sum_n \frac{f_n(z)f_n(z')}{\langle f_n | f_n \rangle} e^{-i\omega_n t} \quad , \quad \forall (z,z') \in C = [0,L]$$

Le frequenze $\omega_n$ dei QNMs sono i poli della funzione di Green nel piano complesso; poiché $\text{Im}\,\omega_n < 0$, questi poli giacciono nel semipiano inferiore.

Difatti, la funzione di Green ha il wronskiano come denominatore e, per le condizioni di risonanza (4.2.1.9), vale la:

$$W(\omega_n) = 0 \quad \text{(A.1.15)}$$

Definita la funzione $\hat{W}(\omega)$, tale che $W(\omega) = (\omega - \omega_n)\hat{W}(\omega)$, questa non presenta singolarità in $\omega = \omega_n$ e soddisfa la:

$$\left(\frac{dW}{d\omega}\right)_{\omega=\omega_n} = \hat{W}(\omega_n) \quad \text{(A.1.16)}$$

La funzione di Green nel dominio del tempo $G(z,z',t)$ si ottiene da quella nel dominio della frequenza $\tilde{G}(z,z',\omega)$ antitrasformandola secondo Fourier [5]:

$$G(z,z',t) = \frac{1}{2\pi}\int\limits_{-\infty}^{+\infty}\tilde{G}(z,z',\omega)e^{-i\omega t}d\omega \quad \text{(A.1.17)}$$

Definita la funzione $\hat{G}(z,z',\omega) = \tilde{G}(z,z',\omega)e^{-i\omega t}$, si chiude l'integrale (A.1.17) all'infinito, lungo il semicerchio inferiore γ, orientato in senso antiorario:

$$G(z,z',t) = \frac{1}{2\pi}\oint\limits_{\gamma}\hat{G}(z,z',\omega)d\omega \quad \text{(A.1.18)}$$

Supposta valida la condizione per la completezza dei QNMs (4.2.1.14), si applica il teorema dei residui [10]; solo $\forall (z,z') \in C$, si può porre:

$$\oint_\gamma \hat{G}(z,z',\omega)d\omega = -2\pi i \sum_n \text{Res}(\omega = \omega_n) \tag{A.1.19}$$

che, inserita nella (A.1.18), fornisce la:

$$G(z,z',t) = -i \sum_n \text{Res}(\omega = \omega_n) \tag{A.1.20}$$

Si calcolano i residui [10], quando $z < z'$, tenendo conto della (A.1.16):

$$\begin{aligned}
\text{Res}(\omega = \omega_n) &= \lim_{\omega \to \omega_n}(\omega - \omega_n)\hat{G}(z,z',\omega) = \lim_{\omega \to \omega_n}(\omega - \omega_n)\tilde{G}(z,z',\omega)e^{-i\omega t} = \\
&= \lim_{\omega \to \omega_n}(\omega - \omega_n)\frac{-g_-(z,\omega)g_+(z',\omega)}{W(\omega)}e^{-i\omega t} = \lim_{\omega \to \omega_n}\frac{-g_-(z,\omega)g_+(z',\omega)}{\hat{W}(\omega)}e^{-i\omega t} = \\
&= -\frac{g_-(z,\omega_n)g_+(z',\omega_n)}{\hat{W}(\omega_n)}e^{-i\omega_n t} = -\frac{g_-(z,\omega_n)g_+(z',\omega_n)}{\left(\dfrac{dW}{d\omega}\right)_{\omega=\omega_n}}e^{-i\omega_n t}
\end{aligned}$$
(A.1.21)

Infine, si inserisce la (A.1.21) nella (A.1.20):

$$G(z,z',t) = i \sum_n \frac{g_-(z,\omega_n)g_+(z',\omega_n)}{\left(\dfrac{dW}{d\omega}\right)_{\omega=\omega_n}}e^{-i\omega_n t} \tag{A.1.22}$$

Si perviene alla tesi, dalla (A.1.22), tenendo conto che, per le autofrequenze $\omega = \omega_n$ dei QNMs, si ha $g_-(z,\omega_n) \propto g_+(z,\omega_n) \cong f_n(z)$ ed ancora $(dW/d\omega)_{\omega=\omega_n} = \langle f_n | f_n \rangle$.

## A.2. Dimostrazioni per la costruzione della norma.

### Prima dimostrazione

Si vuole dimostrare che, utilizzando le condizioni asintotiche (4.2.1.6-7) e di risonanza (4.2.1.9), si perviene alla prima definizione operativa per la norma:

$$\langle f_n | f_n \rangle = 2\omega_n \int_{-R}^{R} \rho(z)f_n^2(z)dz + i\sqrt{\rho_0}[f_n^2(R) + f_n^2(-R)]$$

dove $R \gg L$ e tale che $\rho(\pm R) = \rho_0$.

Si moltiplica per $g_-(z,\omega_n)$ l'equazione (4.2.1.5) di $g_+(z,\omega)$ e per $g_+(z,\omega)$ l'equazione (4.2.1.4) di $g_-(z,\omega)$, calcolata in $\omega = \omega_n$, si sottrae membro a membro, e, dopo opportune manipolazioni, si perviene alla:

$$(\omega^2 - \omega_n^2)\rho(z)g_+(z,\omega)g_-(z,\omega_n) = g_+(z,\omega)\partial_z^2 g_-(z,\omega_n) - g_-(z,\omega_n)\partial_z^2 g_+(z,\omega)$$
(A.2.1)

Si integra la (A.2.1) da $z = -R$ a $z = R$, e si ottiene:

$$(\omega^2 - \omega_n^2)\int_{-R}^{R} \rho(z)g_+(z,\omega)g_-(z,\omega_n)dz =$$
$$= \left[g_+(z,\omega)\partial_z g_-(z,\omega_n) - g_-(z,\omega_n)\partial_z g_+(z,\omega)\right]_{-R}^{R}$$
(A.2.2)

dove si è tenuto conto che:

$$g_+(z,\omega)\partial_z^2 g_-(z,\omega_n) - g_-(z,\omega_n)\partial_z^2 g_+(z,\omega) =$$
$$= \partial_z \left[g_+(z,\omega)\partial_z g_-(z,\omega_n) - g_-(z,\omega_n)\partial_z g_+(z,\omega)\right]$$
(A.2.3)

Si impongono in $R \gg L$, tale che $\rho(R) = \rho_0$, la condizione asintotica (4.2.1.7) per $g_+(z,\omega)$ e quella di risonanza (4.2.1.9) per $g_-(z,\omega)$.

Si deducono le:

$$\begin{cases} \left[\partial_z g_+(z,\omega)\right]_{z=R} = i\sqrt{\rho_0}\,\omega g_+(R,\omega) \\ \left[\partial_z g_-(z,\omega_n)\right]_{z=R} = i\sqrt{\rho_0}\,\omega_n g_-(R,\omega_n) \end{cases}$$
(A.2.4)

Si inseriscono le (A.2.4) nella (A.2.2) e si ottiene:

$$(\omega^2 - \omega_n^2)\int_{-R}^{R} \rho(z)g_+(z,\omega)g_-(z,\omega_n)dz =$$
$$= i\sqrt{\rho_0}(\omega_n - \omega)g_+(R,\omega)g_-(R,\omega_n) + \left[g_-(z,\omega_n)\partial_z g_+(z,\omega) - g_+(z,\omega)\partial_z g_-(z,\omega_n)\right]_{z=-R}$$
(A.2.5)

Si deriva membro a membro la (A.2.5) rispetto ad $\omega$ e si perviene alla:

$$2\omega\int_{-R}^{R} \rho(z)g_+(z,\omega)g_-(z,\omega_n)dz + (\omega^2 - \omega_n^2)\int_{-R}^{R} \rho(z)g_-(z,\omega_n)\partial_\omega g_+(z,\omega)dz =$$
$$= -i\sqrt{\rho_0}\,g_+(R,\omega)g_-(R,\omega_n) + i\sqrt{\rho_0}(\omega_n - \omega)g_-(R,\omega_n)\partial_\omega g_+(z,\omega) +$$
$$+ \left[g_-(z,\omega_n)\partial_{\omega,z} g_+(z,\omega) - \partial_\omega g_+(z,\omega)\partial_z g_-(z,\omega_n)\right]_{z=-R}$$
(A.2.6)

Si calcola la (A.2.6) in $\omega = \omega_n$ e si ottiene:

$$2\omega_n \int_{-R}^{R} \rho(z) g_+(z,\omega_n) g_-(z,\omega_n) dz =$$
$$= -i\sqrt{\rho_0} g_+(R,\omega_n) g_-(R,\omega_n) + \left[ g_-(z,\omega)\partial_{\omega,z} g_+(z,\omega) - \partial_\omega g_+(z,\omega)\partial_z g_-(z,\omega) \right]_{\omega=\omega_n, z=-R}$$
(A.2.7)

Si deriva il wronskiano (4.2.1.12) rispetto ad $\omega$, si riordinano i termini e si perviene alla:

$$g_-(z,\omega)\partial_{\omega,z} g_+(z,\omega) - \partial_\omega g_+(z,\omega)\partial_z g_-(z,\omega) =$$
$$= \frac{dW}{d\omega} + g_+(z,\omega)\partial_{\omega,z} g_-(z,\omega) - \partial_\omega g_-(z,\omega)\partial_z g_+(z,\omega)$$
(A.2.8)

Si impone in $z = -R$, tale che $\rho(-R) = \rho_0$, la condizione asintotica (4.2.1.6) per $g_-(z,\omega)$ e si deduce che:

$$\left[\partial_z g_-(z,\omega)\right]_{z=-R} = -i\sqrt{\rho_0}\,\omega g_-(-R,\omega)$$
(A.2.9)

Si prende la (A.2.9), la si deriva per $\omega$, la si calcola per $\omega = \omega_n$ e si ottiene:

$$\left[\partial_{\omega,z} g_-(z,\omega)\right]_{\omega=\omega_n, z=-R} = -i\sqrt{\rho_0}\, g_-(-R,\omega_n) - i\sqrt{\rho_0}\,\omega_n\left[\partial_\omega g_-(-R,\omega)\right]_{\omega=\omega_n} \quad (A.2.10)$$

Si impone in $z = -R$ la condizione di risonanza (4.2.1.9) per $g_+(z,\omega)$ e si deduce che:

$$\left[\partial_z g_+(z,\omega_n)\right]_{z=-R} = -i\sqrt{\rho_0}\,\omega_n g_+(-R,\omega_n)$$
(A.2.11)

Si inseriscono nella (A.2.7) le (A.2.8), (A.2.10) e (A.2.11), si riordinano i termini e si ottiene:

$$\left(\frac{dW}{d\omega}\right)_{\omega=\omega_n} = 2\omega_n \int_{-R}^{R} \rho(z) g_+(z,\omega_n) g_-(z,\omega) dz +$$
$$+ i\sqrt{\rho_0} \left[ g_+(R,\omega_n) g_-(R,\omega_n) + g_+(-R,\omega_n) g_-(-R,\omega_n) \right]$$
(A.2.12)

Si perviene alla tesi, dalla (A.2.12), tenendo conto che, per le autofrequenze $\omega = \omega_n$ dei QNMs, si ha $g_-(z,\omega_n) \propto g_+(z,\omega_n) \cong f_n(z)$ ed ancora $(dW/d\omega)_{\omega=\omega_n} = \langle f_n | f_n \rangle$.

Utilizzando le condizioni formali di "*outgoing waves*" (4.2.1.10-11), si dimostra la seconda definizione operativa per la norma:

$$\langle f_n | f_n \rangle = 2\omega_n \int_{0^+}^{L^-} \rho(z) f_n^2(z) dz + i\sqrt{\rho_0}[f_n^2(L) + f_n^2(0)]$$

Le condizioni formali (4.2.1.11) di *"outgoing waves"* in $z = L^-$ sono analoghe alle condizioni (A.2.4), asintotica per $g_+(z,\omega)$ e di risonanza per $g_-(z,\omega)$ in $z = R$. Le condizioni formali (4.2.1.10) di *"outgoing waves"* in $z = 0^+$ sono analoghe alla condizione asintotica (A.2.9) per $g_-(z,\omega)$ e alla condizione di risonanza (A.2.11) per $g_+(z,\omega)$ in $z = -R$. La dimostrazione per la II definizione operativa della norma è formalmente analoga alla precedente.

*Seconda dimostrazione*

Si vuole dimostrare che:

$$2\omega_n \int_{-R}^{0^+} \rho(z) f_n^2(z) dz = i\sqrt{\rho_0}\left[f_n^2(0) - f_n^2(-R)\right]$$

Si integra la (A.2.1) da $z = -R$ a $z = 0^+$ e si ottiene:

$$(\omega^2 - \omega_n^2)\int_{-R}^{0^+} \rho(z) g_+(z,\omega) g_-(z,\omega_n) dz = \qquad (A.2.13)$$
$$= \left[g_+(z,\omega)\partial_z g_-(z,\omega_n) - g_-(z,\omega_n)\partial_z g_+(z,\omega)\right]_{-R}^{0^+}$$

Si inseriscono nella (A.2.13) le condizioni formali di *"outgoing waves"* (4.2.1.10) e si ottiene:

$$(\omega^2 - \omega_n^2)\int_{-R}^{0^+} \rho(z) g_+(z,\omega) g_-(z,\omega_n) dz =$$
$$= i\sqrt{\rho_0}(\omega - \omega_n) g_+(0,\omega) g_-(0,\omega_n) + \left[g_-(z,\omega_n)\partial_z g_+(z,\omega) - g_+(z,\omega)\partial_z g_-(z,\omega_n)\right]_{z=-R}$$
(A.2.14)

Si deriva la (A.2.14) per $\omega$, calcolandola per $\omega = \omega_n$, e si ottiene:

$$2\omega_n \int_{-R}^{0^+} \rho(z) g_+(z,\omega_n) g_-(z,\omega_n) dz =$$
$$= i\sqrt{\rho_0} g_+(0,\omega_n) g_-(0,\omega_n) + \left[g_-(z,\omega)\partial_{\omega,z} g_+(z,\omega) - \partial_\omega g_+(z,\omega)\partial_z g_-(z,\omega)\right]_{\omega=\omega_n, z=-R}$$
(A.2.15)

La funzione ausiliaria $g_-(z,\omega)$ soddisfa la condizione asintotica (4.2.1.6), quindi si deduce la (A.2.9); in un intorno della frequenza di risonanza $\omega \cong \omega_n$, la funzione ausiliaria $g_+(z,\omega)$ soddisfa la condizione:

$$\left[\partial_z g_+(z,\omega)\right]_{z=-R} = -i\sqrt{\rho_0}\,\omega g_+(-R,\omega) \qquad (A.2.16)$$

quindi si deduce che:

$$\left[\partial_{\omega,z}g_+(z,\omega)\right]_{\omega=\omega_n,z=-R} = \left[\partial_\omega\left(-i\sqrt{\rho_0}\omega g_+(-R,\omega)\right)\right]_{\omega=\omega_n} =$$
$$= -i\sqrt{\rho_0}g_+(-R,\omega_n) - i\sqrt{\rho_0}\omega_n\left[\partial_\omega g_+(-R,\omega)\right]_{\omega=\omega_n} \quad (A.2.17)$$

Si inseriscono nella (A.2.15) le (A.2.9) e (A.2.17), si ottiene:

$$2\omega_n \int_{-R}^{0^+} \rho(z)g_+(z,\omega_n)g_-(z,\omega_n)dz =$$
$$= i\sqrt{\rho_0}\left[g_+(0,\omega_n)g_-(0,\omega_n) - g_+(-R,\omega_n)g_-(-R,\omega_n)\right] \quad (A.2.18)$$

Si perviene alla tesi, dalla (A.2.18), tenendo conto che, per le autofrequenze $\omega = \omega_n$ dei QNMs, si ha $g_-(z,\omega_n) \propto g_+(z,\omega_n) \cong f_n(z)$.

Si può dimostrare che:

$$2\omega_n \int_{L^-}^{R} \rho(z)f_n^2(z)dz = i\sqrt{\rho_0}\left[f_n^2(L) - f_n^2(R)\right]$$

La dimostrazione è analoga alla precedente: si integra la (A.2.1) da $z = L^-$ ad $z = R$; vi si inseriscono le (A.2.4), dedotte dalla condizione asintotica per $g_+(z,\omega)$ e da quella di risonanza per $g_-(z,\omega)$; si applica all'equazione risultante l'operatore $\partial_\omega$ e la si riferisce alla frequenza di risonanza $\omega = \omega_n$; vi si inseriscono le (4.2.1.11), condizioni formali di *"outgoing waves"* per $z = L^-$.

## A.3. Dimostrazione per lo sviluppo a due componenti.

Si vuole dimostrare che:

$$E(z,t) = \sum_n a_n f_n^N(z)e^{-i\omega_n t} \quad, \quad \forall z \in [0,L]$$

dove:

$$a_n = \frac{i}{2\omega_n}\left\{\int_0^L \left[f_n^N(z)\hat{E}_0(z) + \hat{f}_n^N(z)E_0(z)\right]dz + \sqrt{\rho_0}\left[f_n^N(L)E_0(L) + f_n^N(0)E_0(0)\right]\right\}$$

con:

$$\hat{f}_n^N(z) = -i\omega_n\rho(z)f_n^N(z)$$

Il campo elettrico $E(z,t)$ può essere espresso in termini della funzione di Green $G(z,z',t)$ tramite la [6]:

$$E(z,t) = \int_{-\infty}^{\infty} \left[ G(z,z',t)\hat{E}_0(z') + \partial_t G(z,z',t)\rho(z')E_0(z') \right] dz' \quad \text{(A.3.1)}$$

Si limita l'attenzione all'interno della cavità, quindi $0 < z < L$, e si suddivide l'integrale (A.3.1) nei tre contributi:

$$E(z,t) = I_{z'\in(0,L)} + I_{z'\in(-\infty,0]} + I_{z'\in[L,\infty)} \quad \text{(A.3.2)}$$

Si considera il primo contributo:

$$I_{z'\in(0,L)} = \int_{0^+}^{L^-} \left[ G(z,z',t)\hat{E}_0(z') + \partial_t G(z,z',t)\rho(z')E_0(z') \right] dz' \quad \text{(A.3.3)}$$

Si inserisce la rappresentazione per la funzione di Green in termini dei QNMs (4.2.1.18) nella (A.3.3) e, dopo opportuni passaggi, si ottiene:

$$I_{z'\in(0,L)} = \sum_n a_n^{z'\in(0,L)} f_n^N(z) e^{-i\omega_n t} \quad \text{(A.3.4)}$$

dove:

$$a_n^{z'\in(0,L)} = \frac{i}{2\omega_n} \left\{ \int_{0^+}^{L^-} \left[ f_n^N(z')\hat{E}_0(z') + \hat{f}_n^N(z')E_0(z') \right] dz' \right\} \quad \text{(A.3.5)}$$

Si considera il secondo contributo:

$$I_{z'\in(-\infty,0]} = \int_{-\infty}^{0} \left[ G(z,z',t)\hat{E}_0(z') + \partial_t G(z,z',t)\rho(z')E_0(z') \right] dz' \quad \text{(A.3.6)}$$

Si inseriscono la condizione di *"no tail"* (4.2.3.3), le condizioni di *"outgoing waves"* per il momento coniugato (4.2.3.5) e per la funzione di Green (4.2.1.3) e si tiene conto della condizione radiativa (4.2.3.4), tutte per $z' \leq 0$. Si procede:

$$I_{z'\in(-\infty,0]} = \sqrt{\rho_0} \int_{-\infty}^{0} \left[ G(z,z',t)\partial_{z'}E_0(z') + E_0(z')\partial_{z'}G(z,z',t) \right] dz' =$$

$$= \sqrt{\rho_0} \int_{-\infty}^{0} \partial_{z'} \left[ G(z,z',t)E_0(z') \right] dz' = \sqrt{\rho_0} \left[ G(z,z',t)E_0(z') \right]_{z'=-\infty}^{z'=0} =$$

$$= \sqrt{\rho_0} G(z,0,t)E_0(0)$$

(A.3.7)

Si inserisce la (4.2.1.18) e si ottiene:

$$I_{z'\in(-\infty,0]} = \sum_n a_n^{z'\in(-\infty,0]} f_n^N(z) e^{-i\omega_n t} \quad \text{(A.3.8)}$$

dove:

$$a_n^{z' \in (-\infty, 0]} = \frac{i}{2\omega_n} \sqrt{\rho_0} f_n^N(0) E_0(0) \tag{A.3.9}$$

Si considera il terzo contributo:

$$I_{z' \in [L, +\infty)} = \int_L^{+\infty} \left[ G(z, z', t) \hat{E}_0(z') + \partial_t G(z, z', t) \rho(z') E_0(z') \right] dz' \tag{A.3.10}$$

In modo analogo al precedente, si ottiene:

$$I_{z' \in [L, +\infty)} = \sum_n a_n^{z' \in [L, +\infty)} f_n^N(z) e^{-i\omega_n t} \tag{A.3.11}$$

dove:

$$a_n^{z' \in [L, +\infty)} = \frac{i}{2\omega_n} \sqrt{\rho_0} f_n^N(L) E_0(L) \tag{A.3.12}$$

Si perviene alla tesi, inserendo nella (A.3.2) le (A.3.4-5), le (A.3.7-8) e le (A.3.11-12).

## *A.4. Dimostrazioni relative al prodotto interno.*

### *Prima dimostrazione*

Si vuole dimostrare che:

$$\int_{0^+}^{L^-} f_n^N(z) \rho(z) f_m^N(z) dz = \delta_{n,m} - i \frac{\sqrt{\rho_0}}{\omega_n + \omega_m} [f_n^N(L) f_m^N(L) + f_n^N(0) f_m^N(0)]$$

Si applica la definizione di prodotto interno (4.2.4.2), si tiene conto di quella per i momenti coniugati (4.2.3.8) e si impone l'ortogonalità (4.2.4.5), ottenendo:

$$\begin{aligned}
\left\langle f_n^N, f_m^N \right\rangle &= i \int_{0^+}^{L^-} \left[ f_n^N(z) \hat{f}_m^N(z) + \hat{f}_n^N(z) f_m^N(z) \right] dz + \\
&\quad + i \sqrt{\rho_0} \left[ f_n^N(L) f_m^N(L) + f_n^N(0) f_m^N(0) \right] = \\
&= (\omega_n + \omega_m) \int_{0^+}^{L^-} f_n^N(z) \rho(z) f_m^N(z) dz + \\
&\quad + i \sqrt{\rho_0} \left[ f_n^N(L) f_m^N(L) + f_n^N(0) f_m^N(0) \right] = \\
&= \delta_{n,m} (\omega_n + \omega_m)
\end{aligned} \tag{A.4.1}$$

Si perviene alla tesi riordinando i termini nell'ultima uguaglianza della (A.4.1).

*Seconda dimostrazione*

Si illustra il procedimento dal quale, per strutture simmetriche con perdite minime, si può proporre l'approssimazione:

$$\langle f_n, f_n \rangle \cong \sqrt{\rho_0} \frac{2\omega_n}{\operatorname{Im}\omega_n}$$

Si parte dal precedente integrale di normalizzazione, dove si pone $m = -n$ e si applicano le $\delta_{n,-n} = 0$ ed $\omega_{-n} = -\omega_n^*$, $f_{-n}(z) = f_n^*(z)$ [4].

Si ottiene:

$$I_n = \frac{\sqrt{\rho_0}}{2|\operatorname{Im}\omega_n|}\left[\left|f_n^N(L)\right|^2 + \left|f_n^N(0)\right|^2\right] \tag{A.4.2}$$

dove:

$$I_n = \int_0^L \left|f_n^N(z)\right|^2 \rho(z)dz \tag{A.4.3}$$

Si continua dalla (A.4.2), dove si esplicita la normalizzazione (4.2.1.17) e si tiene conto che $f_n(0) = 1$ [3].

Si ottiene:

$$\left|\langle f_n, f_n \rangle\right| = \sqrt{\rho_0}\left|\frac{\omega_n}{\operatorname{Im}\omega_n}\right|\frac{1 + |f_n(L)|^2}{I_n} \tag{A.4.4}$$

Infine dalla (A.4.4), per perdite minime [9], tali che $I_n \cong 1$, e per cavità simmetriche [3], tali che $f_n(L) = (-1)^n$, si deduce la:

$$\left|\langle f_n, f_n \rangle\right| \cong 2\sqrt{\rho_0}\left|\frac{\omega_n}{\operatorname{Im}\omega_n}\right| \tag{A.4.5}$$

Si propone la tesi, poiché è immediato che l'informazione sulla fase della norma è già contenuta nelle frequenze di risonanza.

## *Bibliografia.*

# Capitolo 5. QNMs TE e TM per i 1D-PBG.
## Autofunzioni ed autofrequenze.

### Paragrafo 5.1. Introduzione.

In questo capitolo, si inizia a studiare il problema elettromagnetico di un'onda che si propaga obliquamente in un PBG unidimensionale, tramite il modello teorico dei QNMs; nello specifico, si determinano le autofunzioni e le autofrequenze sia per i quasi modi TE che TM.

Si riprende un filo logico: nel capitolo 3, è stata estesa la teoria dei QNMs ai 1D-PBG ma solo quando vi è un'onda che incide normalmente; con riferimento ad un 1D-PBG simmetrico a quarto d'onda, sono state graficate nel piano complesso le autofrequenze che, nel caso di incidenza normale, sono le stesse sia per i quasimodi TE che TM [1].

Qui, si introduce un angolo di incidenza: sia per i modi TE che TM, applicando la teoria dei QNM's [2] ed il metodo delle matrici ai 1D-PBG [3], si definisce una procedura di calcolo che permette di determinare le quasi autofunzioni e le quasi autofrequenze, in ipotesi parassiale [4] e per un 1D-PBG simmetrico a quarto d'onda; quindi si graficano le autofunzioni per incidenza normale, sufficienti a caratterizzare i quasi modi, e l'andamento parametrico delle autofrequenze rispetto all'angolo, seppure in approssimazione parassiale, che illustrano le risonanze e le perdite.

### Paragrafo 5.2. Modi quasi normali TE per un 1D-PBG.

In questo paragrafo, si studiano i QNMs di tipo TE in un PBG unidimensionale.

### 5.2.1. Strumenti di analisi.

Qui, con riferimento alla funzione d'onda per i QNMs TE, viene costruita la matrice di trasmissione TE per un 1D-PBG.

### 5.2.1.a. Funzione d'onda per i QNMs TE.

La funzione d'onda $f(z)$ per i QNMs TE soddisfa l'equazione omogenea:

$$\frac{\partial^2 f}{\partial z^2} + \omega^2 \rho(z) f(z) = 0 \qquad (5.2.1.1)$$

dove:

$$\rho(z) = \left(\frac{n(z)\cos\theta(z)}{c}\right)^2 \qquad (5.2.1.2)$$

con $c$ velocità della luce, $n(z)$ profilo dell'indice di rifrazione, $\theta(z)$ profilo dell'angolo di rifrazione (vedi paragrafo 4.2.).

In corrispondenza delle autofrequenze $\omega = \omega_n$ dei QNMs TE, la funzione d'onda dà luogo alle autofunzioni $f_n(z)$ e rispetta entrambe le condizioni asintotiche:

$$\begin{cases} f_n(z) = e^{-iq_0(\omega_n)z} & \text{per } z \to -\infty \\ f_n(z) = e^{+iq_0(\omega_n)z} & \text{per } z \to +\infty \end{cases} \qquad (5.2.1.3)$$

dove:

$$q_0(\omega) = n_0 \frac{\omega}{c} \cos\theta_0 \qquad (5.2.1.4)$$

con $n_0$ indice di rifrazione fuori del mezzo e $\theta_0$ angolo di incidenza.

### 5.2.1.b. Matrice di trasmissione TE per un 1D-PBG.

Si consideri un 1D-PBG tale che la cella elementare sia costituita da due strati, rispettivamente con lunghezza ed indice di rifrazione $h$, $n_h$ ed $l$, $n_l$.

Vale la legge generalizzata di Snell [4]:

$$n_0 \operatorname{sen}\theta_0 = n_h \operatorname{sen}\theta_h = n_l \operatorname{sen}\theta_l \qquad (5.2.1.5)$$

dove $\theta_h$ ed $\theta_l$ sono gli angoli di propagazione nei due strati.

Si suppone che l'onda incidente abbia polarizzazione TE.

La matrice della singola cella è la seguente [3]:

$$M^{(TE)} = \begin{pmatrix} \cos\delta_l & \frac{\sin\delta_l}{q_l} \\ -q_l \sin\delta_l & \cos\delta_l \end{pmatrix} \begin{pmatrix} \cos\delta_h & \frac{\sin\delta_h}{q_h} \\ -q_h \sin\delta_h & \cos\delta_h \end{pmatrix} \qquad (5.2.1.6)$$

dove sono state introdotte le fasi:

$$\begin{cases} \delta_h = q_h h \\ \delta_l = q_l l \end{cases} \quad (5.2.1.7)$$

ed i numeri d'onda:

$$\begin{cases} q_h = n_h \dfrac{\omega}{c} \cos\theta_h \\ q_l = n_l \dfrac{\omega}{c} \cos\theta_l \end{cases} \quad (5.2.1.8)$$

Si sviluppa il prodotto di matrici, ottenendo:

$$M^{(TE)} = \begin{pmatrix} \mu_{11} & \mu_{12} \\ \mu_{21} & \mu_{22} \end{pmatrix} \quad (5.2.1.9)$$

dove:

$$\begin{cases} \mu_{11} = \cos\delta_h \cos\delta_l - \dfrac{q_h}{q_l} \sin\delta_h \sin\delta_l \\ \mu_{12} = \dfrac{1}{q_h} \sin\delta_h \cos\delta_l + \dfrac{1}{q_l} \sin\delta_l \cos\delta_h \\ \mu_{21} = -q_h \sin\delta_h \cos\delta_l - q_l \sin\delta_l \cos\delta_h \\ \mu_{22} = \cos\delta_h \cos\delta_l - \dfrac{q_l}{q_h} \sin\delta_h \sin\delta_l \end{cases} \quad (5.2.1.10)$$

*PBG periodico*

Un PBG periodico è costituito dalla replica della cella elementare per un numero di periodi *N*.

La matrice del PBG periodico è:

$$M^{(TE)}_{PBG-per.} = \left(M^{TE}\right)^N = \begin{pmatrix} \mu_{11} & \mu_{12} \\ \mu_{21} & \mu_{22} \end{pmatrix}^N \quad (5.2.1.11)$$

Se si tiene conto della proprietà di unimodularià, si ottiene [3]:

$$M^{(TE)}_{PBG-per.} = \begin{pmatrix} \mu_{11} U_{N-1}(\vartheta) - U_{N-2}(\vartheta) & \mu_{12} U_{N-1}(\vartheta) \\ \mu_{21} U_{N-1}(\vartheta) & \mu_{22} U_{N-1}(\vartheta) - U_{N-2}(\vartheta) \end{pmatrix} \quad (5.2.1.12)$$

dove sono stati introdotti i polinomi di Chebyshev:

$$U_N(\vartheta) = \dfrac{\sin[(N+1)\vartheta]}{\sin\vartheta} \quad (5.2.1.13)$$

con:

$$\cos\vartheta = \frac{\mu_{11} + \mu_{22}}{2} \tag{5.2.4.14}$$

*PBG simmetrico*

Un PBG simmetrico si ottiene da uno periodico aggiungendo a destra uno strato di lunghezza *h* ed indice di rifrazione $n_h$.

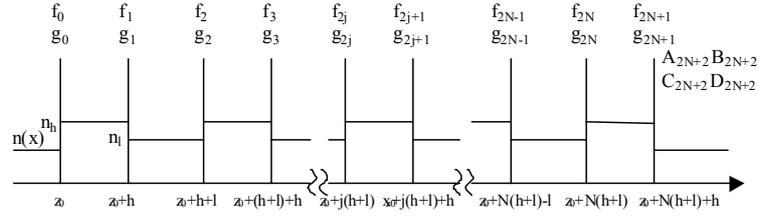

Figura 5.1. Indice di rifrazione n(z) per un 1D-PBG simmetrico.

La matrice del PBG simmetrico si ottiene moltiplicando da sinistra la matrice del PBG periodico per quella del suddetto strato [3]:

$$M_{PBG-simm.}^{(TE)} = \begin{pmatrix} \cos\delta_h & \dfrac{\sin\delta_h}{q_h} \\ -q_h \sin\delta_h & \cos\delta_h \end{pmatrix} M_{PBG-per.}^{(TE)} \tag{5.2.1.15}$$

Si sviluppa il prodotto di matrici, ottenendo:

$$M_{PBG-simm.}^{(TE)} = \begin{pmatrix} m_{11} & m_{12} \\ m_{21} & m_{22} \end{pmatrix} \tag{5.2.1.16}$$

dove:

$$\begin{cases} m_{11} = \cos\delta_h [\mu_{11} U_{N-1}(\vartheta) - U_{N-2}(\vartheta)] + \dfrac{\sin\delta_h}{q_h} \mu_{21} U_{N-1}(\vartheta) \\ m_{12} = \cos\delta_h \mu_{12} U_{N-1}(\vartheta) + \dfrac{\sin\delta_h}{q_h} [\mu_{22} U_{N-1}(\vartheta) - U_{N-2}(\vartheta)] \\ m_{21} = -q_h \sin\delta_h [\mu_{11} U_{N-1}(\vartheta) - U_{N-2}(\vartheta)] + \cos\delta_h \mu_{21} U_{N-1}(\vartheta) \\ m_{22} = -q_h \sin\delta_h \mu_{12} U_{N-1}(\vartheta) + \cos\delta_h [\mu_{22} U_{N-1} - U_{N-2}(\vartheta)] \end{cases} \tag{5.2.1.17}$$

Con riferimento alla figura 5.1, indicata la funzione d'onda del generico QNM TE con $f(z)$ e la sua derivata spaziale con $g(z)$, i loro valori agli estremi del 1D-PBG simmetrico $(f_{2N+1}, g_{2N+1})$ e $(f_0, g_0)$ sono legati dalla [3]:

$$\begin{pmatrix} f_{2N+1} \\ g_{2N+1} \end{pmatrix} = M_{PBG-simm.}^{(TE)} \begin{pmatrix} f_0 \\ g_0 \end{pmatrix} \tag{5.2.1.18}$$

## *5.2.2. Calcolo delle autofunzioni ed autofrequenze.*
## *(1D-PBG simmetrico)*

Qui, viene definita una procedura in cinque passi per il calcolo delle autofunzioni ed autofrequenze QNM's TE relative ad un PBG unidimensionale simmetrico.

*I passo: impostazione*

Risolvendo in ogni strato l'equazione (5.2.1.1), si determina la funzione d'onda per i QNMs TE:

$$\begin{aligned} f(z) = & \left[ A_0(\omega) e^{iq_0(\omega)z} + B_0(\omega) e^{-iq_0(\omega)z} \right] u_{-1}(-z) + \\ & + \sum_{k=0}^{N} \left[ A_{2k+1}(\omega) e^{iq_h(\omega)z} + B_{2k+1}(\omega) e^{-iq_h(\omega)z} \right] u_{-1}[z - k(h+l)] u_{-1}[k(h+l) + h - z] + \\ & + \sum_{k=0}^{N-1} \left[ A_{2k+2}(\omega) e^{iq_l(\omega)z} + B_{2k+2}(\omega) e^{-iq_l(\omega)z} \right] u_{-1}[z - k(h+l) - h] u_{-1}[(k+1)(h+l) - z] + \\ & + \left[ A_{2N+2}(\omega) e^{iq_0(\omega)z} + B_{2N+2}(\omega) e^{-iq_0(\omega)z} \right] u_{-1}(z - N(h+l) - h) \end{aligned}$$
(5.2.2.1)

dove:

$$\begin{cases} q_h(\omega) = n_h(\omega/c)\cos\theta_h \\ q_l(\omega) = n_l(\omega/c)\cos\theta_l \end{cases} \tag{5.2.2.2}$$

e $u_{-1}(z)$ è la funzione gradino unitario.

Se si impongono le condizioni di continuità ad ogni interfaccia per la funzione d'onda e la sua derivata, si determinano le ampiezze $A$ e $B$.

Valgono le:

$$\begin{pmatrix} A_1(\omega) \\ B_1(\omega) \end{pmatrix} = \frac{1}{2} \begin{pmatrix} 1 & \dfrac{1}{n_h \cos\theta_h} \\ 1 & -\dfrac{1}{n_h \cos\theta_h} \end{pmatrix} \begin{pmatrix} 1 & 1 \\ n_0 \cos\theta_0 & -n_0 \cos\theta_0 \end{pmatrix} \begin{pmatrix} A_0(\omega) \\ B_0(\omega) \end{pmatrix} \quad (5.2.2.3)$$

$$\begin{pmatrix} A_{2k}(\omega) \\ B_{2k}(\omega) \end{pmatrix} = \frac{1}{2} \begin{pmatrix} e^{-in_l(\omega/c)\cos\theta_l[(k-1)(h+l)+h]} & \dfrac{1}{n_l \cos\theta_l} e^{-in_l(\omega/c)\cos\theta_l[(k-1)(h+l)+h]} \\ e^{in_l(\omega/c)\cos\theta_l[(k-1)(h+l)+h]} & -\dfrac{1}{n_l \cos\theta_l} e^{in_l(\omega/c)\cos\theta_l[(k-1)(h+l)+h]} \end{pmatrix} \cdot$$
$$\cdot \begin{pmatrix} e^{in_h(\omega/c)\cos\theta_h[(k-1)(h+l)+h]} & e^{-in_h(\omega/c)\cos\theta_h[(k-1)(h+l)+h]} \\ n_h \cos\theta_h e^{in_h(\omega/c)\cos\theta_h[(k-1)(h+l)+h]} & -n_h \cos\theta_h e^{-in_h(\omega/c)\cos\theta_h[(k-1)(h+l)+h]} \end{pmatrix} \begin{pmatrix} A_{2k-1}(\omega) \\ B_{2k-1}(\omega) \end{pmatrix}$$
$$(5.2.2.4)$$

$$\begin{pmatrix} A_{2k+1}(\omega) \\ B_{2k+1}(\omega) \end{pmatrix} = \frac{1}{2} \begin{pmatrix} e^{-in_h(\omega/c)\cos\theta_h[k(h+l)]} & \dfrac{1}{n_h \cos\theta_h} e^{-in_h(\omega/c)\cos\theta_h[k(h+l)]} \\ e^{in_h(\omega/c)\cos\theta_h[k(h+l)]} & -\dfrac{1}{n_h \cos\theta_h} e^{in_h(\omega/c)\cos\theta_h[k(h+l)]} \end{pmatrix} \cdot$$
$$\cdot \begin{pmatrix} e^{in_l(\omega/c)\cos\theta_l[k(h+l)]} & e^{-in_l(\omega/c)\cos\theta_l[k(h+l)]} \\ n_l \cos\theta_l e^{in_l(\omega/c)\cos\theta_l[k(h+l)]} & -n_l \cos\theta_l e^{-in_l(\omega/c)\cos\theta_l[k(h+l)]} \end{pmatrix} \begin{pmatrix} A_{2k}(\omega) \\ B_{2k}(\omega) \end{pmatrix}$$
$$(5.2.2.5)$$

$$\begin{pmatrix} A_{2N+2}(\omega) \\ B_{2N+2}(\omega) \end{pmatrix} = \frac{1}{2} \begin{pmatrix} e^{-in_0(\omega/c)\cos\theta_0[N(h+l)+h]} & \dfrac{1}{n_0 \cos\theta_0} e^{-in_0(\omega/c)\cos\theta_0[N(h+l)+h]} \\ e^{in_0(\omega/c)\cos\theta_0[N(h+l)+h]} & -\dfrac{1}{n_0 \cos\theta_0} e^{in_0(\omega/c)\cos\theta_0[N(h+l)+h]} \end{pmatrix} \cdot$$
$$\cdot \begin{pmatrix} e^{in_h(\omega/c)\cos\theta_h[N(h+l)+h]} & e^{-in_h(\omega/c)\cos\theta_h[N(h+l)+h]} \\ n_h \cos\theta_h e^{in_h(\omega/c)\cos\theta_h[N(h+l)+h]} & -n_h \cos\theta_h e^{-in_h(\omega/c)\cos\theta_h[N(h+l)+h]} \end{pmatrix} \begin{pmatrix} A_{2N+1}(\omega) \\ B_{2N+1}(\omega) \end{pmatrix}$$
$$(5.2.2.6)$$

dove $k \in [1, N]$.

Si ricorda la (5.2.1.18). Si riportano i valori della funzione d'onda e della sua derivata all'interfaccia iniziale $z = 0$; questi sono:

$$\begin{cases} f_0(\omega) = A_0(\omega) + B_0(\omega) \\ g_0(\omega) = iq_0(\omega)[A_0(\omega) - B_0(\omega)] \end{cases} \quad (5.2.2.7)$$

In corrispondenza delle autofrequenze $\omega = \omega_n$ dei QNMs TE, la funzione d'onda $f(z)$ deve essere contropropagante nel semispazio $z \leq 0$, quindi le autofrequenze $\omega_n$ soddisfano la condizione:

$$A_0(\omega) = 0 \quad (5.2.2.8)$$

che, applicata alla (5.2.2.7), fornisce:

$$\begin{cases} f_0(\omega) = B_0(\omega) \\ g_0(\omega) = -iq_0(\omega)B_0(\omega) \end{cases} \qquad (5.2.2.9)$$

Si riportano i valori della funzione d'onda e della sua derivata all'interfaccia finale $z = N(h+l)+h$; questi sono:

$$\begin{cases} f_{2N+1}(\omega) = A_{2N+2}(\omega)e^{iq_0(\omega)[N(h+l)+h]} + B_{2N+2}(\omega)e^{-iq_0(\omega)[N(h+l)+h]} \\ g_{2N+1}(\omega) = iq_0(\omega)A_{2N+2}(\omega)e^{iq_0(\omega)[N(h+l)+h]} - iq_0(\omega)B_{2N+2}(\omega)e^{-iq_0(\omega)[N(h+l)+h]} \end{cases} \qquad (5.2.2.10)$$

Il sistema (5.2.2.10) è di due equazioni nelle due incognite $A_{2N+2}(\omega)$ e $B_{2N+2}(\omega)$. Si determina il coefficiente $B_{2N+2}(\omega)$; si ottiene:

$$B_{2N+2}(\omega) = \frac{iq_0(\omega)f_{2N+1}(\omega) - g_{2N+1}(\omega)}{2iq_0(\omega)} e^{iq_0(\omega)[N(h+l)+h]} \qquad (5.2.2.11)$$

Alle autofrequenze $\omega = \omega_n$ dei QNMs TE, la funzione d'onda deve essere propagante nel semispazio $z \geq N(h+l)+h$; quindi le autofrequenze $\omega_n$ soddisfano anche la condizione:

$$B_{2N+2}(\omega) = 0 \qquad (5.2.2.12)$$

*II passo → sistema di equazioni*

Tenendo conto delle (5.2.1.16) e (5.2.1.17), si inserisce la (5.2.1.18) nella (5.2.2.11) e si ottiene:

$$\begin{aligned} B_{2N+2} = &\Big\{\Big[(if_0q_0\cos\delta_h + f_0q_h\sin\delta_h)\mu_{11} + (ig_0q_0\cos\delta_h + g_0q_h\operatorname{sen}\delta_h)\mu_{12} + \\ &+ (if_0\frac{q_0}{q_h}\sin\delta_h - f_0\cos\delta_h)\mu_{21} + (ig_0\frac{q_0}{q_h}\sin\delta_h - g_0\cos\delta_h)\mu_{22}\Big]U_{N-1}(\vartheta) + \\ &+ \Big[-if_0q_0\cos\delta_h - f_0q_h\sin\delta_h + g_0\cos\delta_h - ig_0\frac{q_0}{q_h}\sin\delta_h\Big]U_{N-2}(\vartheta)\Big\}\frac{e^{iq_0[N(h+l)+h]}}{2iq_0} \end{aligned}$$

$$(5.2.2.13)$$

Si inserisce la (5.2.2.9) nella (5.2.2.13) e si ottiene:

$$\begin{aligned} B_{2N+2} = &\Big\{\Big[(iq_0\cos\delta_h + q_h\sin\delta_h)\mu_{11} - iq_0(iq_0\cos\delta_h + q_h\sin\delta_h)\mu_{12} + \\ &+ (i\frac{q_0}{q_h}\sin\delta_h - \cos\delta_h)\mu_{21} - iq_0(i\frac{q_0}{q_h}\sin\delta_h - \cos\delta_h)\mu_{22}\Big]U_{N-1}(\vartheta) + \\ &- \Big[2iq_0\cos\delta_h + \Big(q_h + \frac{q_0^2}{q_h}\Big)\sin\delta_h\Big]U_{N-2}(\vartheta)\Big\} B_0 \frac{e^{iq_0[N(h+l)+h]}}{2iq_0} \end{aligned}$$

$$(5.2.2.14)$$

Ora, si impone la (5.2.2.12) e si ottiene:

$$\left[(iq_0 \cos\delta_h + q_h \sin\delta_h)\mu_{11} - iq_0(iq_0 \cos\delta_h + q_h \sin\delta_h)\mu_{12} + \right.$$
$$\left. +(i\frac{q_0}{q_h}\sin\delta_h - \cos\delta_h)\mu_{21} - iq_0(i\frac{q_0}{q_h}\sin\delta_h - \cos\delta_h)\mu_{22}\right]U_{N-1}(\vartheta) - \quad (5.2.2.15)$$
$$-\left[2iq_0 \cos\delta_h + \left(q_h + \frac{q_0^2}{q_h}\right)\sin\delta_h\right]U_{N-2}(\vartheta) = 0$$

Si prende la definizione dei polinomi di Chebyshev (5.2.1.13), si applica la formula di Eulero per il seno, si effettuano alcune manipolazioni e si utilizza la serie geometrica [5]:

$$U_N(\vartheta) = \frac{\sin[(N+1)\vartheta]}{\sin\vartheta} = \frac{e^{i(N+1)\vartheta} - e^{-i(N+1)\vartheta}}{e^{i\vartheta} - e^{-i\vartheta}} =$$
$$= e^{-iN\vartheta}\frac{1-e^{2i(N+1)\vartheta}}{1-e^{2i\vartheta}} = \left(e^{-i\vartheta}\right)^N \sum_{k=0}^{N}\left(e^{i2\vartheta}\right)^k \quad (5.2.2.16)$$

Si prende la posizione relativa ai polinomi di Chebyshev (5.2.1.14), si applica la formula di Eulero per il coseno [5] e si ottiene la condizione:

$$e^{i\vartheta} + e^{-i\vartheta} = \mu_{11} + \mu_{22} \quad (5.2.2.17)$$

Si inserisce la (5.2.2.16) nella (5.2.2.15) e si ottiene l'equazione:

$$\left[(iq_0 \cos\delta_h + q_h \sin\delta_h)\mu_{11} - iq_0(iq_0 \cos\delta_h + q_h \sin\delta_h)\mu_{12} + \right.$$
$$\left. +(i\frac{q_0}{q_h}\sin\delta_h - \cos\delta_h)\mu_{21} - iq_0(i\frac{q_0}{q_h}\sin\delta_h - \cos\delta_h)\mu_{22}\right]\sum_{k=0}^{N-1}\left(e^{i2\vartheta}\right)^k - \quad (5.2.2.18)$$
$$-\left[2iq_0 \cos\delta_h + \left(q_h + \frac{q_0^2}{q_h}\right)\sin\delta_h\right]e^{i\vartheta}\sum_{k=0}^{N-2}\left(e^{i2\vartheta}\right)^k = 0$$

Quindi, posto:

$$x = e^{i\vartheta} \quad (5.2.2.19)$$

si perviene al sistema costituito dalla condizione (5.2.2.17), che diviene:

$$x^2 - \gamma x + 1 = 0 \quad (5.2.2.20)$$

e dalla equazione (5.2.2.18), che diviene:

$$\alpha\sum_{k=0}^{N-1}(x^2)^k - \beta x\sum_{k=0}^{N-2}(x^2)^k = 0 \quad (5.2.2.21)$$

dove la dipendenza dalla frequenza é nei coefficienti, tali che:

$$\gamma = \mu_{11} + \mu_{22} \quad (5.2.2.22)$$

$$\frac{iq_0\beta}{2} = 2iq_0\cos\delta_h + \left(q_h + \frac{q_0^2}{q_h}\right)\sin\delta_h \tag{5.2.2.23}$$

$$\begin{aligned}\frac{iq_0\alpha}{2} &= (iq_0\cos\delta_h + q_h\sin\delta_h)\mu_{11} - iq_0(iq_0\cos\delta_h + q_h\sin\delta_h)\mu_{12} + \\ &+ (i\frac{q_0}{q_h}\sin\delta_h - \cos\delta_h)\mu_{21} - iq_0(i\frac{q_0}{q_h}\sin\delta_h - \cos\delta_h)\mu_{22}\end{aligned} \tag{5.2.2.24}$$

*III passo → equazione fondamentale*

Si utilizza la [1]:

$$\sum_{k=0}^{N}(x')^{N-k}(y')^k = \sum_{k=0}^{[N/2]}\frac{(-1)^k}{k!}\frac{(N-k)!}{(N-2k)!}(x'+y')^{N-2k}(x'y')^k \tag{5.2.2.25}$$

Definite le applicazioni $x' \to 1$ ed $y' \to x^2$, la (5.2.2.25) diviene:

$$\sum_{k=0}^{N}(x^2)^k = \sum_{k=0}^{[N/2]}\frac{(-1)^k}{k!}\frac{(N-k)!}{(N-2k)!}(1+x^2)^{N-2k}(x^2)^k \tag{5.2.2.26}$$

Si inserisce la (5.2.2.20) nella (5.2.2.26) e si ottiene:

$$\begin{aligned}\sum_{k=0}^{N}(x^2)^k &= \sum_{k=0}^{[N/2]}\frac{(-1)^k}{k!}\frac{(N-k)!}{(N-2k)!}(\gamma x)^{N-2k}(x^2)^k = \\ &= x^N \sum_{k=0}^{[N/2]}\frac{(-1)^k}{k!}\frac{(N-k)!}{(N-2k)!}\gamma^{N-2k}\end{aligned} \tag{5.2.2.27}$$

Quindi, si inserisce la (5.2.2.27) nella (5.2.2.21) e si ottiene l'equazione fondamentale per le autofrequenze dei QNMs per un 1D-PBG simmetrico:

$$\alpha \sum_{k=0}^{\left[\frac{N-1}{2}\right]}\frac{(-1)^k}{k!}\frac{(N-1-k)!}{(N-1-2k)!}\gamma^{N-1-2k} - \beta \sum_{k=0}^{\left[\frac{N-2}{2}\right]}\frac{(-1)^k}{k!}\frac{(N-2-k)!}{(N-2-2k)!}\gamma^{N-2-2k} = 0 \tag{5.2.2.28}$$

*IV passo → coefficienti α, β, γ dell'equazione*

Si inseriscono nella (5.2.2.23) la (5.2.1.4) e la prima delle (5.2.2.2); dopo alcune semplificazioni si ottiene:

$$\frac{i\beta}{2} = 2i\cos\delta_h + \left(\frac{n_h\cos\theta_h}{n_0\cos\theta_0} + \frac{n_0\cos\theta_0}{n_h\cos\theta_h}\right)\sin\delta_h \tag{5.2.2.29}$$

Si inseriscono nella (5.2.2.22) la prima e l'ultima delle (5.2.1.10) e si tiene conto delle (5.2.2.2); dopo alcuni passaggi, si ottiene:

$$\gamma = 2\cos\delta_h \cos\delta_l - \frac{n_h^2 \cos^2\theta_h + n_l^2 \cos^2\theta_l}{n_h n_l \cos\theta_h \cos\theta_l} \sin\delta_h \sin\delta_l \qquad (5.2.2.30)$$

Si inseriscono nella (5.2.2.24) le (5.2.1.10); dopo alcuni passaggi e semplificazioni, si ottiene:

$$\frac{iq_0 \alpha}{2} = 2iq_0 \cos 2\delta_h \cos\delta_l - iq_0 \left(\frac{q_h}{q_l} + \frac{q_l}{q_h}\right) \sin 2\delta_h \sin\delta_l + \left(q_h + \frac{q_0^2}{q_h}\right) \sin 2\delta_h \cos\delta_l +$$
$$- \left(\frac{q_h^2}{q_l} + q_0^2 \frac{q_l}{q_h}\right) \frac{1-\cos 2\delta_h}{2} \sin\delta_l + \left(q_l + \frac{q_0^2}{q_l}\right) \frac{1+\cos 2\delta_h}{2} \sin\delta_l$$

(5.2.2.31)

Si inseriscono nella (5.2.2.31) la (5.2.1.4) e le (5.2.2.2); dopo alcune semplificazioni, si ottiene:

$$i\alpha = i\left[4\cos 2\delta_h \cos\delta_l - 2\left(\frac{n_h \cos\theta_h}{n_l \cos\theta_l} + \frac{n_l \cos\theta_l}{n_h \cos\theta_h}\right) \sin 2\delta_h \sin\delta_l\right] +$$
$$+ 2\left(\frac{n_h \cos\theta_h}{n_0 \cos\theta_0} + \frac{n_0 \cos\theta_0}{n_h \cos\theta_h}\right) \sin 2\delta_h \cos\delta_l +$$
$$- 2\left(\frac{n_h^2 \cos^2\theta_h}{n_0 n_l \cos\theta_0 \cos\theta_l} + \frac{n_0 n_l \cos\theta_0 \cos\theta_l}{n_h^2 \cos^2\theta_h}\right) \frac{1-\cos 2\delta_h}{2} \sin\delta_l +$$
$$+ 2\left(\frac{n_l \cos\theta_l}{n_0 \cos\theta_0} + \frac{n_0 \cos\theta_0}{n_l \cos\theta_l}\right) \frac{1+\cos 2\delta_h}{2} \sin\delta_l$$

(5.2.2.32)

Si applicano le formule di Eulero [5], per il seno ed il coseno delle fasi, nelle formule (5.2.2.29), (5.2.2.30) e (5.2.2.32); si ottengono le:

$$\gamma = \frac{1}{4 n_h n_l \cos\theta_h \cos\theta_l} \Big[ (n_h \cos\theta_h + n_l \cos\theta_l)^2 e^{i(\delta_h+\delta_l)} - (n_h \cos\theta_h - n_l \cos\theta_l)^2 e^{i(\delta_h-\delta_l)} +$$
$$- (n_h \cos\theta_h - n_l \cos\theta_l)^2 e^{-i(\delta_h-\delta_l)} + (n_h \cos\theta_h + n_l \cos\theta_l)^2 e^{-i(\delta_h+\delta_l)} \Big]$$

(5.2.2.33)

$$\beta = \left[2 - \left(\frac{n_h \cos\theta_h}{n_0 \cos\theta_0} + \frac{n_0 \cos\theta_0}{n_h \cos\theta_h}\right)\right] e^{i\delta_h} + \left[2 + \left(\frac{n_h \cos\theta_h}{n_0 \cos\theta_0} + \frac{n_0 \cos\theta_0}{n_h \cos\theta_h}\right)\right] e^{-i\delta_h}$$

(5.2.2.34)

$$\alpha = \frac{1}{4}\Bigg\{\Bigg[4-2\left(\frac{n_h \cos\theta_h}{n_0 \cos\theta_0}+\frac{n_0 \cos\theta_0}{n_h \cos\theta_h}\right)-\left(\frac{n_l \cos\theta_l}{n_0 \cos\theta_0}+\frac{n_0 \cos\theta_0}{n_l \cos\theta_l}\right)+$$

$$+2\left(\frac{n_h \cos\theta_h}{n_l \cos\theta_l}+\frac{n_l \cos\theta_l}{n_h \cos\theta_h}\right)-\left(\frac{n_h^2 \cos^2\theta_h}{n_0 n_l \cos\theta_0 \cos\theta_l}+\frac{n_0 n_l \cos\theta_0 \cos\theta_l}{n_h^2 \cos^2\theta_h}\right)\Bigg]e^{i(2\delta_h+\delta_l)}+$$

$$+\Bigg[4-2\left(\frac{n_h \cos\theta_h}{n_0 \cos\theta_0}+\frac{n_0 \cos\theta_0}{n_h \cos\theta_h}\right)+\left(\frac{n_l \cos\theta_l}{n_0 \cos\theta_0}+\frac{n_0 \cos\theta_0}{n_l \cos\theta_l}\right)+$$

$$-2\left(\frac{n_h \cos\theta_h}{n_l \cos\theta_l}+\frac{n_l \cos\theta_l}{n_h \cos\theta_h}\right)+\left(\frac{n_h^2 \cos^2\theta_h}{n_0 n_l \cos\theta_0 \cos\theta_l}+\frac{n_0 n_l \cos\theta_0 \cos\theta_l}{n_h^2 \cos^2\theta_h}\right)\Bigg]e^{i(2\delta_h-\delta_l)}+$$

$$-2\Bigg[\left(\frac{n_l \cos\theta_l}{n_0 \cos\theta_0}+\frac{n_0 \cos\theta_0}{n_l \cos\theta_l}\right)-\left(\frac{n_h^2 \cos^2\theta_h}{n_0 n_l \cos\theta_0 \cos\theta_l}+\frac{n_0 n_l \cos\theta_0 \cos\theta_l}{n_h^2 \cos^2\theta_h}\right)\Bigg]e^{i\delta_l}+$$

$$+2\Bigg[\left(\frac{n_l \cos\theta_l}{n_0 \cos\theta_0}+\frac{n_0 \cos\theta_0}{n_l \cos\theta_l}\right)-\left(\frac{n_h^2 \cos^2\theta_h}{n_0 n_l \cos\theta_0 \cos\theta_l}+\frac{n_0 n_l \cos\theta_0 \cos\theta_l}{n_h^2 \cos^2\theta_h}\right)\Bigg]e^{-i\delta_l}+$$

$$+\Bigg[4+2\left(\frac{n_h \cos\theta_h}{n_0 \cos\theta_0}+\frac{n_0 \cos\theta_0}{n_h \cos\theta_h}\right)-\left(\frac{n_l \cos\theta_l}{n_0 \cos\theta_0}+\frac{n_0 \cos\theta_0}{n_l \cos\theta_l}\right)+$$

$$-2\left(\frac{n_h \cos\theta_h}{n_l \cos\theta_l}+\frac{n_l \cos\theta_l}{n_h \cos\theta_h}\right)-\left(\frac{n_h^2 \cos^2\theta_h}{n_0 n_l \cos\theta_0 \cos\theta_l}+\frac{n_0 n_l \cos\theta_0 \cos\theta_l}{n_h^2 \cos^2\theta_h}\right)\Bigg]e^{-i(2\delta_h-\delta_l)}+$$

$$+\Bigg[4+2\left(\frac{n_h \cos\theta_h}{n_0 \cos\theta_0}+\frac{n_0 \cos\theta_0}{n_h \cos\theta_h}\right)+\left(\frac{n_l \cos\theta_l}{n_0 \cos\theta_0}+\frac{n_0 \cos\theta_0}{n_l \cos\theta_l}\right)+$$

$$+2\left(\frac{n_h \cos\theta_h}{n_l \cos\theta_l}+\frac{n_l \cos\theta_l}{n_h \cos\theta_h}\right)+\left(\frac{n_h^2 \cos^2\theta_h}{n_0 n_l \cos\theta_0 \cos\theta_l}+\frac{n_0 n_l \cos\theta_0 \cos\theta_l}{n_h^2 \cos^2\theta_h}\right)\Bigg]e^{-i(2\delta_h+\delta_l)}\Bigg\}$$

(5.2.2.35)

*Risoluzione in approssimazione parassiale per un 1D-PBG a λ/4*

Il numero d'onda lungo la direzione $z$ è il seguente [vedi eq. (4.2.6)]:

$$q(z) = k n(z) \cos\theta(z) \qquad (5.2.2.36)$$

dove:

$$k = \frac{\omega}{c} \qquad (5.2.2.37)$$

Si può definire una lunghezza d'onda effettiva per la propagazione lungo la direzione $z$ come [3]:

$$\lambda_z = \frac{2\pi}{q(z)} = \frac{2\pi/k}{n(z)\cos\theta(z)} \qquad (5.2.2.38)$$

Con riferimento ad una lunghezza d'onda $\lambda_{rif}$, quelle negli strati di spessore $h$ ed $l$ sono rispettivamente:

$$\begin{cases} \lambda_h = \dfrac{\lambda_{rif}}{n_h \cos\theta_h} \\ \lambda_l = \dfrac{\lambda_{rif}}{n_l \cos\theta_l} \end{cases} \quad (5.2.2.39)$$

Si suppone che gli spessori geometrici siano tali che:

$$\begin{cases} h = m_h \dfrac{\lambda_h}{4} \\ l = m_l \dfrac{\lambda_l}{4} \end{cases} \Rightarrow \begin{cases} n_h h = m_h \dfrac{\lambda_{rif}}{4\cos\theta_h} \\ n_l l = m_l \dfrac{\lambda_{rif}}{4\cos\theta_l} \end{cases} \quad \text{con} \quad m_h, m_l \in N \quad (5.2.2.40)$$

Se si applica l'approssimazione parassiale [4]:

$$\cos\theta_h \cong \cos\theta_l \cong 1 \quad (5.2.2.41)$$

alle equazioni (5.2.2.40), si può parlare di un 1D-PBG a quarto d'onda:

$$n_h h \cong n_l l \cong \dfrac{\lambda_{rif}}{4} \quad (5.2.2.42)$$

Le fasi accumulate negli strati di spessore $h$ ed $l$ sono rispettivamente:

$$\begin{cases} \delta_h = q_h h = k n_h h \cos\theta_h \\ \delta_l = q_l l = k n_l l \cos\theta_l \end{cases} \quad (5.2.2.43)$$

Se si inseriscono le condizioni (5.2.2.40) nelle equazioni (5.2.2.43), si ottiene:

$$\begin{cases} \delta_h = m_h \delta \\ \delta_l = m_l \delta \end{cases} \quad \text{con} \quad m_h, m_l \in N \quad (5.2.2.44)$$

dove:

$$\delta = k \dfrac{\lambda_{rif}}{4} \quad (5.2.2.45)$$

L'equazione (5.2.2.28), per le autofrequenze dei QNMs TE in un 1D-PBG simmetrico, diviene un'equazione polinomiale a coefficienti reali di grado $N(m_l + m_h) + m_h$ nella variabile composta $e^{2i\delta}$.

Si deduce che:

a) i QNMs sono suddivisi in famiglie indipendenti di numero: $N_F = N(m_l + m_h) + m_h$;

b) le autofrequenze di ogni famiglia hanno la parte immaginaria uguale e si ripetono periodicamente in direzione dell'asse reale con passo:
$\Delta = 2\pi c m_h / n_h h \cos\theta_h$

I grafici seguenti sono per un 1D-PBG a *λ/4*, con numero dei periodi $N = 4$, indici di rifrazione $n_h = 2$, $n_l = 1$ e lunghezza di riferimento $\lambda_{rif} = 1\mu m$; sono rappresentati i moduli delle autofunzioni per i QNMs TE quando l'incidenza è normale (si fa riferimento alle equazioni (5.2.2.1-2)).

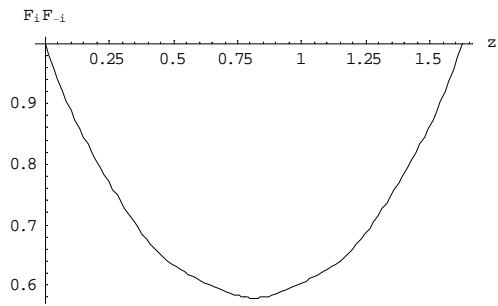
Grafico 5.1. QNM$_0$ TE.

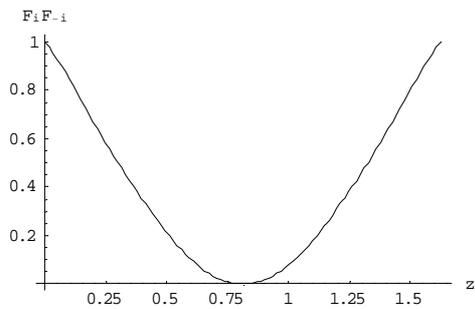
Grafico 5.2 QNM$_1$ TE.

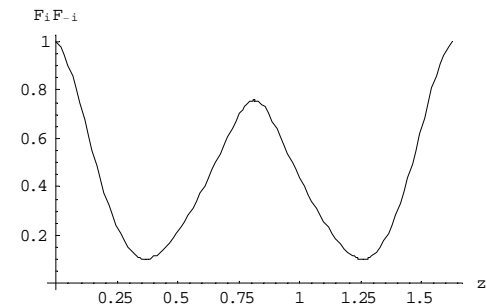
Grafico 5.3. QNM$_2$ TE.

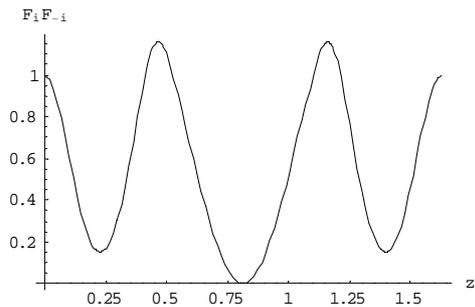
Grafico 5.4. QNM$_3$ TE.

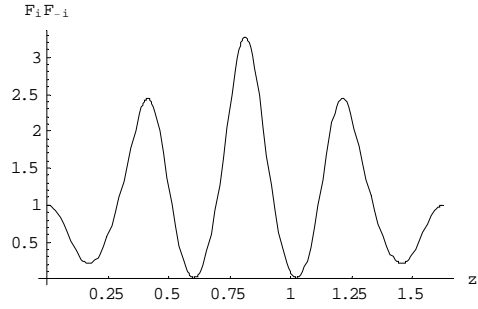
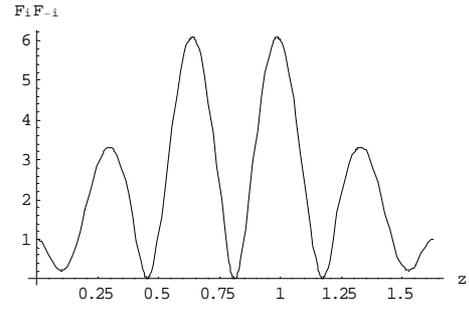

Grafico 5.5. QNM$_4$ TE.    Grafico 5.6. QNM$_5$ TE.

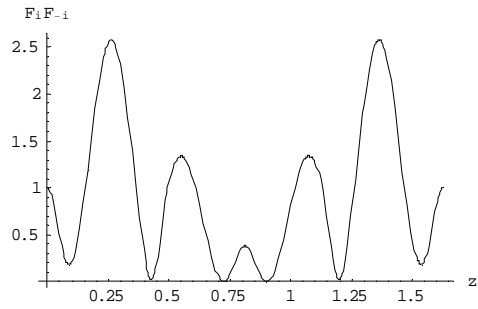
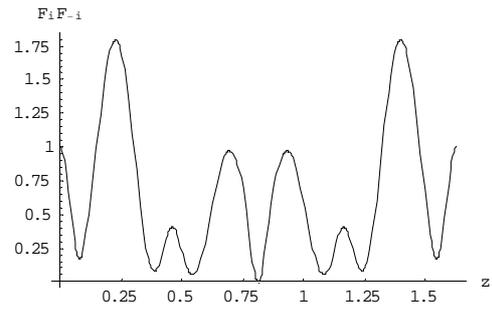

Grafico 5.7. QNM$_6$ TE.    Grafico 5.8. QNM$_7$ TE.

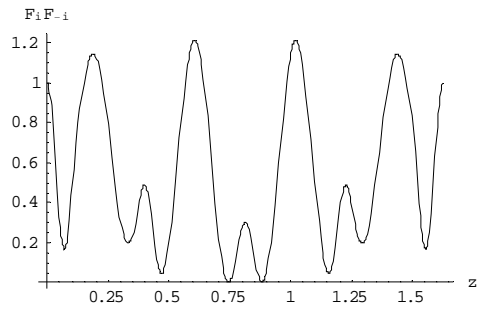
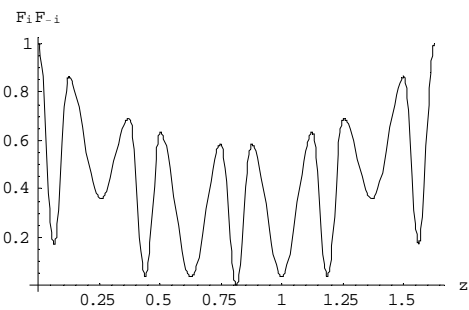

Grafico 5.9. QNM$_8$ TE.    Grafico 5.10. QNM$_9$ TE.

Si osserva che le autofunzioni corrispondono effettivamente ai QNMs in quanto sono diverse da zero alle estremità del PBG; in particolare $f_n(0)=1$ e, per la simmetria, $f_n(L)=(-1)^n f_n(0)$.

Inoltre, l'autofunzione $f_n(z)$ presenta $n$ minimi lungo il PBG.

Nell'appendice alla tesi, qualora una pompa incida normalmente sul 1D-PBG, si determinerà il campo elettrico dentro al 1D-PBG come sviluppo di queste autofunzioni QNM e si confronterà il risultato con quello ottenuto tramite metodi numerici basati sulla matrice di trasmissione.

I grafici seguenti sono per un 1D-PBG a *λ/4*, con numero dei periodi $N = 4$, indici di rifrazione $n_h = 2$, $n_l = 1.5$ e lunghezza di riferimento $\lambda_{rif} = 1\mu m$. Si ritiene valida l'approssimazione parassiale finché l'angolo di incidenza rimane $\theta_0 \leq 30$. Si fa riferimento all'eq. (5.2.2.28) con coeff. (5.2.2.33-35).

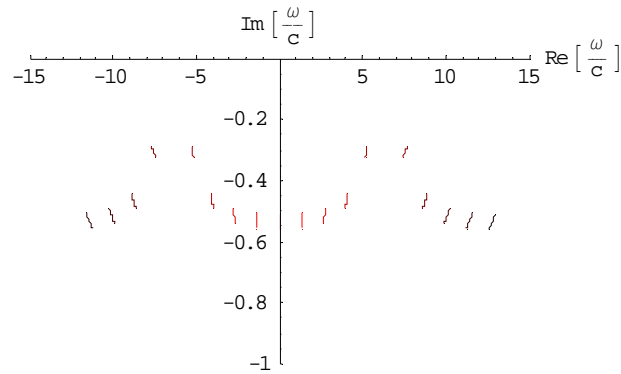

Grafico 5.11. Luogo dei QNMs TE nel piano complesso quando, in ipotesi di approssimazione parassiale, l'angolo di incidenza varia da 0 a 30 gradi.

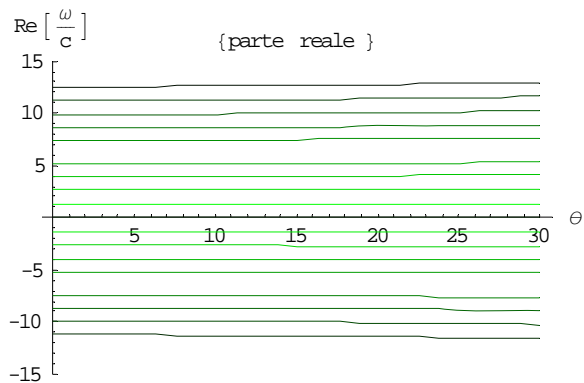

Grafico 5.12. Parte reale di ogni QNM TE in funzione dell'angolo di incidenza (approssimazione parassiale: 0-30 gradi).

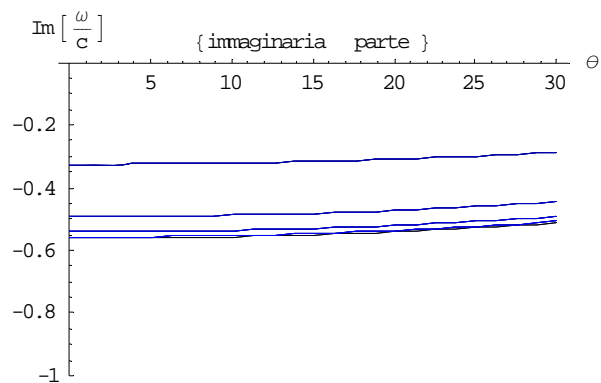

Grafico 5.13. Parte immaginaria di ogni QNM TE in funzione dell'angolo di incidenza (approssimazione parassiale: 0-30 gradi).

Dal grafico 5.11, i QNMs TE sono simmetrici all'asse immaginario, cioè ad ogni parte immaginaria corrispondono due parti reali uguali ed opposte, e salgono per incidenze crescenti, pur rimanendo nel semipiano negativo.

Dal grafico 5.12, le parti reali dei QNMs TE sono pressoché costanti al variare dell'angolo; difatti, le risonanze sono determinate dalla struttura.

Dal grafico 5.13, le parti immaginarie dei QNMs TE diminuiscono in modulo quando l'angolo aumenta; quindi, le perdite si riducono per incidenze più grandi: infatti, in un cono di raggi ottici, i più esterni sono a minor contenuto energetico e, quando investono la cavità aperta, questa dà luogo a perdite relative più basse.

Nell'appendice alla tesi, dato un 1D-PBG su cui una pompa incide normalmente, si verificherà che i QNM hanno, come parti reali, le frequenze di risonanza e, come parti immaginarie, le ampiezze di oscillazione che si evincono dallo spettro di trasmissione, ottenuto tramite i metodi numerici.

### *Paragrafo 5.3. Modi quasi normali TM per un 1D-PBG.*

In questo paragrafo, si studiano i QNMs di tipo TM in un PBG unidimensionale.

### *5.3.1. Strumenti di analisi.*

Qui, con riferimento alla funzione d'onda per i QNMs TM, viene costruita la matrice di trasmissione TM per un 1D-PBG.

### *5.3.1.a. Funzione d'onda per i QNMs TM.*

Indicata la funzione d'onda effettiva per i QNMs TM con $f(z)$, si introduce la trasformazione [3]:

$$f^{(b)}(z) = \frac{n_0}{n(z)} f(z) \qquad (5.3.1.1)$$

cosicché l'equazione d'onda per i QNMs TM diviene formalmente analoga a quella per i QNMs TE [3]:

$$\frac{\partial^2 f^{(b)}}{\partial z^2} + q_b^2(z) f^{(b)}(z) = 0 \qquad (5.3.1.2)$$

dove:

$$q_b^2(z) = q^2(z) - n(z) \frac{d^2}{dz^2}\left(\frac{1}{n(z)}\right) \qquad (5.3.1.3)$$

con:

$$q(z) = k n(z) \cos\theta(z) \tag{5.3.1.4}$$

$$k = \frac{\omega}{c} \tag{5.3.1.5}$$

In corrispondenza delle autofrequenze $\omega = \omega_n$ dei QNMs TM, la funzione d'onda $f^{(b)}(z)$ dà luogo alle autofunzioni $f_n^{(b)}(z)$ e rispetta entrambe le condizioni asintotiche:

$$\begin{cases} f_n^{(b)}(z) = e^{-i q_0(\omega_n) z} & \text{per } z \to -\infty \\ f_n^{(b)}(z) = e^{+i q_0(\omega_n) z} & \text{per } z \to +\infty \end{cases} \tag{5.3.1.6}$$

dove:

$$q_0(\omega) = n_0 \frac{\omega}{c} \cos\theta_0 \tag{5.3.1.7}$$

### 5.3.1.b. Matrice di trasmissione TM per un 1D-PBG.

Si consideri un 1D-PBG tale che la cella elementare sia costituita da due strati, rispettivamente con lunghezza ed indice di rifrazione $h$, $n_h$ ed $l$, $n_l$.

Si suppone che l'onda incidente abbia polarizzazione TM.

La matrice della singola cella è del tipo [3]:

$$M^{(TM)} = \begin{pmatrix} \mu'_{11} & \mu'_{12} \\ \mu'_{21} & \mu'_{22} \end{pmatrix} \tag{5.3.1.8}$$

dove:

$$\begin{cases} \mu'_{11} = \cos\delta_h \cos\delta_l - \dfrac{Q_h}{Q_l} \sin\delta_h \sin\delta_l \\ \mu'_{12} = \dfrac{1}{Q_h} \sin\delta_h \cos\delta_l + \dfrac{1}{Q_l} \sin\delta_l \cos\delta_h \\ \mu'_{21} = -Q_h \sin\delta_h \cos\delta_l - Q_l \sin\delta_l \cos\delta_h \\ \mu'_{22} = \cos\delta_h \cos\delta_l - \dfrac{Q_l}{Q_h} \sin\delta_h \sin\delta_l \end{cases} \tag{5.3.1.9}$$

con:

$$\begin{cases} Q_h = q_h / n_h^2 \\ Q_l = q_l / n_l^2 \end{cases} \tag{5.3.1.10}$$

*PBG periodico*

Un PBG periodico è costituito dalla replica della cella elementare per un numero di periodi *N*.

Con riferimento al caso TM, valgono ancora le (5.2.1.11-14).

*PBG simmetrico*

Un PBG simmetrico si ottiene da uno periodico aggiungendo a destra uno strato di lunghezza $h$ ed indice di rifrazione $n_h$.

La matrice del PBG simmetrico si ottiene moltiplicando da sinistra la matrice del PBG periodico per quella del suddetto strato [3]. Si ottiene:

$$M_{PBG-simm.}^{(TM)} = \begin{pmatrix} m'_{11} & m'_{12} \\ m'_{21} & m'_{22} \end{pmatrix} \qquad (5.3.1.11)$$

dove:

$$\begin{cases} m'_{11} = \cos\delta_h [\mu_{11} U_{N-1}(\vartheta) - U_{N-2}(\vartheta)] \frac{\sin\delta_h}{Q_h} \mu_{21} U_{N-1}(\vartheta) \\ m'_{12} = \cos\delta_h \mu_{12} U_{N-1}(\vartheta) + \frac{\sin\delta_h}{Q_h} [\mu_{22} U_{N-1}(\vartheta) - U_{N-2}(\vartheta)] \\ m'_{21} = -Q_h \sin\delta_h [\mu_{11} U_{N-1}(\vartheta) - U_{N-2}(\vartheta)] + \cos\delta_h \mu_{21} U_{N-1}(\vartheta) \\ m'_{22} = -Q_h \sin\delta_h \mu_{12} U_{N-1}(\vartheta) + \cos\delta_h [\mu_{22} U_{N-1} - U_{N-2}(\vartheta)] \end{cases} \qquad (5.3.1.12)$$

Con riferimento alla figura 5.1, indicata la funzione d'onda del generico QNM TE con $f(z)$ e introdotta la $g(z)$ tale che $n^2(z)g(z) = df/dz$, i loro valori agli estremi del 1D-PBG simmetrico $(f_{2N+1}, g_{2N+1})$ e $(f_0, g_0)$ sono legati dalla [3]:

$$\begin{pmatrix} f_{2N+1} \\ g_{2N+1} \end{pmatrix} = M_{PBG-simm.}^{(TM)} \begin{pmatrix} f_0 \\ g_0 \end{pmatrix} \qquad (5.3.1.13)$$

## 5.3.2. Calcolo delle autofunzioni ed autofrequenze. (1D-PBG simmetrico)

*Impostazione*

Tenendo conto della trasformazione (5.3.1.1), si risolve l'equazione (5.3.1.2) in ogni strato, dove $q_b(z) = q(z)$ poiché $n(z) = cost$ (vedi eq. 5.3.1.3).

Si ottiene la funzione d'onda per i QNMs TM:

$$f(z) = \left[ A_0(\omega)e^{iq_0(\omega)z} + B_0(\omega)e^{-iq_0(\omega)z} \right] u_{-1}(-z) +$$

$$+ \frac{n_h}{n_0} \sum_{k=0}^{N} \left[ A_{2k+1}(\omega)e^{iq_h(\omega)z} + B_{2k+1}(\omega)e^{-iq_h(\omega)z} \right] u_{-1}[z-k(h+l)] u_{-1}[k(h+l)+h-z] +$$

$$+ \frac{n_l}{n_0} \sum_{k=0}^{N-1} \left[ A_{2k+2}(\omega)e^{iq_l(\omega)z} + B_{2k+2}(\omega)e^{-iq_l(\omega)z} \right] u_{-1}[z-k(h+l)-h] u_{-1}[(k+1)(h+l)-z] +$$

$$+ \left[ A_{2N+2}(\omega)e^{iq_0(\omega)z} + B_{2N+2}(\omega)e^{-iq_0(\omega)z} \right] u_{-1}(z-N(h+l)-h)$$

(5.3.2.1)

Se si impongono le condizioni di continuità ad ogni interfaccia per le funzioni $f(z)$ e $g(z)$, si determinano le ampiezze $A$ e $B$.

Valgono le:

$$\begin{pmatrix} A_1(\omega) \\ B_1(\omega) \end{pmatrix} = \frac{1}{2} \begin{pmatrix} \dfrac{1}{n_h} & \dfrac{1}{\cos\theta_h} \\ \dfrac{1}{n_h} & -\dfrac{1}{\cos\theta_h} \end{pmatrix} \begin{pmatrix} 1 & 1 \\ n_0\cos\theta_0 & -n_0\cos\theta_0 \end{pmatrix} \begin{pmatrix} A_0(\omega) \\ B_0(\omega) \end{pmatrix} \quad (5.3.2.2)$$

$$\begin{pmatrix} A_{2k}(\omega) \\ B_{2k}(\omega) \end{pmatrix} = \frac{1}{2} \begin{pmatrix} \dfrac{1}{n_l}e^{-in_l(\omega/c)\cos\theta_l[(k-1)(h+l)+h]} & \dfrac{1}{\cos\theta_l}e^{-in_l(\omega/c)\cos\theta_l[(k-1)(h+l)+h]} \\ \dfrac{1}{n_l}e^{in_l(\omega/c)\cos\theta_l[(k-1)(h+l)+h]} & -\dfrac{1}{\cos\theta_l}e^{in_l(\omega/c)\cos\theta_l[(k-1)(h+l)+h]} \end{pmatrix} \cdot$$

$$\cdot \begin{pmatrix} n_h e^{in_h(\omega/c)\cos\theta_h[(k-1)(h+l)+h]} & n_h e^{-in_h(\omega/c)\cos\theta_h[(k-1)(h+l)+h]} \\ \cos\theta_h e^{in_h(\omega/c)\cos\theta_h[(k-1)(h+l)+h]} & -\cos\theta_h e^{-in_h(\omega/c)\cos\theta_h[(k-1)(h+l)+h]} \end{pmatrix} \begin{pmatrix} A_{2k-1}(\omega) \\ B_{2k-1}(\omega) \end{pmatrix}$$

(5.3.2.3)

$$\begin{pmatrix} A_{2k+1}(\omega) \\ B_{2k+1}(\omega) \end{pmatrix} = \frac{1}{2} \begin{pmatrix} \dfrac{1}{n_h}e^{-in_h(\omega/c)\cos\theta_h[k(h+l)]} & \dfrac{1}{\cos\theta_h}e^{-in_h(\omega/c)\cos\theta_h[k(h+l)]} \\ \dfrac{1}{n_h}e^{in_h(\omega/c)\cos\theta_h[k(h+l)]} & -\dfrac{1}{\cos\theta_h}e^{in_h(\omega/c)\cos\theta_h[k(h+l)]} \end{pmatrix} \cdot$$

$$\cdot \begin{pmatrix} n_l e^{in_l(\omega/c)\cos\theta_l[k(h+l)]} & n_l e^{-in_l(\omega/c)\cos\theta_l[k(h+l)]} \\ \cos\theta_l e^{in_l(\omega/c)\cos\theta_l[k(h+l)]} & -\cos\theta_l e^{-in_l(\omega/c)\cos\theta_l[k(h+l)]} \end{pmatrix} \begin{pmatrix} A_{2k}(\omega) \\ B_{2k}(\omega) \end{pmatrix}$$

(5.3.2.4)

$$\begin{pmatrix} A_{2N+2}(\omega) \\ B_{2N+2}(\omega) \end{pmatrix} = \frac{1}{2} \begin{pmatrix} e^{-in_0(\omega/c)\cos\theta_0[N(h+l)+h]} & \dfrac{1}{n_0\cos\theta_0}e^{-in_0(\omega/c)\cos\theta_0[N(h+l)+h]} \\ e^{in_0(\omega/c)\cos\theta_0[N(h+l)+h]} & -\dfrac{1}{n_0\cos\theta_0}e^{in_0(\omega/c)\cos\theta_0[N(h+l)+h]} \end{pmatrix} \cdot$$

$$\cdot \begin{pmatrix} n_h e^{in_h(\omega/c)\cos\theta_h[N(h+l)+h]} & n_h e^{-in_h(\omega/c)\cos\theta_h[N(h+l)+h]} \\ \cos\theta_h e^{in_h(\omega/c)\cos\theta_h[N(h+l)+h]} & -\cos\theta_h e^{-in_h(\omega/c)\cos\theta_h[N(h+l)+h]} \end{pmatrix} \begin{pmatrix} A_{2N+1}(\omega) \\ B_{2N+1}(\omega) \end{pmatrix}$$

(5.3.2.5)

In corrispondenza delle autofrequenze $\omega = \omega_n$ dei QNMs TM, la funzione d'onda $f(z)$ deve essere contropropagante nel semispazio $z \leq 0$, quindi le autofrequenze $\omega_n$ soddisfano la condizione:

$$A_0(\omega) = 0 \tag{5.3.2.6}$$

Analogamente al caso TE, si ricavano le funzioni $f(z)$ e $g(z)$ all'interfaccia $z = 0$. Si ottengono le:

$$\begin{cases} f_0(\omega) = B_0(\omega) \\ g_0(\omega) = -(i/n_0^2)q_0(\omega)B_0(\omega) \end{cases} \tag{5.3.2.7}$$

Alle autofrequenze $\omega = \omega_n$ dei QNMs TM, la funzione d'onda deve essere propagante nel semispazio $z \geq N(h+l)+h$. Analogamente al caso TE, si ricava l'altra condizione per le autofrequenze $\omega_n$. Si ottiene:

$$B_{2N+2}(\omega) = \frac{iq_0(\omega)f_{2N+1}(\omega) - n_0^2 g_{2N+1}(\omega)}{2iq_0(\omega)} e^{iq_0(\omega)[N(h+l)+h]} = 0 \tag{5.3.2.8}$$

*Equazione fondamentale e suoi coefficienti*

Procedendo in modo analogo al caso TE, si ottiene l'equazione fondamentale per le autofrequenze dei QNMs per un 1D-PBG simmetrico:

$$\alpha \sum_{k=0}^{\left[\frac{N-1}{2}\right]} \frac{(-1)^k}{k!} \frac{(N-1-k)!}{(N-1-2k)!} \gamma^{N-1-2k} - \beta \sum_{k=0}^{\left[\frac{N-2}{2}\right]} \frac{(-1)^k}{k!} \frac{(N-2-k)!}{(N-2-2k)!} \gamma^{N-2-2k} = 0$$

$$\tag{5.3.2.9}$$

quando la radiazione è di tipo TM:

$$\gamma = \frac{1}{4n_h n_l \cos\theta_h \cos\theta_l}\Big[(n_l\cos\theta_h + n_h\cos\theta_l)^2 e^{i(\delta_h+\delta_l)} - (n_l\cos\theta_h - n_h\cos\theta_l)^2 e^{i(\delta_h-\delta_l)} +$$
$$-(n_l\cos\theta_h - n_h\cos\theta_l)^2 e^{-i(\delta_h-\delta_l)} + (n_l\cos\theta_h + n_h\cos\theta_l)^2 e^{-i(\delta_h+\delta_l)}\Big]$$

$$\tag{5.3.2.10}$$

$$\beta = \left[2 - \left(\frac{n_0\cos\theta_h}{n_h\cos\theta_0} + \frac{n_h\cos\theta_0}{n_0\cos\theta_h}\right)e^{i\delta_h}\right] + \left[2 + \left(\frac{n_0\cos\theta_h}{n_h\cos\theta_0} + \frac{n_h\cos\theta_0}{n_0\cos\theta_h}\right)e^{-i\delta_h}\right]$$

$$\tag{5.3.2.11}$$

$$\alpha = \frac{1}{4}\left\{\left[4 - 2\left(\frac{n_0 \cos\theta_h}{n_h \cos\theta_0} + \frac{n_h \cos\theta_0}{n_0 \cos\theta_h}\right) - \left(\frac{n_0 \cos\theta_l}{n_l \cos\theta_0} + \frac{n_l \cos\theta_0}{n_0 \cos\theta_l}\right) + \right.\right.$$

$$+ 2\left(\frac{n_l \cos\theta_h}{n_h \cos\theta_l} + \frac{n_h \cos\theta_l}{n_l \cos\theta_h}\right) - \left(\frac{n_0 n_l \cos^2\theta_h}{n_h^2 \cos\theta_0 \cos\theta_l} + \frac{n_h^2 \cos\theta_0 \cos\theta_l}{n_0 n_l \cos^2\theta_h}\right)\Bigg] e^{i(2\delta_h+\delta_l)} +$$

$$+ \left[4 - 2\left(\frac{n_0 \cos\theta_h}{n_h \cos\theta_0} + \frac{n_h \cos\theta_0}{n_0 \cos\theta_h}\right) + \left(\frac{n_0 \cos\theta_l}{n_l \cos\theta_0} + \frac{n_l \cos\theta_0}{n_0 \cos\theta_l}\right) + \right.$$

$$\left. - 2\left(\frac{n_l \cos\theta_h}{n_h \cos\theta_l} + \frac{n_h \cos\theta_l}{n_l \cos\theta_h}\right) + \left(\frac{n_0 n_l \cos^2\theta_h}{n_h^2 \cos\theta_0 \cos\theta_l} + \frac{n_h^2 \cos\theta_0 \cos\theta_l}{n_0 n_l \cos^2\theta_h}\right)\right] e^{i(2\delta_h-\delta_l)} +$$

$$- 2\left[\left(\frac{n_0 \cos\theta_l}{n_l \cos\theta_0} + \frac{n_l \cos\theta_0}{n_0 \cos\theta_l}\right) - \left(\frac{n_0 n_l \cos^2\theta_h}{n_h^2 \cos\theta_0 \cos\theta_l} + \frac{n_h^2 \cos\theta_0 \cos\theta_l}{n_0 n_l \cos^2\theta_h}\right)\right] e^{i\delta_l} +$$

$$+ 2\left[\left(\frac{n_0 \cos\theta_l}{n_l \cos\theta_0} + \frac{n_l \cos\theta_0}{n_0 \cos\theta_l}\right) - \left(\frac{n_0 n_l \cos^2\theta_h}{n_h^2 \cos\theta_0 \cos\theta_l} + \frac{n_h^2 \cos\theta_0 \cos\theta_l}{n_0 n_l \cos^2\theta_h}\right)\right] e^{-i\delta_l} +$$

$$+ \left[4 + 2\left(\frac{n_0 \cos\theta_h}{n_h \cos\theta_0} + \frac{n_h \cos\theta_0}{n_0 \cos\theta_h}\right) - \left(\frac{n_0 \cos\theta_l}{n_l \cos\theta_0} + \frac{n_l \cos\theta_0}{n_0 \cos\theta_l}\right) + \right.$$

$$\left. - 2\left(\frac{n_l \cos\theta_h}{n_h \cos\theta_l} + \frac{n_h \cos\theta_l}{n_l \cos\theta_h}\right) - \left(\frac{n_0 n_l \cos^2\theta_h}{n_h^2 \cos\theta_0 \cos\theta_l} + \frac{n_h^2 \cos\theta_0 \cos\theta_l}{n_0 n_l \cos^2\theta_h}\right)\right] e^{-i(2\delta_h-\delta_l)} +$$

$$+ \left[4 + 2\left(\frac{n_0 \cos\theta_h}{n_h \cos\theta_0} + \frac{n_h \cos\theta_0}{n_0 \cos\theta_h}\right) + \left(\frac{n_0 \cos\theta_l}{n_l \cos\theta_0} + \frac{n_l \cos\theta_0}{n_0 \cos\theta_l}\right) + \right.$$

$$\left.\left. + 2\left(\frac{n_l \cos\theta_h}{n_h \cos\theta_l} + \frac{n_h \cos\theta_l}{n_l \cos\theta_h}\right) + \left(\frac{n_0 n_l \cos^2\theta_h}{n_h^2 \cos\theta_0 \cos\theta_l} + \frac{n_h^2 \cos\theta_0 \cos\theta_l}{n_0 n_l \cos^2\theta_h}\right)\right] e^{-i(2\delta_h+\delta_l)}\right\}$$

(5.3.2.12)

*Risoluzione dell'equazione e rappresentazione dei risultati*

Procedendo come nel caso TE, l'equazione (5.3.2.9) viene risolta, in approssimazione parassiale, per un 1D-PBG a quarto d'onda.

I grafici seguenti sono per un 1D-PBG a $\lambda/4$, con numero dei periodi $N = 4$, indici di rifrazione $n_h = 2$, $n_l = 1$ e lunghezza di riferimento $\lambda_{rif} = 1\mu m$; sono rappresentati i moduli delle autofunzioni per i QNMs TM quando l'incidenza è normale (si fa riferimento all'equazione (5.3.2.1)).

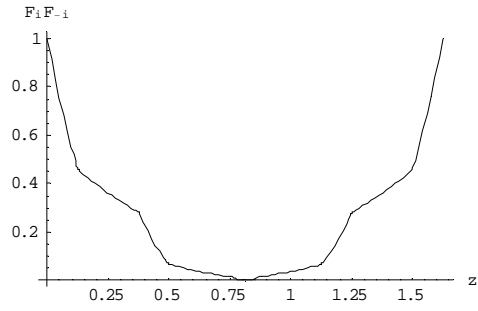
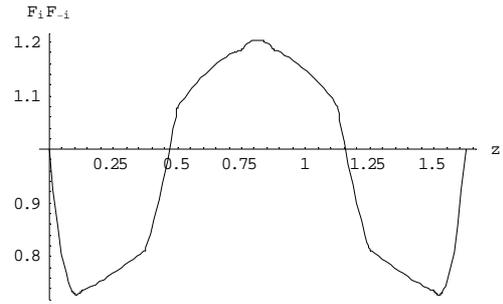

Grafico 5.14. $QNM_0$ TM.           Grafico 5.15. $QNM_1$ TM.

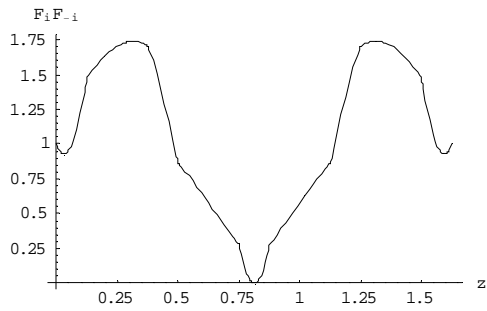
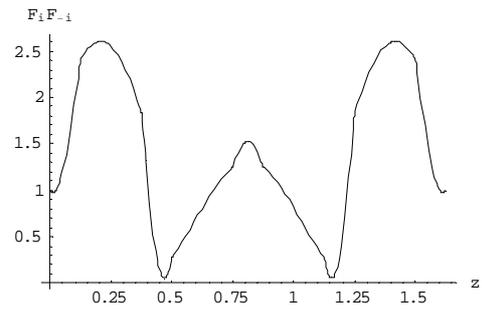

Grafico 5.16. $QNM_2$ TM.           Grafico 5.17. $QNM_3$ TM.

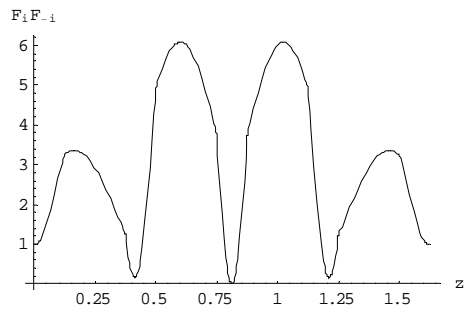
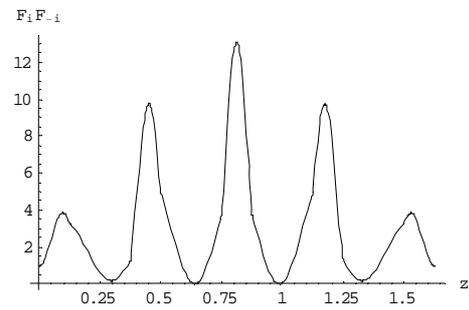

Grafico 5.18. $QNM_4$ TM.           Grafico 5.19. $QNM_5$ TM.

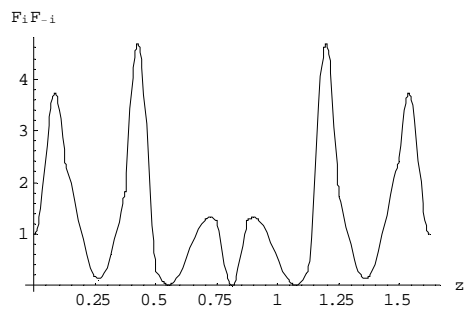
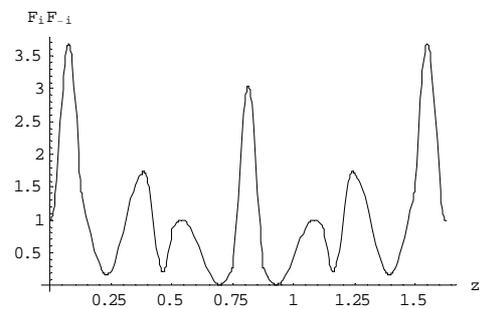

Grafico 5.20. $QNM_6$ TM.           Grafico 5.21. $QNM_7$ TM.

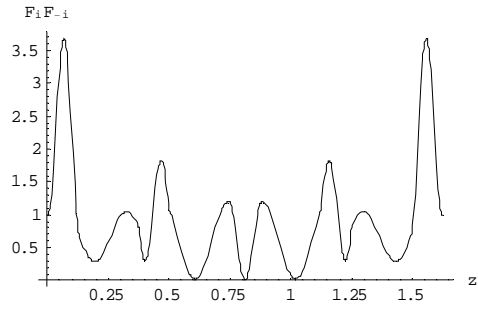 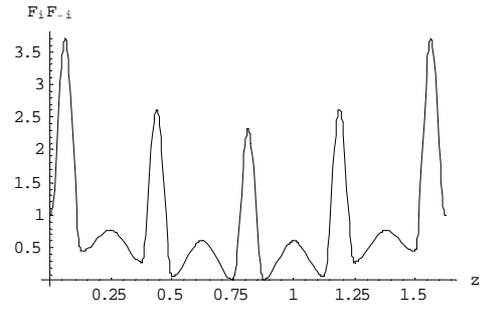

Grafico 5.22. QNM$_8$ TM.  Grafico 5.23. QNM$_9$ TM.

Le autofunzioni dei QNMs TM sono diverse da quelle dei QNMs TE anche per incidenza normale, in quanto le prime sono autofunzioni per l'induzione magnetica mentre le seconde lo sono per il campo elettrico.
Per il resto, valgono commenti analoghi al caso TE.

I grafici seguenti sono per un 1D-PBG a $\lambda/4$, con numero dei periodi $N = 4$, indici di rifrazione $n_h = 2$, $n_l = 1.5$ e lunghezza di riferimento $\lambda_{rif} = 1\mu m$. Si ritiene valida l'approssimazione parassiale finché l'angolo di incidenza rimane $\theta_0 \leq 30$. Si fa riferimento all'eq. (5.3.2.9) con coeff. (5.3.2.10-12).

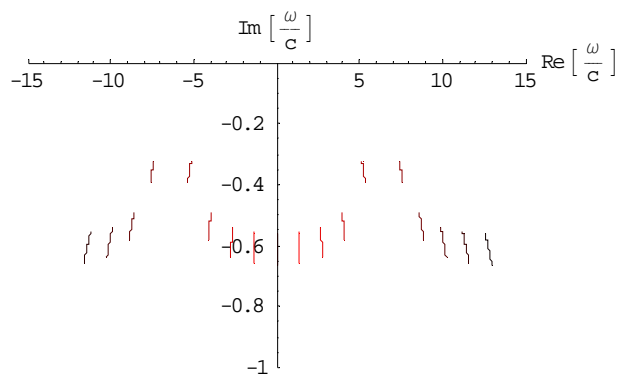

Grafico 5.24. Luogo dei QNMs TM nel piano complesso quando, in ipotesi di approssimazione parassiale, l'angolo di incidenza varia da 0 a 30 gradi.

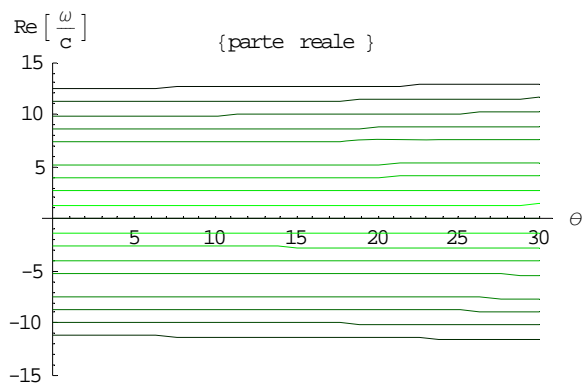

Grafico 5.25. Parte reale di ogni QNM TM in funzione dell'angolo di incidenza (approssimazione parassiale: 0-30 gradi).

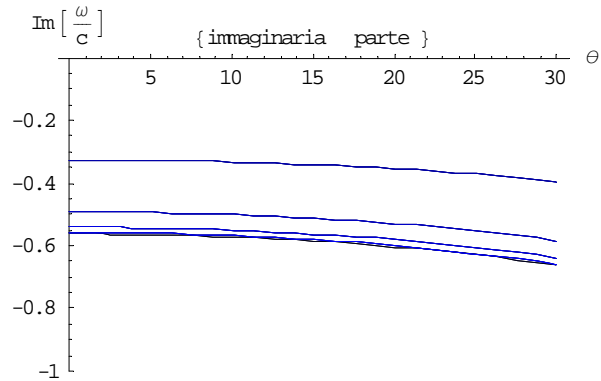

Grafico 5.26. Parte immaginaria di ogni QNM TM in funzione dell'angolo di incidenza (approssimazione parassiale: 0-30 gradi).

Dai grafici 5.11 e 5.24, si nota che i QNMs TM e TE non coincidono; per incidenze crescenti, rispetto all'asse reale, i QNMs TM scendono, mentre i QNMs TE salgono.

Dai grafici 5.12 e 5.25, si nota che le parti reali dei QNMs TE e TM sono pressoché identiche tra loro e costanti al variare dell'angolo; sotto le condizioni poste dalla parassialità sul salto d'indice, le risonanze sono quasi indipendenti dalla polarizzazione e sono determinate prevalentemente dalla geometria della struttura.

Dai grafici 5.13 e 5.26 si nota che, in approssimazione parassiale, rispetto al caso TE, le parti immaginarie dei QNMs TM aumentano in modulo quando l'angolo viene incrementato; quindi, le perdite si riducono per incidenze più piccole. Difatti, se è vero che, applicando il principio di dualità, si ricavano formalmente i modi TM dai modi TE [6], così, almeno in approssimazione parassiale, questa dualità si trasferisce per i modi TE e TM al livello di caratteristiche trasmissive [6]. Infine, si noti l'analogia: come per un'onda polarizzata TM esiste un angolo di Brewster per cui il coefficiente di riflessione si annulla [6], così, per ogni QNM TM, dato che per un'onda radente ($\theta_0 \to \pi/2$) non vi possono essere perdite ($\text{Im}\,\omega_n \to 0$), esiste un angolo intermedio $\theta_n^M$ per cui le perdite sono massime ($|\text{Im}\,\omega_n| \to \max$).

## *Appendice B.*

In questa appendice, si svolge una verifica sulla coerenza interna della teoria QNM's; posta un'eccitazione elementare in un 1D-PBG, la sua evoluzione deve rispettare le condizioni di *"outgoing waves"* e sulle perdite.

*Base di teoria*

Si consideri una radiazione con polarizzazione TE che si propaga normalmente alla cavità aperta $C = [0, L]$.

I QNM's sono le coppie di autofrequenze ed autofunzioni, $[\omega_n, f_n(z)]$ con $\text{Im}\,\omega_n < 0$, che, assieme alle condizioni asintotiche [1], soddisfano l'equazione:

$$\frac{d^2 f_n}{dz^2} + \omega_n^2 \rho(z) f_n(z) = 0 \tag{B.1}$$

dove:

$$\rho(z) = \left(\frac{n(z)}{c}\right)^2 \tag{B.2}$$

con $n(z)$ profilo dell'indice di rifrazione e $c$ velocità della luce.

Si definisce il momento coniugato di $f_n(z)$ come [2]:

$$\hat{f}_n(z) = -i\omega_n \rho(z) f_n(z) \tag{B.3}$$

Si è interessati all'evoluzione della radiazione nella cavità, quando non vi è un pompaggio esterno.

Il campo elettrico abbia una distribuzione iniziale:

$$E_0(z) = E(z, t = 0) \tag{B.4}$$

con momento coniugato [2]:

$$\hat{E}_0(z) = \rho(z) \left(\frac{\partial E}{\partial t}\right)_{t=0} \tag{B.5}$$

In condizioni di completezza dei QNMs [2], il campo elettrico evolve secondo la [2]:

$$E(z, t) = \sum_n a_n f_n^N(z) e^{-i\omega_n t} \tag{B.6}$$

dove i coefficienti sono forniti da [1]:

$$a_n = \frac{i}{2\omega_n} \left\{ \int_0^L \left[ f_n^N(z) \hat{E}_0(z) + \hat{f}_n^N(z) E_0(z) \right] dz + \sqrt{\rho_0} \left[ f_n^N(L) E_0(L) + f_n^N(0) E_0(0) \right] \right\} \tag{B.7}$$

*Eccitazione iniziale*

Si suppone che l'eccitazione iniziale sia di natura impulsiva e che la cavità sia un 1D-PBG simmetrico a quarto d'onda; in particolare, l'impulso sia confinato nello strato centrale, con spessore $h$ ed indice di rifrazione $n_h$.

L'impulso nello strato centrale inizia a diffondere, attenuandosi e propagandosi a destra e a sinistra.

Per $t \to 0$, si propone il modello:

$$E(z,t) = \frac{1}{2}\left[E_+(z,t) + E_-(z,t)\right] \tag{B.8}$$

dove:

$$E_\pm(z,t) = A \frac{h}{\sqrt{2\pi}\sigma(t)} \exp\left[-\frac{(z - L/2 \pm v \cdot t)^2}{2\sigma^2(t)}\right] \tag{B.9}$$

con:

$$\sigma(t) = \sigma_0 + v \cdot t \tag{B.10}$$

$$2\sigma_0 \ll h \tag{B.11}$$

$$v = \frac{c}{n_h} \tag{B.12}$$

Dal modello descritto, si possono dedurre, nell'istante iniziale, l'eccitazione:

$$E_\pm(z, t=0) = A \frac{h}{\sqrt{2\pi}\sigma_0} \exp\left[-\frac{(z - L/2)^2}{2\sigma_0^2}\right] \tag{B.13}$$

$$E_0(z) = E(z, t=0) = \frac{1}{2}\left[E_+(z,t=0) + E_-(z,t=0)\right] = E_\pm(z, t=0) \tag{B.14}$$

ed il suo momento coniugato:

$$\left.\frac{\partial_t E_\pm}{\partial t}\right|_{t=0} = E_0(z) \frac{v}{\sigma_0}\left[-1 + \frac{z - L/2}{\sigma_0}\left(\pm 1 + \frac{z - L/2}{\sigma_0}\right)\right] \tag{B.15}$$

$$\hat{E}_0(z) = \rho(z) \left.\frac{\partial_t E}{\partial t}\right|_{t=0} = \frac{1}{2}\rho(z)\left[\left.\frac{\partial_t E_+}{\partial t}\right|_{t=0} + \left.\frac{\partial_t E_-}{\partial t}\right|_{t=0}\right] =$$
$$= E_0(z)\rho(z)\frac{v}{\sigma_0}\left[-1 + \left(\frac{z - L/2}{\sigma_0}\right)^2\right] \tag{B.16}$$

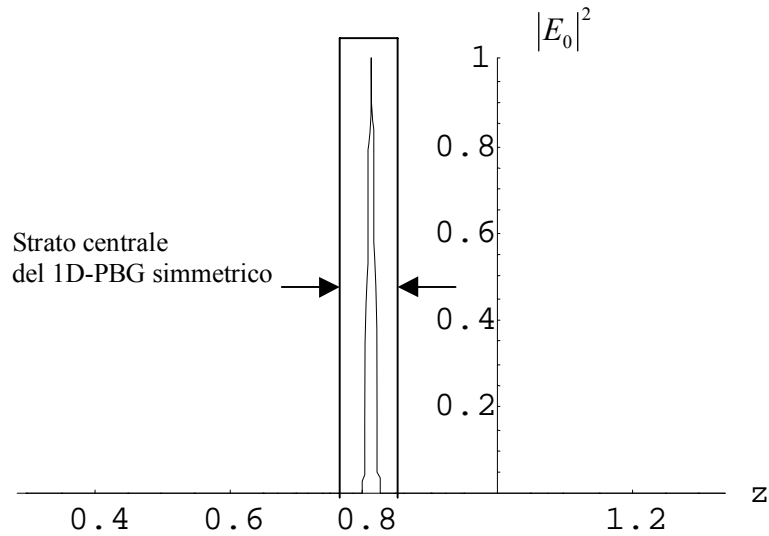

Grafico 5.27. Impulso di eccitazione nello strato centrale di un 1D-PBG simmetrico a $\lambda/4$. L'impulso ha ampiezza unitaria e larghezza $2\sigma_0 = h/n$, con $n=10$. L'1D-PBG ha parametri: $N=4$, $n_h=2$, $n_l=1$ e $\lambda_{rif}=1\mu m$. Il riferimento è alle equazioni (B.13-14).

## *Evoluzione nel tempo*

Quando la radiazione supera i confini posti dallo strato centrale, il campo nel 1D-PBG viene determinato come sovrapposizione dei QNMs, tramite le (B.6) e (B.7), e le condizioni sull'eccitazione iniziale sono quelle dedotte dal modello sopra proposto, vale a dire le (B.13-14) e (B.16).

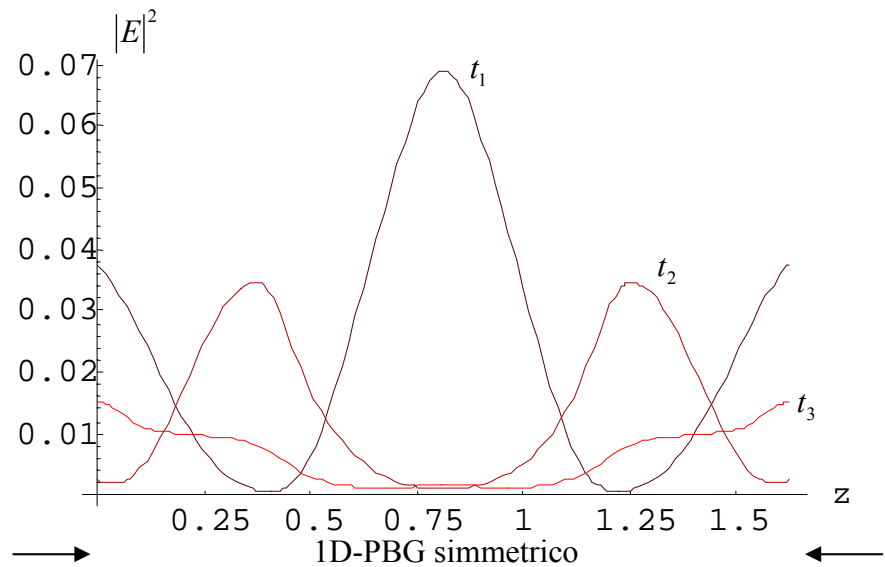

Grafico 5.28. Evoluzione naturale dell'impulso TE all'interno dell'1D-PBG simmetrico a λ/4, campionata in tre istanti successivi $t_1 < t_2 < t_3$. I rif. sono alle eq. (B.6-7) e (B.13-14), (B.16). I parametri sono quelli del grafico 5.27.

Si osserva come l'impulso si suddivide in altri due impulsi che si propagano a destra e a sinistra, come previsto dal modello sull'eccitazione iniziale e compatibilmente con le condizioni di *"outgoing waves"*, implicite nelle (B.6) e (B.7); inoltre gli impulsi si attenuano fortemente, poiché il 1D-PBG è una cavità aperta che introduce perdite rilevanti ($|\text{Im}\,\omega_n| > \text{Re}\,\omega_n$).

## *Bibliografia*

# *Capitolo 6. QNMs TE e TM per i 1D-PBG.*
# *Densità dei modi.*

## *Paragrafo 6.1. Introduzione.*

In questo capitolo, si continua a studiare il problema elettromagnetico di un'onda che si propaga obliquamente in un PBG unidimensionale, tramite il modello teorico dei QNMs [1]; nello specifico, si discute il comportamento della densità dei modi [2].

Nel lavoro di tesi [3], è stata riportata l'estensione della teoria QNM's alle strutture 1D-PBG per incidenza normale. In quello stesso lavoro, è stata affrontata, in modo euristico, la discussione circa la densità dei modi nell'1D-PBG, proponendone una rappresentazione come somma di opportune lorenziane; l'ipotesi è stata confermata con simulazioni ma non accertata formalmente.

Lo scopo di questo capitolo è di discutere quanto proposto nel lavoro [3] e di formalizzarlo in situazioni specifiche sia per la configurazione geometrica del 1D-PBG, che la polarizzazione del campo e.m. nella cavità. La densità dei modi descrive le proprietà del campo in interazione con un mezzo materiale; questa assume forme e comportamenti differenti a seconda del mezzo materiale e della radiazione che vi si propaga.

Nel seguito si discuterà sulla densità dei modi secondo il formalismo dei QNMs, per differenti condizioni di eccitazione nell'1D-PBG. Nel caso di pompaggio esterno assente, si propone un modello di calcolo per la densità dei modi dentro una cavità aperta, applicandolo ad un 1D-PBG simmetrico a quarto d'onda. Quando vi è una pompa esterna che incide obliquamente sulla cavità, per entrambe le polarizzazioni TE e TM, si utilizza un formalismo (a due componenti) che prende spunto dall'articolo [4], determinando la funzione di autocorrelazione per il campo e.m. dentro la cavità. Da qui, viene dimostrato che, per una struttura simmetrica, la densità dei modi è una somma di lorenziane, ma solo sotto l'ipotesi di risonanze strette e perdite minime. Infine, con riferimento ad un 1D-PBG simmetrico a quarto d'onda, vengono confrontati i risultati della DOM senza pompa e con pompa esterna a incidenza normale e, nell'ipotesi di parassialità [5] in cui l'equazione per le autofrequenze QNM's è stata risolta [capitolo 5], si confrontano la DOM ottenuta secondo il modello QNM's con quella ottenuta tramite la matrice di trasmissione [6].

## *Paragrafo 6.2. DOM in assenza di pompaggio.*

In questo paragrafo, si inizia lo studio della DOM in assenza di pompa esterna. Questa descrive la distribuzione spettrale dei modi propagativi del campo elettromagnetico che è determinata dalla sola legge di dispersione del mezzo. Il modello non é rigoroso, ma si pongono le basi di ulteriori sviluppi.

### *6.2.1. Premesse.*

Si consideri una cavità aperta $C = [0, L]$; non vi sia alcun pompaggio esterno.

Il campo elettrico, comunque presente nella cavità, sia un processo stocastico [7], insieme di tutte le radiazioni che possono aver luogo nella struttura.

In condizioni di completezza, il campo elettrico può essere sviluppato in termini dei QNMs all'interno della cavità [1]:

$$E(z,t) = \sum_n a_n(t) f_n^N(z) \quad , \quad \forall z \in C \tag{6.2.1.1}$$

Ciascun coefficiente dello sviluppo ha carattere aleatorio e le sue medie statistiche informano sulla distribuzione energetica del QNM corrispondente.

In assenza di pompaggio [1]:

$$a_n(t) = a_n e^{-i\omega_n t} \tag{6.2.1.2}$$

Si suppone che le variabili aleatorie $a_n$ siano indipendenti [7] ed a valor medio nullo, per cui:

$$\langle a_n a_m^* \rangle = \delta_{n,m} \langle |a_n|^2 \rangle \tag{6.2.1.3}$$

dove si è indicato il simbolo di Kronecker con $\delta_{n,m}$.

Il campo è costituito da fotoni. Il valor medio dell'energia per ciascun fotone segue la distribuzione di Bose [8]:

$$\left\langle \mathcal{E}_{fotone} \right\rangle = \frac{\hbar\omega}{2} + \frac{\hbar\omega}{e^{\beta\hbar\omega}-1} \qquad (6.2.1.4)$$

dove $\beta = 1/k_B T$, con $k_B$ costante di Boltzman, e $\hbar = h/2\pi$, con $h$ costante di Planck.

Se $\beta\hbar\omega \ll 1$, allora:

$$\left\langle \mathcal{E}_{fotone} \right\rangle \cong \frac{\hbar\omega}{2}\frac{1}{1-e^{-\beta\hbar\omega}} \qquad (6.2.1.5)$$

Inoltre, ogni QNM sia costituito in media da un solo fotone, per cui il peso energetico di ogni $\omega_n$ segue la distribuzione di Bose:

$$\left\langle |a_n|^2 \right\rangle = K \left\langle \mathcal{E}_{fotone} \right\rangle \qquad (6.2.1.6)$$

dove la costante $K$ restituisce le dimensioni fisiche corrette all'espressione.

Ora si vuole determinare la funzione di autocorrelazione per il campo e.m. dentro la cavità, che si renderà necessaria per ottenere la distribuzione spazio-frequenziale dei modi e quindi la densità spettrale dei modi.

### 6.2.2. Funzione di autocorrelazione.

Passando al dominio di Fourier, lo sviluppo del campo elettrico nei QNMs all'interno della cavità si esprime come:

$$\tilde{E}(z,\omega) = \sum_n \tilde{a}_n(\omega) f_n^N(z) \qquad (6.2.2.1)$$

dove, tenendo conto del legame fra le trasformate di Laplace e di Fourier [9], i coefficienti dello sviluppo sono:

$$\tilde{a}_n(\omega) = \tilde{a}_n(s = i\omega) = \frac{a_n}{i(\omega - \omega_n)} \qquad (6.2.2.2)$$

Da qui, si deduce che la correlazione [7] dei coefficienti $\tilde{a}_n(\omega)$ è la:

$$\left\langle \tilde{a}_n(\omega)\tilde{a}_m^*(\omega) \right\rangle = \frac{\left\langle a_n a_m^* \right\rangle}{(\omega - \omega_n)(\omega - \omega_m^*)} \qquad (6.2.2.3)$$

L'autocorrelazione per il campo elettrico è definita come [5]:

$$F(z, z', \omega) = \left\langle \tilde{E}(z,\omega)\tilde{E}^*(z',\omega) \right\rangle \qquad (6.2.2.4)$$

Inserendo le (6.2.2.1) e (6.2.2.3) nella (6.2.2.4), si ottiene:

$$F(z,z',\omega) = \sum_{n,m} \frac{\langle a_n a_m^* \rangle}{(\omega-\omega_n)(\omega-\omega_m^*)} f_n^N(z) \left[ f_n^N(z') \right]^* \qquad (6.2.2.5)$$

Tenendo conto delle (6.2.1.3), (6.2.1.6) e (6.2.1.5), si deduce:

$$F(z,z',\omega) = K \frac{\hbar\omega/2}{1-e^{-\beta\hbar\omega}} \sum_n \frac{f_n^N(z)\left[f_n^N(z')\right]^*}{(\omega-\omega_n)(\omega-\omega_m^*)} \qquad (6.2.2.6)$$

### 6.2.3. Distribuzione spazio-frequenziale dei modi.

In assenza di pompaggio, se la distribuzione spaziale dei QNMs è indicata con $N_{QNMs}(z,\omega)$, si definisce la distribuzione spazio-frequenziale dei "modi" come:

$$d(z,\omega) = k_{rif} \frac{\partial N_{QNMs}(z,\omega)}{\partial \omega} \qquad (6.2.3.1)$$

dove si è introdotto un numero d'onda di riferimento $k_{rif}$.

Poiché ogni QNM è costituito in media da un solo fotone, si può riscrivere la (6.2.3.1) come:

$$d(z,\omega) = k_{rif} \frac{\partial N_{fotoni}(z,\omega)}{\partial \omega} \qquad (6.2.3.2)$$

Si osservi che l'energia infinitesima del campo ha espressioni equivalenti:

$$\delta \mathcal{E}_{campo} \propto \left\langle \left| \tilde{E}(\omega) \right|^2 \right\rangle \delta\omega = \left\langle \mathcal{E}_{fotone} \right\rangle \delta N_{fotoni} \qquad (6.2.3.3)$$

In accordo, la distribuzione spazio-frequenziale (6.2.3.2) può essere calcolata come:

$$d(z,\omega) = K' k_{rif} \frac{\left\langle \left| \tilde{E}(z,\omega) \right|^2 \right\rangle}{\left\langle \mathcal{E}_{fotone} \right\rangle} = K' k_{rif} \frac{F(z,z,\omega)}{\left\langle \mathcal{E}_{fotone} \right\rangle} \qquad (6.2.3.4)$$

dove la costante $K'$ restituisce le dimensioni fisiche corrette all'espressione.

Si inseriscono le (6.2.1.5) e (6.2.2.6) nella (6.2.3.4) e si ottiene:

$$d(z,\omega) = K'' k_{rif} \sum_n \frac{\left| f_n^N(z) \right|^2}{(\omega-\omega_n)(\omega-\omega_n^*)} \qquad (6.2.3.5)$$

con $K'' = K'K$.

### *6.2.4. Densità spettrale dei modi.*

In accordo con quanto proposto nel lavoro [3], la densità spettrale dei modi si calcola come:

$$d(\omega) = \int_0^L d(z,\omega)\rho(z)dz \qquad (6.2.4.1)$$

Si inserisce la (6.2.3.5) nella (6.2.4.1) e si ottiene:

$$d(\omega) = K''k_{rif} \sum_n \frac{I_n}{(\omega-\omega_n)(\omega-\omega_n^*)} \qquad (6.2.4.2)$$

dove:

$$I_n = \int_0^L \left|f_n^N(z)\right|^2 \rho(z)dz \qquad (6.2.4.3)$$

*Perdite minime*

Quando le perdite sono minime $|Im\,\omega_n| \ll Re\,\omega_n$, allora è lecita l'approssimazione $I_n \cong 1$ e la formula operativa per la densità spettrale dei modi (6.2.4.2) si semplifica:

$$d(\omega) \cong K'' \sum_n \frac{k_{rif}}{(\omega-\omega_n)(\omega-\omega_n^*)} \qquad (6.2.4.4)$$

Poiché $\omega_n^* = -\omega_{-n}$, la densità spettrale dei modi (6.2.4.4) diviene:

$$d(\omega) = K'' \sum_{n \geq 0} \frac{k_{rif}}{(\omega-\omega_n)(\omega+\omega_{-n})} \qquad (6.2.4.5)$$

*Rappresentazione della DOM per un 1D-PBG simmetrico a λ/4*

Tenendo conto che $k_{rif} = \omega_{rif}/c$ e posto $K'' = \omega_{rif}\overline{K}$, la densità dei modi adimensionata $d_a = cd$, con argomento $\overline{\omega} = \omega/\omega_{rif}$, ha espressione:

$$d_a(\overline{\omega}) = \overline{K} \sum_{n \geq 0} \frac{1}{(\overline{\omega}-\overline{\omega}_n)(\overline{\omega}+\overline{\omega}_n)} \qquad (6.2.4.6)$$

dove la costante $\overline{K}$ è tale che la DOM tende ad uno nelle bande passanti.

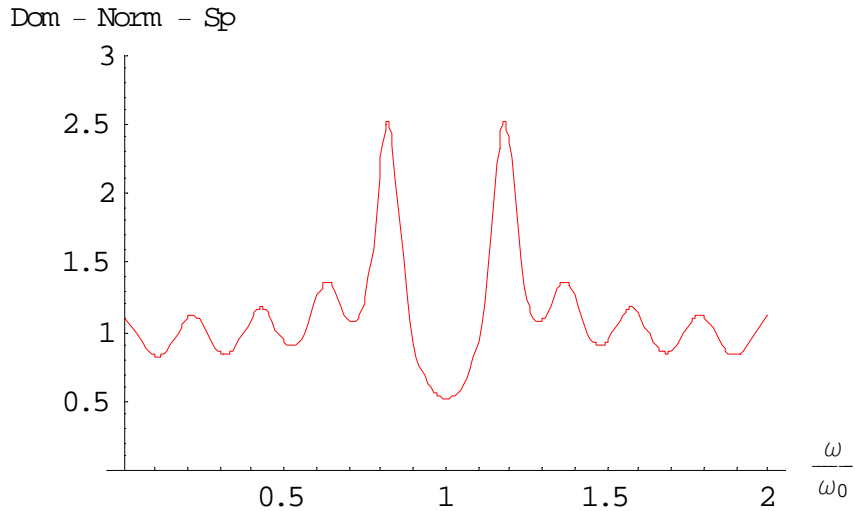

Grafico 6.1. Densità dei modi, in assenza di pompaggio, per un 1D-PBG simmetrico a *λ/4*, con parametri: $N=4$, $n_h=1.5$, $n_l=1$ e $\lambda_{rif}=1\mu m$. Si fa riferimento all'equazione (6.2.4.6).

## *Paragrafo 6.3. DOM TE e TM per una pompa.*

In questo paragrafo, sia per la polarizzazione TE che TM, si determina rigorosamente la DOM, quando una pompa incide obliquamente e da sinistra su una cavità aperta, quale un 1D-PBG simmetrico a quarto d'onda.

### *6.3.1. Polarizzazione TE.*

La pompa esterna, che incide obliquamente sulla struttura, abbia polarizzazione di tipo TE.

### *6.3.1.a. Formalismo a due componenti.*

In precedenza (capitolo 4), asserita l'ipotesi di parassialità [5], si è applicata la legge generalizzata di Snell [5], per cui, introdotto il profilo per l'angolo di rifrazione $\theta(z)$, si è definita la funzione:

$$\rho(z) = \left[\frac{n(z)\cos\theta(z)}{c}\right]^2 \qquad (6.3.1.1)$$

tale che il campo elettrico soddisfa l'equazione:

$$\left[\partial_z^2 - \rho(z)\partial_t^2\right]E(z,t) = 0 \qquad (6.3.1.2)$$

mentre per il suo momento coniugato:

$$\hat{E}(z,t) = \rho(z)\partial_t E(z,t) \tag{6.3.1.3}$$

Anche in presenza di una pompa esterna, qualora sussista l'ipotesi si completezza [1], il campo elettrico può essere rappresentato in termini dei QNMs all'interno della struttura [4]:

$$E(z,t) = \sum_n a_n(t) f_n^N(z) \quad \forall z \in C = [0, L] \tag{6.3.1.4}$$

dove [10]:

$$\hat{f}_n^N(z) = -i\omega_n \rho(z) f_n^N(z) \tag{6.3.1.5}$$

I coefficienti dello sviluppo sono forniti da [4]:

$$a_n(t) = \frac{1}{2\omega_n} \left\langle f_n^N, E(t) \right\rangle \tag{6.3.1.6}$$

dove, tenendo conto della definizione di prodotto interno per i QNMs TE, si ottiene:

$$\begin{aligned} a_n(t) = &\frac{i}{2\omega_n} \int_0^L \left[ f_n^N(z)\hat{E}(z,t) + \hat{f}_n^N(z)E(z,t) \right] dz + \\ &+ \frac{i}{2\omega_n} \sqrt{\rho_0} \left[ f_n^N(L)E(L,t) + f_n^N(0)E(0,t) \right] \end{aligned} \tag{6.3.1.7}$$

*Condizioni ai bordi per il campo*

Si suppone che:

$$\rho(z) = \rho_0 \quad \text{per} \quad z \leq 0 \text{ e } z \geq L \tag{6.3.1.8}$$

Si considera il semispazio $z \leq 0$; si osservano un campo di pompa ed un campo riflesso:

$$E(z,t) = E_P(z,t) + E_R(z,t) \tag{6.3.1.9}$$

$$\hat{E}(z,t) = \hat{E}_P(z,t) + \hat{E}_R(z,t) \tag{6.3.1.10}$$

che soddisfano rispettivamente le condizioni di *"incomig e outgoing waves"*:

$$\partial_z E_P = -\sqrt{\rho_0}\partial_t E_P \tag{6.3.1.11}$$

$$\partial_z E_R = \sqrt{\rho_0}\partial_t E_R \tag{6.3.1.12}$$

Si prende l'equazione (6.3.1.10), esplicitando i momenti coniugati del campo di pompa e riflesso secondo la definizione (6.3.1.3) e tenendo conto della (6.3.1.8), si inseriscono le condizioni di *"outgoing waves"* (6.3.1.11-12) e si tiene conto della (6.3.1.9); si ottiene:

$$\hat{E}(z,t) = \sqrt{\rho_0}\partial_z E(z,t) - 2\sqrt{\rho_0}\partial_z E_p(z,t) \qquad (6.3.1.13)$$

Definito il termine di pompa come:

$$b(t) = -2\sqrt{\rho_0}\partial_z E_p(z,t)\big|_{z=0} \qquad (6.3.1.14)$$

l'equazione (6.3.1.13), calcolata per $z = 0$, dà luogo alla prima condizione ai bordi:

$$\hat{E}(0,t) = \sqrt{\rho_0}\partial_z E(z,t)\big|_{z=0} + b(t) \qquad (6.3.1.15)$$

che, altrimenti, si può scrivere come:

$$\partial_t E(0,t) = \frac{1}{\sqrt{\rho_0}}\partial_z E(z,t)\big|_{z=0} + \frac{1}{\rho_0}b(t) \qquad (6.3.1.16)$$

Si considera il semispazio $z \geq L$; si osserva un campo trasmesso:

$$E(z,t) = E_T(z,t) \qquad (6.3.1.17)$$

che soddisfa la condizione di "*outgoing wave*":

$$\partial_z E_T = -\sqrt{\rho_0}\partial_t E_T \qquad (6.3.1.18)$$

Si prende l'equazione (6.3.1.18), si inserisce la (6.3.1.17) e la si calcola per $z = L$; si ottiene la seconda condizione ai bordi:

$$\partial_t E(L,t) = -\frac{1}{\sqrt{\rho_0}}\partial_z E(z,t)\big|_{z=L} \qquad (6.3.1.19)$$

*Equazione dinamica per i coefficienti dello sviluppo*

Si prende la (6.3.1.7), si inseriscono le (6.3.1.3) e (6.3.1.5); si ottiene la:

$$\begin{aligned}a_n(t) = &\frac{i}{2\omega_n}\int_0^L f_n^N(z)\rho(z)\dot{E}(z,t)dz + \frac{1}{2}\int_0^L f_n^N(z)\rho(z)E(z,t)dz + \\ &+ \frac{i}{2\omega_n}\sqrt{\rho_0}\left[f_n^N(L)E(L,t) + f_n^N(0)E(0,t)\right]\end{aligned} \qquad (6.3.1.20)$$

dove si è indicato l'operatore di derivata rispetto al tempo con il punto.

Si deriva rispetto al tempo la (6.3.1.20) e si ottiene la:

$$\dot{a}_n(t) = \frac{i}{2\omega_n}\int_0^L f_n^N(z)\rho(z)\ddot{E}(z,t)dz + \frac{1}{2}\int_0^L f_n^N(z)\rho(z)\dot{E}(z,t)dz +$$
$$+ \frac{i}{2\omega_n}\sqrt{\rho_0}\left[f_n^N(L)\dot{E}(L,t) + f_n^N(0)\dot{E}(0,t)\right] \quad (6.3.1.21)$$

Si inseriscono nella (6.3.1.21) le (6.3.1.2), (6.3.1.16) e (6.3.1.19); si ottiene:

$$\dot{a}_n(t) = \frac{i}{2\omega_n}\int_0^L f_n^N(z)\partial_z^2 E(z,t)dz + \frac{1}{2}\int_0^L f_n^N(z)\rho(z)\dot{E}(z,t)dz +$$
$$- \frac{i}{2\omega_n}\left[f_n^N(L)\partial_z E(z,t)\big|_{z=L} - f_n^N(0)\partial_z E(z,t)\big|_{z=0}\right] + \frac{i}{2\omega_n\sqrt{\rho_0}}f_n^N(0)b(t)$$

$$(6.3.1.22)$$

Si integra per parti il primo integrale della (6.3.1.22), ottenendo:

$$\int_0^L f_n^N(z)\partial_z^2 E(z,t)dz = \left[f_n^N(L)\partial_z E(z,t)\big|_{z=L} - f_n^N(0)\partial_z E(z,t)\big|_{z=0}\right] - \int_0^L \frac{df_n^N}{dz}\partial_z E(z,t)dz$$

$$(6.3.1.23)$$

per cui la (6.3.1.22) diviene:

$$\dot{a}_n(t) = -\frac{i}{2\omega_n}\int_0^L \frac{df_n^N}{dz}\partial_z E(z,t)dz + \frac{1}{2}\int_0^L f_n^N(z)\rho(z)\dot{E}(z,t)dz + \frac{i}{2\omega_n\sqrt{\rho_0}}f_n^N(0)b(t)$$

$$(6.3.1.24)$$

Si integra per parti il primo integrale della (6.3.1.24), che diviene:

$$\dot{a}_n(t) = -\frac{i}{2\omega_n}\left[\frac{df_n^N}{dz}\bigg|_{z=L}E(L,t) - \frac{df_n^N}{dz}\bigg|_{z=0}E(0,t)\right] + \frac{i}{2\omega_n}\int_0^L \frac{d^2 f_n^N}{dz^2}E(z,t)dz +$$
$$+ \frac{1}{2}\int_0^L f_n^N(z)\rho(z)\dot{E}(z,t)dz + \frac{i}{2\omega_n\sqrt{\rho_0}}f_n^N(0)b(t)$$

$$(6.3.1.25)$$

Si inseriscono l'equazione agli autovalori per i QNMs:

$$\frac{d^2 f_n^N}{dz^2} + \omega_n^2 \rho(z) f_n^N(z) = 0 \quad (6.3.1.26)$$

e le condizioni formali di *"outgoing waves"* per le autofunzioni dei QNMs:

$$\begin{cases} \dfrac{df_n^N}{dz}\bigg|_{z=0} = -i\omega_n\sqrt{\rho_0}\,f_n^N(0) \\ \dfrac{df_n^N}{dz}\bigg|_{z=L} = i\omega_n\sqrt{\rho_0}\,f_n^N(L) \end{cases} \quad (6.3.1.27)$$

per cui la (6.3.1.25) diviene:

$$\dot{a}_n(t) = \frac{1}{2}\int_0^L f_n^N(z)\rho(z)\dot{E}(z,t)dz - \frac{i\omega_n}{2}\int_0^L f_n^N(z)\rho(z)E(z,t)dz + \\ + \frac{1}{2}\sqrt{\rho_0}\left[f_n^N(L)E(L,t) + f_n^N(0)E(0,t)\right] + \frac{i}{2\omega_n\sqrt{\rho_0}}f_n^N(0)b(t)$$

(6.3.1.28)

Confrontando le (6.3.1.20) e (6.3.1.28), si deduce l'equazione dinamica:

$$\dot{a}_n(t) + i\omega_n a_n(t) = \frac{i}{2\omega_n\sqrt{\rho_0}}f_n^N(0)b(t) \qquad (6.3.1.29)$$

A regime, i coefficienti dello sviluppo sono:

$$a_n(t) = \frac{i}{2\omega_n\sqrt{\rho_0}}f_n^N(0)\int_{-\infty}^t b(\tau)e^{i\omega_n(\tau-t)}d\tau \qquad (6.3.1.30)$$

### *6.3.1.b. Funzione di autocorrelazione.*

In primo luogo, si osserva che qualsiasi campo elettrico $E(z,t)$ è una grandezza fisica nel campo reale, quindi con trasformata di Fourier $\tilde{E}(z,\omega)$ tale che [11]:

$$\left[\tilde{E}(z,\omega)\right]^* = \tilde{E}(z,-\omega) \qquad (6.3.1.31)$$

*Per il campo di pompa*

La funzione di autocorrelazione per il campo di pompa è definita come [5]:

$$F_P(z,z',\omega) = \left\langle \tilde{E}_P(z,\omega)\tilde{E}_P(z',-\omega)\right\rangle \qquad (6.3.1.32)$$

Il campo di pompa soddisfa la condizione di *"incoming wave"*:

$$\partial_z E_p(z,t)\big|_{z=0} = -\sqrt{\rho_0}\partial_t E_p(0,t) \qquad (6.3.1.33)$$

per cui il termine di pompa (6.3.1.14) può essere riscritto come:

$$b(t) = 2\rho_0\partial_t E_p(0,t) \qquad (6.3.1.34)$$

che, trasformato secondo Fourier, diviene:

$$\tilde{b}(\omega) = 2\rho_0[-i\omega\tilde{E}_P(0,\omega)] \qquad (6.3.1.35)$$

La varianza [7] per il termine di pompa è definita come:

$$\left\langle \tilde{b}(\omega)\tilde{b}(-\omega)\right\rangle = 4\rho_0^2\omega^2\left\langle \tilde{E}_P(0,\omega)\tilde{E}_P(0,-\omega)\right\rangle \qquad (6.3.1.36)$$

che, tenendo conto della (6.3.1.32), si traduce nella:

$$\left\langle \tilde{b}(\omega)\tilde{b}(-\omega) \right\rangle = 4\rho_0^2 \omega^2 F_P(z=z'=0,\omega) \qquad (6.3.1.37)$$

Si propone il modello per l'autocorrelazione di pompa [4]:

$$F_P(z,z',\omega) = K \frac{\exp\left[i\frac{\omega}{c}(z-z')\cos\theta_0\right]}{\omega[1-\exp(-\beta\hbar\omega)]} \qquad (6.3.1.38)$$

dove sono stati introdotti una costante opportuna $K$ e l'angolo di incidenza $\theta_0$.

Dalla (6.3.1.37), lo stesso modello prevede per la varianza di pompa:

$$\left\langle \tilde{b}(\omega)\tilde{b}(-\omega) \right\rangle = K\rho_0^2 \frac{4\omega}{1-\exp(-\beta\hbar\omega)} \qquad (6.3.1.39)$$

*Per il campo dentro la struttura*

La funzione di autocorrelazione per il campo dentro la struttura è definita come:

$$F(z,z',\omega) = \left\langle \tilde{E}(z,\omega)\tilde{E}(z',-\omega) \right\rangle \qquad (6.3.1.40)$$

Si trasforma nel dominio di Fourier la (6.3.1.4):

$$\tilde{E}(z,\omega) = \sum_n \tilde{a}_n(\omega) f_n^N(z) \qquad (6.3.1.41)$$

e si inserisce la (6.3.1.41) nella (6.3.1.40):

$$F(z,z',\omega) = \sum_{n,m} \left\langle \tilde{a}_n(\omega)\tilde{a}_m(-\omega) \right\rangle f_n^N(z) f_m^N(z') \qquad (6.3.1.42)$$

Si trasforma nel dominio di Fourier la (6.3.1.30):

$$\tilde{a}_n(\omega) = \frac{f_n^N(0)}{2\omega_n\sqrt{\rho_0}} \frac{\tilde{b}(\omega)}{\omega-\omega_n} \qquad (6.3.1.43)$$

se ne determina la correlazione:

$$\left\langle \tilde{a}_n(\omega)\tilde{a}_m(-\omega) \right\rangle = \frac{f_n^N(0)f_m^N(0)}{4\rho_0\omega_n\omega_m(\omega_n-\omega)(\omega_m+\omega)} \left\langle \tilde{b}(\omega)\tilde{b}(-\omega) \right\rangle \qquad (6.3.1.44)$$

e si introduce la (6.3.1.44) nella (6.3.1.42):

$$F(z,z',\omega) = \left\langle \tilde{b}(\omega)\tilde{b}(-\omega) \right\rangle \sum_{n,m} \frac{f_n^N(0)f_m^N(0)}{4\rho_0\omega_n\omega_m(\omega_n-\omega)(\omega_m+\omega)} f_n^N(z) f_m^N(z') \qquad (6.3.1.45)$$

Infine, inserendo la (6.3.1.39) nella (6.3.1.45), si ottiene la formula definitiva per l'autocorrelazione del campo dentro la struttura:

$$F(z,z',\omega) = K\rho_0 \frac{\omega}{1-\exp(-\beta\hbar\omega)} \sum_{n,m} \frac{f_n^N(0)f_m^N(0)}{\omega_n\omega_m(\omega_n-\omega)(\omega_m+\omega)} f_n^N(z)f_m^N(z')$$

(6.3.1.46)

### 6.3.1.c. Densità dei modi.

In accordo con il lavoro [4], la densità locale degli stati $d(z,\omega)$ si ottiene dalla funzione di correlazione $F(z,z',\omega)$, per frequenze $\omega$ reali e positive:

$$d(z,\omega) = \frac{\omega}{\pi}[1-\exp(-\beta\hbar\omega)]F(z,z,\omega) \tag{6.3.1.47}$$

Si inserisce la (6.3.1.46) nella (6.3.1.47) e si ottiene:

$$d(z,\omega) = K\rho_0 \frac{\omega^2}{\pi} \sum_{n,m} \frac{f_n^N(0)f_m^N(0)}{\omega_n\omega_m(\omega_n-\omega)(\omega_m+\omega)} f_n^N(z)f_m^N(z) \tag{6.3.1.48}$$

Applicando la relazione di completezza debole per i QNMs [1]:

$$\sum_n \frac{\left[f_n^N(z)\right]^2}{\omega_n} = 0 \tag{6.3.1.49}$$

la (6.3.1.48) si semplifica come:

$$d(z,\omega) = K\frac{\rho_0}{\pi} \sum_{n,m} \frac{f_n^N(0)f_m^N(0)}{(\omega-\omega_n)(\omega+\omega_m)} f_n^N(z)f_m^N(z) \tag{6.3.1.50}$$

*Approssimazione di singola risonanza*

Nell'approssimazione di singola risonanza $m=-n$, tenendo conto che $(\omega_n^*, f_n^*) = (-\omega_{-n}, f_{-n})$, la densità locale dei modi (6.3.1.50) diviene:

$$d_n(z,\omega) = K\frac{\rho_0}{\pi} \frac{\left|f_n^N(0)f_n^N(z)\right|^2}{(\omega-\text{Re}\,\omega_n)^2 + \text{Im}^2\,\omega_n} \tag{6.3.1.51}$$

La densità dei modi per singola risonanza si calcola come:

$$d_n(\omega) = \int_0^L d_n(z,\omega)\rho(z)dz \tag{6.3.1.52}$$

Si inserisce la (6.3.1.51) nella (6.3.1.52) e si ottiene la lorenziana:

$$d_n(\omega) = K\frac{\rho_0}{\pi} I_n \frac{\left|f_n^N(0)\right|^2}{(\omega-\text{Re}\,\omega_n)^2 + \text{Im}^2\,\omega_n} \tag{6.3.1.53}$$

dove:

$$I_n = \int_0^L \left|f_n^N(z)\right|^2 \rho(z)dz \qquad (6.3.1.54)$$

*Somma di lorenziane*

Per risonanze sufficientemente strette, i QNMs non interferiscono quasi tra loro, la densità dei modi corrispondente ad ogni QNM non subisce troppo il fenomeno dell'aliasing [12] dovuto a quelle adiacenti, quindi si può approssimare la densità dei modi globale come semplice risultante:

$$d(\omega) = \sum_n d_n(\omega) = 2\sum_{n>0} d_n(\omega) \qquad (6.3.1.55)$$

Si inserisce la (6.3.1.53) nella (6.3.1.55) e si ottiene:

$$d(\omega) = 2K\frac{\rho_0}{\pi}\sum_{n>0} I_n \frac{\left|f_n^N(0)\right|^2}{(\omega - \operatorname{Re}\omega_n)^2 + \operatorname{Im}^2 \omega_n} \qquad (6.3.1.56)$$

In precedenza, si è dimostrato che (cap.4):

$$I_n = \frac{\sqrt{\rho_0}}{2|\operatorname{Im}\omega_n|}\left[\left|f_n^N(L)\right|^2 + \left|f_n^N(0)\right|^2\right] \qquad (6.3.1.57)$$

Si consideri una struttura simmetrica:

$$f_n^N(L) = (-1)^n f_n^N(0) \qquad (6.3.1.58)$$

per cui la (6.3.1.57) si riduce:

$$\left|f_n^N(0)\right|^2 = \frac{1}{\sqrt{\rho_0}} I_n |\operatorname{Im}\omega_n| \qquad (6.3.1.59)$$

e la densità dei modi (6.3.1.56) diviene:

$$d(\omega) = 2K\frac{\sqrt{\rho_0}}{\pi}\sum_{n>0} I_n^2 \frac{|\operatorname{Im}\omega_n|}{(\omega - \operatorname{Re}\omega_n)^2 + \operatorname{Im}^2 \omega_n} \qquad (6.3.1.60)$$

Ora, si inserisce nella definizione (6.3.1.54) la normalizzazione:

$$f_n^N(z) = f_n(z)\sqrt{\frac{2\omega_n}{\langle f_n, f_n\rangle}} \qquad (6.3.1.61)$$

e si ottiene:

$$I_n = \left|\frac{2\omega_n}{\langle f_n, f_n\rangle}\right|\int_0^L |f_n(z)|^2 \rho(z)dz \qquad (6.3.1.62)$$

Poi, si inserisce nella (6.3.1.62) le definizione di norma:

$$\langle f_n, f_n \rangle = 2\omega_n \int_0^L f_n^2(z)\rho(z)dz + i\sqrt{\rho_0}[f_n^2(L) + f_n^2(0)] \qquad (6.3.1.63)$$

e si ottiene:

$$I_n = \frac{\int_0^L |f_n(z)|^2 \rho(z)dz}{\left|\int_0^L f_n^2(z)\rho(z)dz + i\frac{\sqrt{\rho_0}}{2\omega_n}[f_n^2(L) + f_n^2(0)]\right|} \qquad (6.3.1.64)$$

Si possono effettuare le maggiorazioni:

$$I_n \leq \frac{\int_0^L |f_n(z)|^2 \rho(z)dz}{\left|\int_0^L f_n^2(z)\rho(z)dz\right|} \leq 1 \qquad (6.3.1.65)$$

Le perdite sono quantificate dai coefficienti $I_n$; se le perdite sono minime:

$$|\mathrm{Im}\,\omega_n| \ll |\mathrm{Re}\,\omega_n| \qquad (6.3.1.66)$$

è lecita l'approssimazione:

$$I_n \cong 1 \qquad (6.3.1.67)$$

che inserita nella (6.3.1.56), dà luogo alla densità dei modi quale somma di lorenziane:

$$d(\omega) = \frac{K_d}{\pi} \sum_{n>0} \frac{|\mathrm{Im}\,\omega_n|}{(\omega - \mathrm{Re}\,\omega_n)^2 + \mathrm{Im}^2\,\omega_n} \qquad (6.3.1.68)$$

dove:

$$K_d = 2K\sqrt{\rho_0} \qquad (6.3.1.69)$$

### 6.3.1.d. Risultati DOM per un 1D-PBG simmetrico a λ/4.

La densità dei modi adimensionata $d_a = cd$, con argomento $\overline{\omega} = \omega/\omega_{rif}$, dove $\omega_{rif}$ è una frequenza di riferimento, ha espressione:

$$d_a(\overline{\omega}) = \frac{K_a}{\pi} \sum_n \frac{|\mathrm{Im}\,\overline{\omega}_n|}{(\overline{\omega} - \mathrm{Re}\,\overline{\omega}_n)^2 + \mathrm{Im}^2\,\overline{\omega}_n} \qquad (6.3.1.70)$$

dove si è introdotta la costante adimensionale:

$$K_a = \frac{K_d}{\omega_{rif}/c} \qquad (6.3.1.71)$$

Se il cammino ottico nella struttura è $L_{ott}$, si può definire la DOM normalizzata:

$$d_a^N(\overline{\omega}) = (L/L_{ott})d_a(\overline{\omega}) \qquad (6.3.1.72)$$

La costante $K_a$ deve essere tale che la DOM normalizzata $d_a^N(\overline{\omega})$ abbia un valore medio prossimo ad *1* nelle bande passanti.

Tutti i grafici seguenti si riferiscono ad un 1D-PBG simmetrico a *λ/4*, con parametri: $N=4$, $n_h = 1.5$, $n_l = 1$ e $\lambda_{rif} = 1\mu m$.

*Confronto con la DOM senza pompaggio*

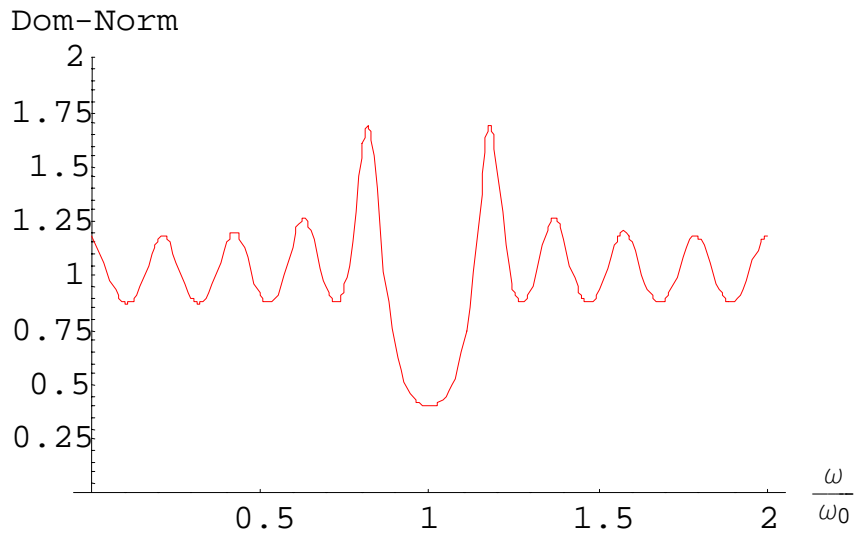

Grafico 6.2. DOM ad incidenza normale. Vedi equazione (6.3.1.70).

Il confronto è ancora in una fase di studio che non ha ottenuto risultati definitivi. Comunque si osserva la somiglianza tra la DOM in assenza di pompaggio (grafico 6.1.) e quella con pompaggio ad incidenza normale (grafico 6.2.).

*Confronto con la DOM TE per una pompa dal metodo delle matrici*

Tramite il metodo delle matrici [6], data una pompa polarizzata TE che incide sul 1D-PBG, si può determinare il coefficiente di trasmissione TE:

$$T(\omega) = x(\omega) + iy(\omega) \tag{6.3.1.72}$$

e quindi la DOM TE [2]:

$$d(\omega) = \frac{1}{L} \frac{x(\omega)y'(\omega) - x'(\omega)y(\omega)}{|T(\omega)|^2} \tag{6.3.1.73}$$

Al solito, con argomento $\bar{\omega} = \omega/\omega_{rif}$, si definisce la densità dei modi adimensionata $d_a = cd$ e si grafica la DOM normalizzata $d_a^N = (L/L_{ott})d_a$.

Si osserva che, per il 1D-PBG definito, si ottiene la stessa DOM ad incidenza normale, sia dal metodo delle matrici che dalla teoria QNM's.

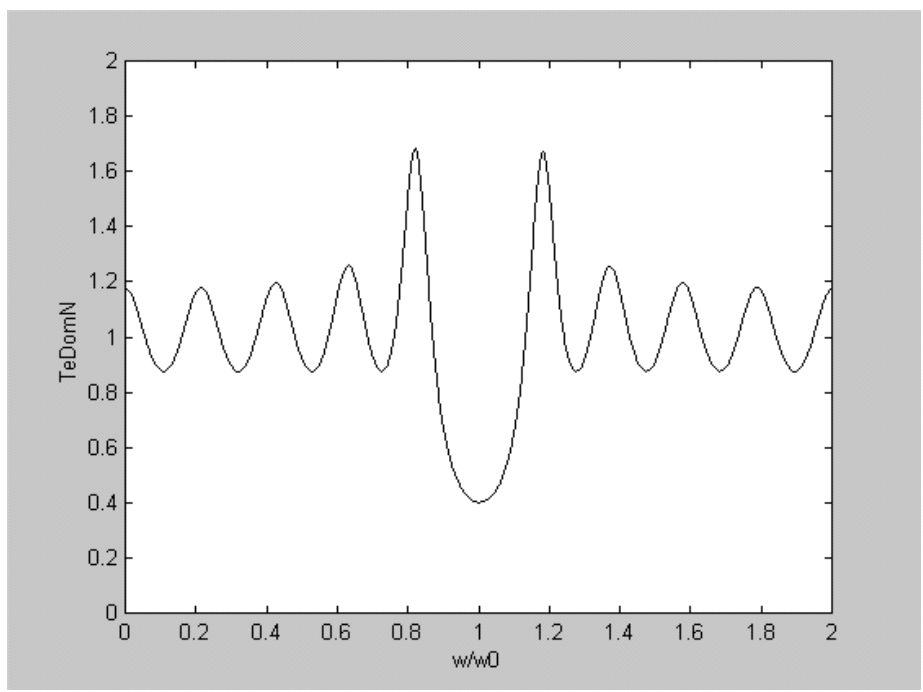

Grafico 6.3. DOM ad incidenza normale dal metodo delle matrici. Vedi equazione (6.3.1.73).

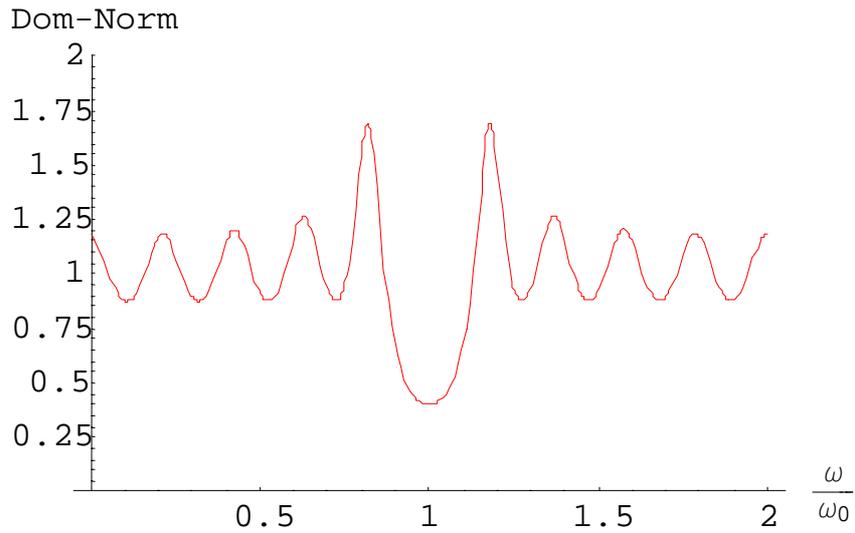

Grafico 6.4. DOM ad incidenza normale dalla teoria QNM's. Vedi equazione (6.3.1.70).

Si verifica, in approssimazione parassiale [5], condizione sotto cui è stata risolta l'equazione delle autofrequenze QNM's TE, che si ottiene, per il 1D-PBG definito, la stessa DOM TE, sia dal metodo delle matrici che dalla teoria QNM's.

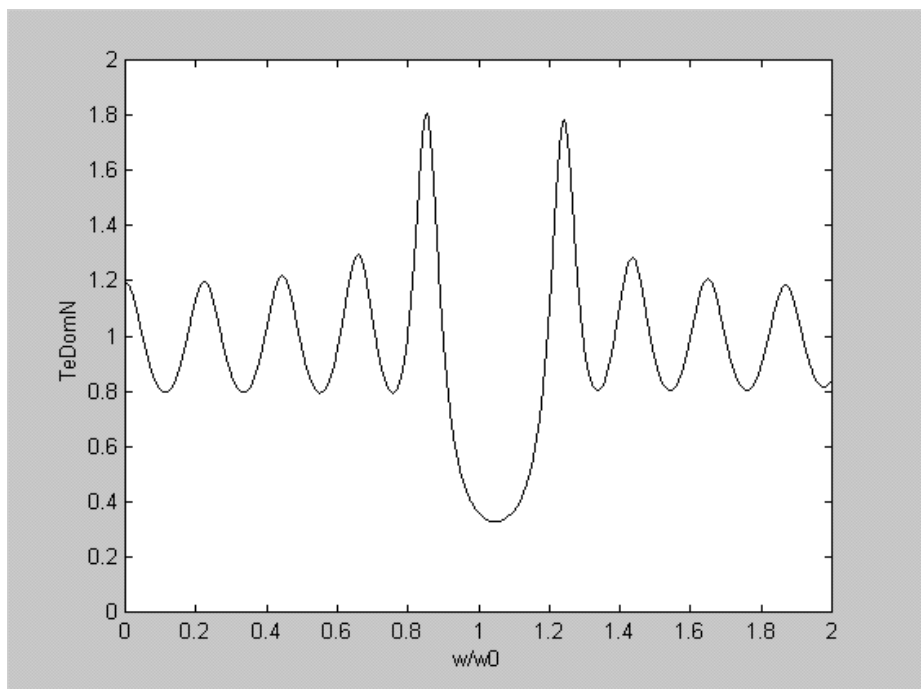

Grafico 6.5. DOM TE a 30 gradi dal metodo delle matrici. Vedi equazione (6.3.1.73).

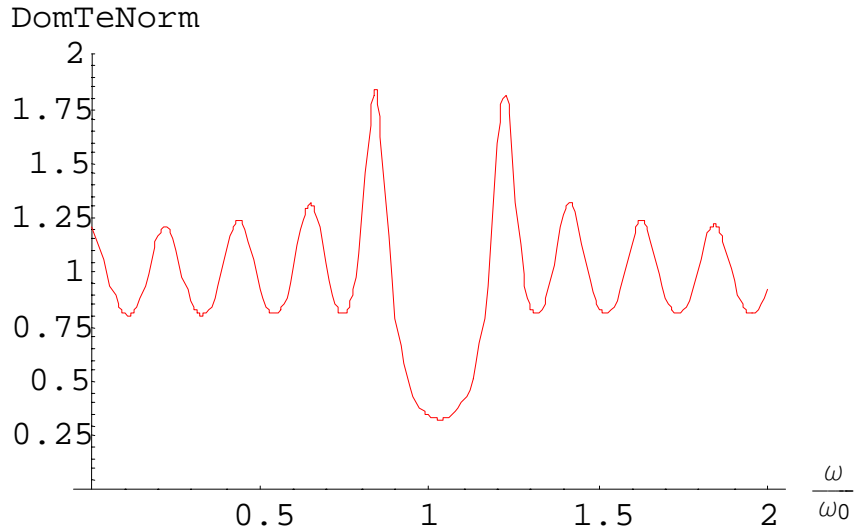

Grafico 6.6. DOM TE a 30 gradi dalla teoria QNM's. Vedi equazione (6.3.1.70).

## 6.3.2. Polarizzazione TM.

La pompa esterna, che incide obliquamente sulla struttura, abbia polarizzazione di tipo TM.

Procedendo come nel caso TE, si dimostra che, sotto le ipotesi:

a) frequenze di risonanza strette;
b) strutture simmetriche;
c) perdite minime;

la densità dei modi TM può essere espressa come somma di tante lorenziane quanti sono i QNMs TM, ciascuna con parametri forniti dalle parti reale ed immaginarie del QNM TM corrispondente:

$$d^{TM}(\omega) = \frac{K_d^{TM}}{\pi} \sum_n \frac{\left|\text{Im}\,\omega_n^{TM}\right|}{(\omega - \text{Re}\,\omega_n^{TM})^2 + \text{Im}^2\,\omega_n^{TM}} \qquad (6.3.2.1)$$

dove la costante $K_d^{TM}$ è di normalizzazione.

### 6.3.2.a. Risultati DOM per un 1D-PBG simmetrico a $\lambda_0/4$.

Come nel caso TE, introdotto l'argomento $\bar{\omega} = \omega/\omega_{rif}$, si definisce la densità dei modi adimensionata $d_a = cd$ e si grafica la DOM normalizzata $d_a^N = (L/L_{ott})d_a$.

La cavità aperta sia un 1D-PBG simmetrico a *λ/4*, con parametri: $N = 4$, $n_h = 1.5$, $n_l = 1$ e $\lambda_{rif} = 1\mu m$.

*Confronto con la DOM TM ad una pompa dal metodo delle matrici*

Tramite il metodo delle matrici, data una pompa polarizzata TM che incide sul 1D-PBG, si può determinare il coefficiente di trasmissione TM [6] e quindi la DOM TM [2].

Si verifica, in approssimazione parassiale [5], condizione sotto cui è stata risolta l'equazione delle autofrequenze QNM's TM, che si ottiene, per il 1D-PBG definito, la stessa DOM TM, sia dal metodo delle matrici che dalla teoria QNM's.

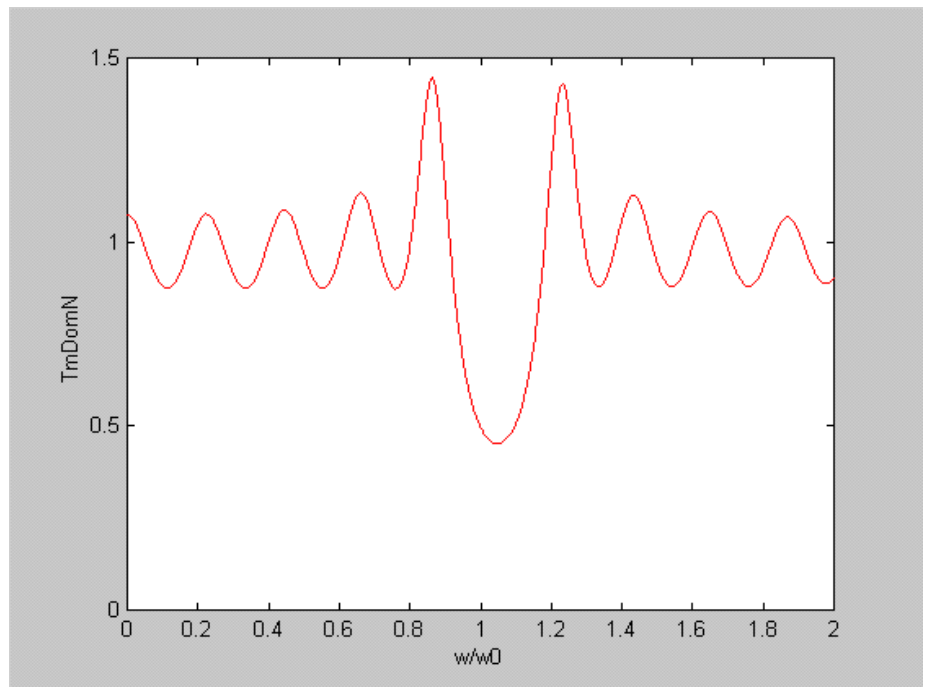

Grafico 6.7. DOM TM a 30 gradi dal metodo delle matrici. Vedi equazione (6.3.1.73).

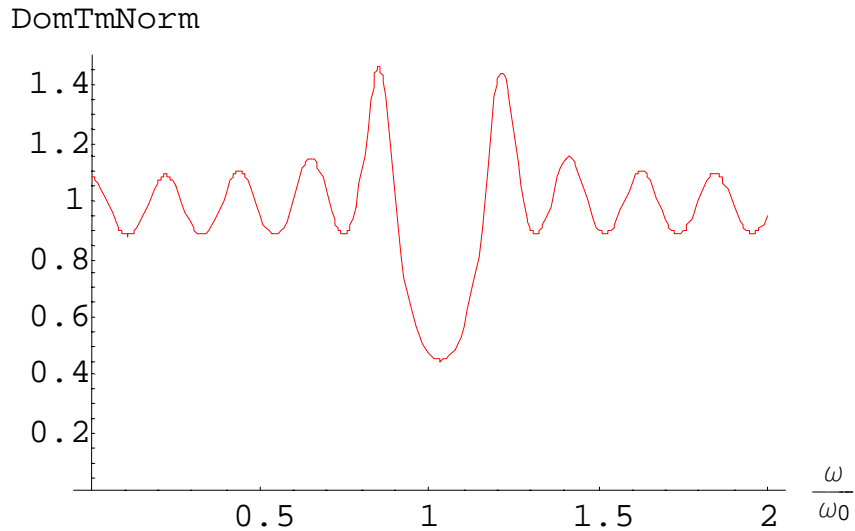

Grafico 6.8. DOM TM a 30 gradi dalla teoria QNM's. Vedi equazione (6.3.2.1).

## *Bibliografia*

## *Appendice C.*

In questa appendice, qualora un'onda piana monocromatica incida normalmente su un 1D-PBG simmetrico a quarto d'onda, si illustra come il modello dei QNM permette di prevedere su base teorica le frequenze di risonanza, con le relative larghezze di banda e la distribuzione di campo all'interno della cavità. Inoltre si riportano dei risultati, ottenuti tramite metodi numerici, quali lo spettro di trasmissione e il campo e.m. alle frequenze di bordo banda.

Con riferimento al paragrafo 3.3.2, se il 1D-PBG simmetrico è a quarto d'onda, nelle equazioni (3.3.2.5) va posto $m_h = m_l = 1$, per cui, se il 1D-PBG è costituito da N celle, l'equazione delle frequenze QNM (3.3.2.3-4) è polinomiale di grado $2N+1$. Le frequenze QNM di base sono $2N+1$, con parte immaginaria negativa $\text{Im}\,\omega_n < 0$, di cui una sull'asse immaginario $\text{Re}\,\omega_0 = 0$ e le altre $2N$, intervallate da un gap, sul semipiano destro $\text{Re}\,\omega_n > 0$. Escludendo la frequenza QNM sull'asse immaginario, che non può corrispondere ad una oscillazione, si è verificato che il gap tra i due gruppi di frequenze QNM nel semipiano destro corrisponde al gap nello spettro di trasmissione; che le altre $2N$ frequenze corrispondono ai picchi di risonanza nello spettro: nel senso che la $\text{Re}\,\omega_n$ coincide con il picco *n*-esimo, mentre la $|\text{Im}\,\omega_n|$ è una misura della larghezza di banda della risonanza *n*-esima.

Se la lunghezza d'onda di riferimento è $\lambda_{rif}$, la frequenza centrale del gap è $\omega_{rif} = 2\pi c / \lambda_{rif}$, avendo indicato con *c* la velocità della luce.

Le lunghezze d'onda per i picchi di risonanza sono fornite da:

$$\lambda_n = \frac{\lambda_{rif}}{\text{Re}(\omega_n/\omega_0)} \tag{C.1}$$

Mentre una misura in lunghezza d'onda per la larghezza di banda della risonanza *n*-esima è la seguente:

$$\Delta\lambda_n = \frac{\lambda_0}{\text{Re}^2(\omega_n/\omega_{rif})}\left|\text{Im}(\frac{\omega_n}{\omega_{rif}})\right| \tag{C.2}$$

Se il 1D-PBG simmetrico a quarto d'onda ha come parametri $\lambda_{rif} = 1\mu m$, $N = 10$, $n_h = 3$, $n_l = 2$, $n_0 = 1$, le lunghezze d'onda per i due bordi banda sono $\lambda_1 = 0.86876 \mu m$ e $\lambda_2 = 1.17795 \mu m$, mentre le rispettive larghezze di banda sono $\Delta\lambda_1 = 4.21438 nm$ e $\Delta\lambda_2 = 7.74794 nm$. Questi dati sono pressoché in accordo con quelli ottenuti tramite metodi numerici.

Ora, si fa riferimento al paragrafo 6.3.1.b., per evidenziare come, data un'onda piana monocromatica che incide normalmente su un 1D-PBG, la teoria dei QNM consente di determinare la distribuzione di campo elettrico dentro la cavità.

L'onda piana monocromatica, proveniente da sinistra, che incide normalmente sul 1D-PBG, ha espressione:

$$\widetilde{E}_p(z,\omega) = A e^{i n_0 \frac{\omega}{c} z} \tag{C.3}$$

per cui il termine di pompa (6.3.1.35) diviene:

$$\widetilde{b}(\omega) = -2i\left(\frac{n_0}{c}\right)^2 \omega \widetilde{E}_p(0,\omega) = -2i\left(\frac{n_0}{c}\right)^2 A \omega \tag{C.4}$$

Il campo elettrico dentro la cavità può essere sviluppato nelle autofunzioni dei QNM:

$$\widetilde{E}(z,\omega) = \sum_n \widetilde{a}_n(\omega) f_n^N(z) \tag{C.5}$$

dove ciascun coefficiente dello sviluppo (6.3.1.43) assume l'espressione:

$$\widetilde{a}_n(\omega) = \frac{f_n^N(0)}{2\omega_n(n_0/c)} \frac{\widetilde{b}(\omega)}{\omega - \omega_n} = -iA\left(\frac{n_0}{c}\right) f_n^N(0) \frac{\omega}{\omega_n(\omega - \omega_n)} \tag{C.6}$$

Se il 1D-PBG simmetrico a quarto d'onda ha come parametri $\lambda_{rif} = 1\mu m$, $N = 10$, $n_h = 3$, $n_l = 2$, $n_0 = 1$, quando l'incidenza è normale, si determinano le autofrequenze ed autofunzioni dei QNM e quindi il campo elettrico nella cavità per i vari picchi di trasmissione: a meno di errori puramente numerici, sono stati ottenuti gli stessi risultati del metodo basato sulla matrice di trasmissione.

Di seguito, si riportano i risultati della stessa applicazione, con parametri invariati, ottenuti tramite metodi numerici.

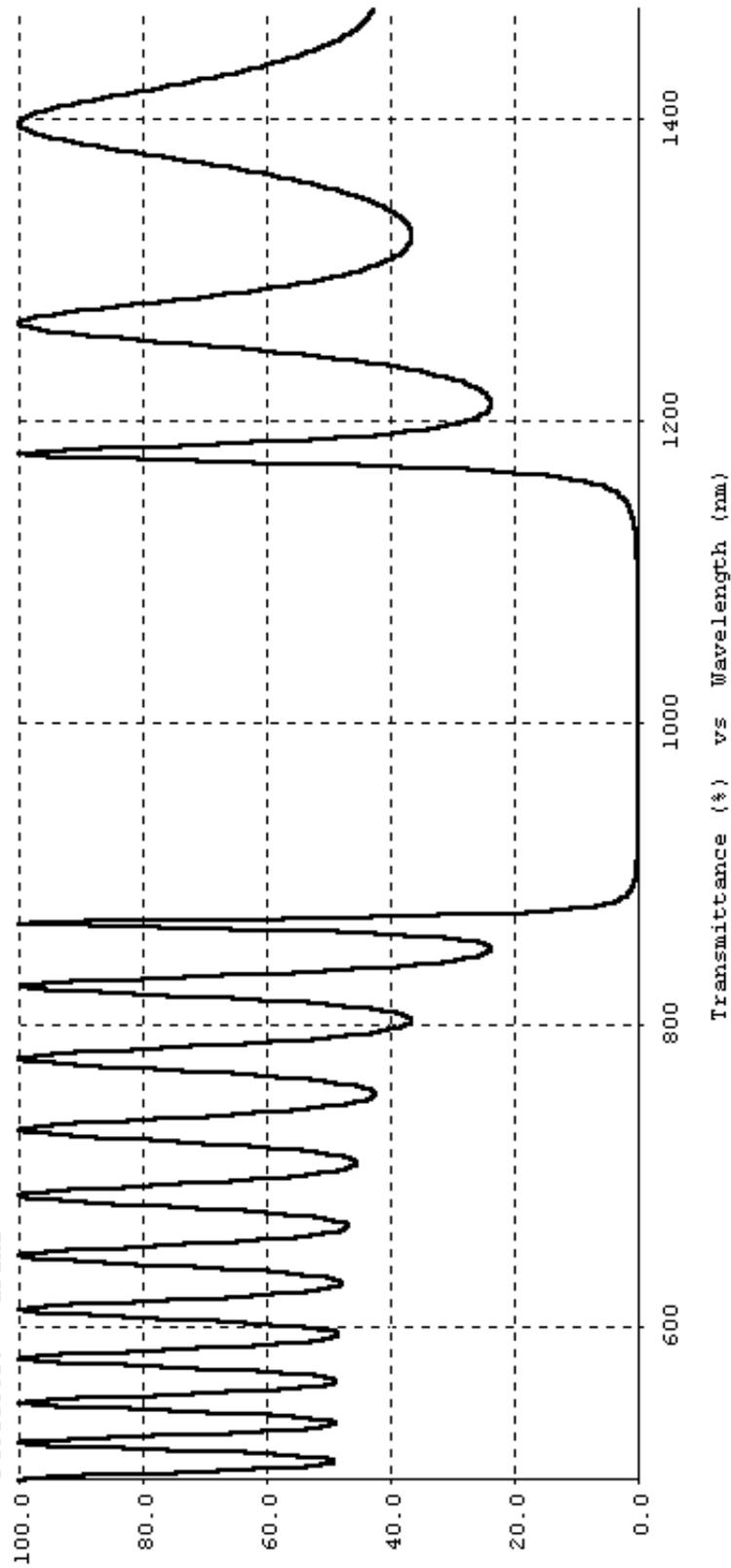

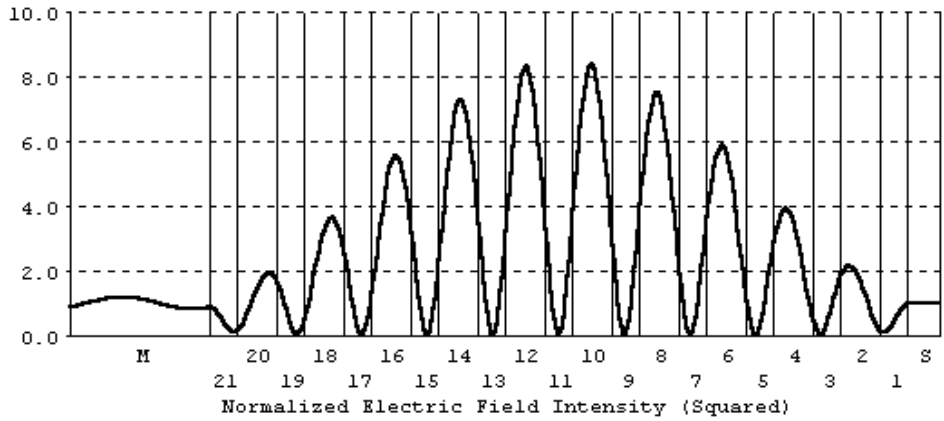

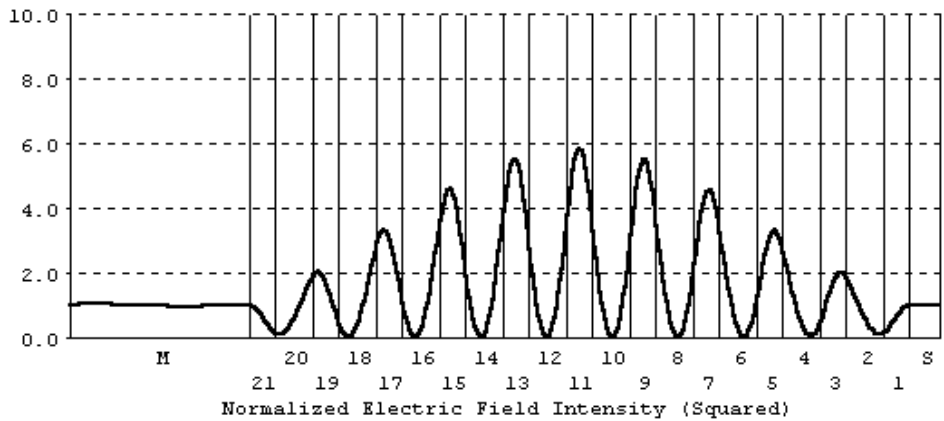

## *Conclusioni*

In questo lavoro di tesi, è stato studiato il comportamento del campo elettromagnetico, alle frequenze ottiche, nei cristalli fotonici unidimensionali, tramite la teoria dei "Quasi-Normal-Modes".

Le strutture unidimensionali a band-gap fotonico (1D-PBG) sono delle cavità ottiche, riempite di un mezzo stratificato con indice di rifrazione pseudo periodico e aperte verso lo spazio esterno. Non si è ritenuto possibile studiare i 1D-PBG tramite la teoria di Bloch che impone delle condizioni al contorno periodiche, come se la struttura fosse infinita; inoltre, si sono prese le distanze da altre metodologie che, quando non erano puramente numeriche, comunque sviluppavano il campo e.m. in un set di modi normali, e solo per le frequenze di risonanza, trattando la struttura come se fosse isolata dall'esterno.

Si è applicata la teoria dei QNM, secondo cui ogni cavità aperta viene trattata come una struttura finita e immersa in uno spazio illimitato. La rinuncia alla conservazione dell'energia per il sistema in esame comporta il passaggio da autofrequenze reali a quasi autofrequenze complesse. L'operatore del sistema non è hermitiano, quindi i modi non sono più normali, ma quasi normali. Infatti, utilizzando un metodo basato sulla funzione di Green, viene recuperata la completezza per i QNM, almeno all'interno della cavità, ma solo sotto condizioni opportune sull'indice di rifrazione all'estremità. Inoltre, tramite un formalismo che rappresenta ogni QNM come un vettore a due componenti, viene generalizzata, rispetto al caso conservativo, la definizione di prodotto interno, che risulta complesso e con uno o più termini aggiuntivi, così da recuperare anche l'ortogonalità per i QNM.

Si è svolto lo studio, suddividendolo in passi successivi. Dapprima, è stata definita la teoria dei QNM con riferimento alle cavità omogenee semiaperte; poi, nel caso che la radiazione incida solo normalmente sulla struttura, la stessa teoria è stata estesa alle cavità omogenee doppiamente aperte ed è stata applicata ai 1D-PBG; infine, trattando il caso di incidenza anche obliqua, sono stati caratterizzati i modi quasi normali, di entrambe le polarizzazioni TE e TM, nelle cavità omogenee e quindi nei 1D-PBG.

Si è pervenuti a dei risultati importanti. Sono state introdotte delle condizioni, chiamate "outgoing waves", che formalizzano le condizioni ai bordi per il campo e.m. e consentono di generalizzare la definizione di prodotto interno, così da recuperare l'ortogonalità dei QNM. Per quanto riguarda i 1D-PBG, si é sottolineato che le autofunzioni dei QNM non si annullano alle interfaccie, come invece accade nel caso conservativo; si è osservato che le autofrequenze dei QNM non sono distribuite con uniformità nel piano complesso, ma si dispongono secondo bande permesse e proibite, in accordo con le caratteristiche fisiche del mezzo: in particolare, si è verificato che i QNM hanno, come parti reali, le frequenze di risonanza e, come parti immaginarie, le ampiezze di oscillazione che si evincono dallo spettro di trasmissione. Sempre con riferimento ai 1D-PBG, si è utilizzata la teoria dei QNM per determinare il campo elettrico dentro la cavità e si è verificato che fosse quello effettivo tramite simulazioni. Infine, si è determinata la densità dei modi (DOM) sia in assenza di pompaggio che per una pompa. La DOM in assenza di pompaggio è un'esclusiva della teoria QNM's, in quanto non è stata ottenuta da altri metodi; è il primo passo per lo studio dell'emissione spontanea nella struttura. La DOM per una pompa può essere ottenuta anche tramite altri metodi; si è verificata la coincidenza dei risultati.